\documentclass[11pt]{article}
\pdfoutput=1

\usepackage[top = 1.0in, bottom = 1.0in, left = 1.0in, right = 1.0in]{geometry}

\usepackage{amsmath}
\usepackage{amssymb}
\usepackage{xspace}
\usepackage{theorem}
\usepackage{graphicx}
\usepackage{subfig}
\usepackage{epsfig}
\usepackage{ifpdf}
\usepackage{url,hyperref}
\usepackage{latexsym}
\usepackage{euscript}
\usepackage{xspace}
\usepackage{color}
\usepackage{makeidx}
\usepackage{picins}
\usepackage{wrapfig}
\usepackage{stackrel}
\usepackage{amscd}
\usepackage[all]{xy}
\usepackage{multirow}

\long\def\remove#1{}

\newtheorem{thm}{Theorem}

\newtheorem{lem}[thm]{Lemma}

\newtheorem{cor}[thm]{Corollary}

\theorembodyfont{\normalfont}
\newtheorem{definition}[thm]{Definition}
\newtheorem{hyp}[thm]{Hypothesis}
\newtheorem{remark}[thm]{Remark}
\newtheorem{observation}[thm]{Observation}

\newenvironment{proof}[1][{}]{
  \begin{trivlist}\item[]\textit{Proof #1}\quad}%
  {\hfill\hspace*{\fill}~$\square$\end{trivlist}}

\definecolor{turquoise}{cmyk}{0.65,0,0.1,0.1}

\newcommand{\rawdef}[1]{\emph{#1}} 
\newcommand{\defn}[1]{\rawdef{#1}\index{#1}}

\newcommand{\Lemref}[1]{Lemma~\ref{#1}}

\DeclareMathOperator{\argmin}{argmin}

\DeclareMathOperator{\convh}{conv}
\newcommand{\convhull}[1]{\convh(#1)}

\newcommand{\reel}{\mathbb{R}}
\newcommand{\R}{\mathbb{R}}

\newcommand{\rdee}{\reel^d}
\newcommand{\rem}{\reel^m}

\newcommand{\norm}[1]{\lVert#1\rVert}

\newcommand{\tanspace}[2]{T_{#1}{#2}} 
\newcommand{\normspace}[2]{N_{#1}{#2}} 

\newcommand{\simplex}[1]{[#1]} 

\newcommand{\seg}[2]{#1#2} 

\newcommand{\gdist}{d} 
\newcommand{\dist}[2]{\gdist(#1,#2)}

\DeclareMathOperator{\aff}{aff} 
\newcommand{\affhull}[1]{\aff(#1)}

\newcommand{\angleop}[2]{\angle(#1,#2)}

\newcommand{\pts}{\mathsf{P}}

\DeclareMathOperator{\vol}{vol}
\newcommand{\man}{\mathcal{M}}

\DeclareMathOperator{\vor}{Vor}

\DeclareMathOperator{\del}{Del}
\newcommand{\wit}{{\rm Wit}}

\newcommand{\st}{{\rm st}}

\newcommand{\pdelta}{\check{\delta}} 

\newcommand{\sing}[2]{s_{#1}(#2)}

\newcommand{\splxs}{\sigma}

\newcommand{\splxjoin}[2]{{#1}*{#2}}

\newcommand{\thickness}[1]{\Upsilon(#1)}

\newcommand{\splxalt}[2]{D(#1,#2)} 
\newcommand{\longedge}[1]{\Delta(#1)}

\newcommand{\reach}{\mathrm{rch}(\man)}

\newcommand{\e}{\varepsilon}
\newcommand{\M}{\man}

\newcommand{\sphere}[2]{S(#1,#2)} 
\newcommand{\glift}{\phi}
\newcommand{\lift}[1]{\glift(#1)}

\newcommand{\hplane}[2]{{\mathcal H}^{#1}_{#2}}

\newcommand{\sliverbnd}{\Gamma_{0}}

\usepackage{algorithmic}
\usepackage{algorithm}

\title{
Only distances are required to reconstruct submanifolds
}

\author{
Jean-Daniel Boissonnat
\footnote{
This research has been partially supported by the 7th Framework Programme for 
Research of the European Commission, under FET-Open grant number 255827 
(CGL Computational Geometry Learning) and by the European Research Council (ERC) 
under  the  European  Union's  Seventh  Framework  Programme  (FP/2007-2013) ERC  
Grant  Agreement No. 339025 GUDHI (Algorithmic Foundations of Geometry Understanding 
in Higher Dimensions).}\\
DataShape Group\\
INRIA Sophia Antipolis -- M\'{e}diterran\'{e}e, France\\
\url{Jean-Daniel.Boissonnat@inria.fr}
\and
Ramsay Dyer
\footnotemark[1]\\
DataShape Group\\
INRIA Sophia Antipolis -- M\'{e}diterran\'{e}e, France\\
\url{yasmar@gmail.com}
\and
\newline
Arijit Ghosh
\footnote{Arijit Ghosh is supported by Ramanujan Fellowship (No. SB/S2/RJN-064/2015).
Part of this work was done when Arijit Ghosh was a Researcher
at Max-Planck-Institute for Informatics, Germany supported by
the Indo-German Max Planck Center for Computer Science (IMPECS).
Part of this work was also done when Arijit Ghosh was a
Visiting Scientist
at Advanced Computing and Microelectronics Unit,
Indian Statistical Institute, Kolkata, India.
}\\
Advanced Computing and Microelectronic Unit\\
Indian Statistical Institute, Kolkata, India 
\and
Steve Y. Oudot
\footnotemark[1]\\
DataShape Group\\
INRIA Saclay -- \^{I}le-de-France, France\\
\url{Steve.Oudot@inria.fr}
}

\begin{document}

\maketitle

\begin{abstract}
  In this paper, we give the first algorithm that outputs a
  faithful reconstruction of a submanifold of Euclidean space without
  maintaining or even constructing complicated data structures such as
  Voronoi diagrams or Delaunay complexes. Our algorithm uses the
  witness complex and relies on the stability of {\em power protection},
  a notion introduced in this paper.
  The complexity of the algorithm depends exponentially on the
  intrinsic dimension of the manifold, rather than the dimension of
  ambient space, and linearly on the dimension of the ambient space.
  Another interesting feature of this work is that no explicit coordinates
  of the points in the point sample is needed.
  The algorithm only needs the {\em distance matrix} as input, i.e.,
  only distance between points in the point sample as input.

\end{abstract}



%

\pagenumbering{roman}

\clearpage

\tableofcontents

\clearpage

\pagenumbering{arabic}


\section{Introduction}
We present an algorithm for reconstructing a submanifold of Euclidean
space, from an input point sample,
that does not require Delaunay complexes, unlike previous
algorithms, which either had to maintain a subset of the Delaunay
complex in the ambient space~\cite{cheng2005,boissonnat2009}, or a
family of $m$-dimensional Delaunay
complexes~\cite{boissonnat2011tancplx}.  Maintaining these highly
structured data structures is challenging and in addition, the methods
are limited as they require explicit coordinates of the points in the
input point sample. One of the
goals of this work was to develop a procedure to reconstruct
submanifolds that only uses elementary data structures.


We use the witness complex to achieve this goal.  The witness complex
was introduced by Carlsson and de
Silva~\cite{wit-carlsson-silva2004}. Given a point cloud $W$, their
idea was to carefully select a subset $L$ of landmarks on top of which
the witness complex is built, and to use the remaining data
points to drive the complex construction. More precisely, a point
$w\in W$ is called a {\em witness} for a simplex $\sigma \in 2^L$ if
no point of $L \setminus \sigma$ is closer to $w$ than are the
vertices of $\sigma$, i.e., if there is a closed ball centered at $w$ that
includes the vertices of $\sigma$, but contains no other points of $L$
in its interior. The
witness complex is then the largest abstract simplicial complex that
can be assembled using only witnessed simplices. The geometric test
for being a witness can be viewed as a simplified version of the
classical Delaunay predicate, and its great advantage is to only
require mere comparisons of (squared) distances. As a result, witness
complexes can be built in arbitrary metric spaces, and the
construction time is bound to the size of the input point cloud rather
than to the dimension $d$ of the ambient space.

Since its introduction, the witness complex has attracted interest,
which can be explained by its close connection to the Delaunay
triangulation and the restricted Delaunay
complex~\cite{attali2007,boissonnat2009,Carlsson06onthe,chazalOtpbres08,wit-carlsson-silva2004,guibas2008}.
In his seminal paper~\cite{deSilva2008}, de Silva showed that the
witness complex is always a subcomplex of the Delaunay triangulation
$\del(L)$, provided that the data points lie in some Euclidean space
or more generally in some Riemannian manifold of constant sectional
curvature. With applications to reconstruction in mind, 
Attali, Edelsbrunner, and Mileyko~\cite{attali2007},
and Guibas and Oudot~\cite{guibas2008}
considered the case where the data points lie on or close to some
$m$-submanifold of $\R^d$. They showed that the witness complex is
equal to the restricted Delaunay complex when $m=1$, and a subset of
it when $m=2$. Unfortunately, the case of $3$-manifolds is once again
problematic, and it is now a well-known fact that the restricted
Delaunay and witness complexes may differ significantly (no respective
inclusion, different topological types, etc) when $m\geq
3$~\cite{boissonnat2009}. To overcome this issue, Boissonnat, Guibas
and Oudot~\cite{boissonnat2009} resorted to the sliver removal
technique on some superset of
the witness complex, whose construction incurs an exponential
dependence on $d$, the dimension of the ambient space.
The state of affairs as of now is that the
complexity of witness complex based manifold reconstruction is
exponential in $d$, and whether it could be made only polynomial in $d$
(while still exponential in $m$) was an open question, which this paper
answers affirmatively.

\subsection*{Our contributions}



Our paper builds on recent results on the stability of Delaunay
triangulations~\cite{boissonnat2012sdt} which we extend in the context
of Laguerre geometry where points are weighted. We introduce the
notion of power protection of Delaunay simplices and show that the
weighting mechanism already used in~\cite{cheng2000,cheng2005} and
\cite{boissonnat2009} can be adapted to our context. As a result, we
get an algorithm that constructs a (weighted) witness complex that is
a {\em faithful reconstruction}, i.e. homeomorphic and a close geometric
approximation,
of the manifold.  Differently from previous
reconstruction
algorithms~\cite{cheng2005,boissonnat2009,boissonnat2011tancplx},
our algorithm can be
simply adapted to work when we don't have explicit coordinates of the
points but  just the interpoint distance matrix.

\section{Definitions and preliminaries}
\label{sec:definitions-preliminaries}
\label{ssec-simplices}



\subsection{General notations}
\label{ssec-general-notations}

We will mainly work in $d$-dimensional Euclidean space $\R^{d}$ with
the standard $\ell_{2}$-norm, $\| \cdot \|$.  The distance between $p \in \R^{d}$
and a set $X \subset \R^{d}$, is
$$
	\dist{p}{X} = \inf_{x \in X} \|x-p\|.
$$
We refer to the
distance between two points $a$ and $b$ as $\| b-a \|$ or $\dist{a}{b}$ as convenient.

A ball $B(c, r) = \{ x : \dist{x}{c} < r\}$ is open,
and $\overline{B}(c, r)
= \{ x : \dist{x}{c} \leq r \}$ is closed.




Generally, we denote the convex hull of a set $X$
by $\convhull{X}$, and the affine hull by $\aff(X)$.
The cardinality of $X$, and not its measure, is denoted by $\# X$.
If $X \subseteq \R$, $\mu(X)$ denotes the
standard Lebesgue measure of $X$.


For given vectors $u$ and $v$ in $\R^{d}$, $\langle u,\, v \rangle$ denotes the
{\em Euclidean inner product} of the vectors $u$ and $v$.

For given $U$ and $V$ vector spaces of $\R^{d}$, with $\dim U \leq \dim V$, the
{\it angle} between them is defined by
$$
	\angle (U, V) = \max_{u \in U} \min_{v \in V} \angle (u, v).
$$
By angle between affine spaces, we mean the angle between
corresponding parallel vector subspaces.

The following result is a simple consequence of the above definition.
For a proof refer to~\cite{boissonnat2011tancplx}.

\begin{lem}\label{lem-affine-space}
  Let $U$ and $V$ be
vector subspaces
of $\R^{d}$ with
  $\dim(U) \leq \dim(V)$.
  \begin{enumerate}
  \item If $U^{\perp}$ and $V^{\perp}$ are the orthogonal complements
    of $U$ and $V$ in $\R^{d}$, then
    $\angle ({U}, {V}) = \angle ({V^{\perp}}, {U^{\perp}})$.
  \item If $\dim(U) = \dim(V)$ then $\angle (U, V) = \angle (V, U)$.
  \end{enumerate}
\end{lem}

Let $\sing{i}{A}$ denote the $i^{th}$ singular value of
matrix $A$. The singular values are non-negative and
ordered by decreasing order of magnitude.
The largest singular value $\sing{1}{A}$ is equal to the
norm $\| A \|$ of the matrix, i.e.,
$$
  \sing{1}{A} = \|A \|=\sup_{\norm{x}=1} \norm{Ax}.
$$

If $A$ is an $r \times c$ matrix, its smallest singular value is
$$
	\sing{j}{A} = \inf_{\norm{x}=1}\norm{Ax},\;\; \mbox{where} \;\; j=\min\{r,c\}.
$$
It is easy to see that:

\begin{lem}\label{lem-singular-value-1-m-matrix-inverse}
If $A$ is an invertible $j \times j$ matrix, then 
$$
	\sing{1}{A^{-1}} = \sing{j}{A}^{-1}.
$$
\end{lem}

\subsection{Simplices}
\label{ssec-simplices-definition-background}

Given a set of $j+1$ points $p_{0}, \, \dots, \, p_{j}$ in $\reel^{d}$,
a {\em $j$-simplex}, or just simplex, $\sigma = [p_{0}, \, \dots, \, p_{j}]$
denotes the set $\{ p_{0}, \, \dots, \, p_{j}\}$.
The points $p_{i}$ are called the {\em vertices} of $\sigma$ and $j$ denotes
the {\em combinatorial dimension} of the simplex
$\sigma$. Sometimes we will use
an additional superscript, like $\sigma^{j}$, to denote a $j$-simplex. A
simplex $\sigma^{j}$ is called {\em degenerate} if $j> \dim \aff(\sigma)$.


We will denote by $R(\sigma)$, $L(\sigma)$, $\Delta(\sigma)$ the lengths of the
smallest circumradius,
the smallest edge, and the longest edge of the simplex $\sigma$ respectively.
The circumcentre of the simplex
$\sigma$ will be denoted by $C(\sigma)$ and $N(\sigma)$ denotes the affine space,
passing through $C(\sigma)$ and
of dimension $d - \dim \aff (\sigma)$, orthogonal to $\aff (\sigma)$.

Any subset $\{p_{i_{0}}, \, \dots, \, p_{i_{k}}\}$ of $\{p_{0}, \, \dots, \, p_{j}\}$
defines a $k$-simplex which we call a {\em face} of $\sigma$.
We will write $\tau \leq \sigma$
if $\tau$ is a face of $\sigma$, and $\tau < \sigma$ if $\tau$ is a {\em proper face} of $\sigma$.


\begin{wrapfigure}{r}{0.50\textwidth}
  \vspace{-10pt}
  \begin{center}
    \includegraphics[width=0.450\textwidth]{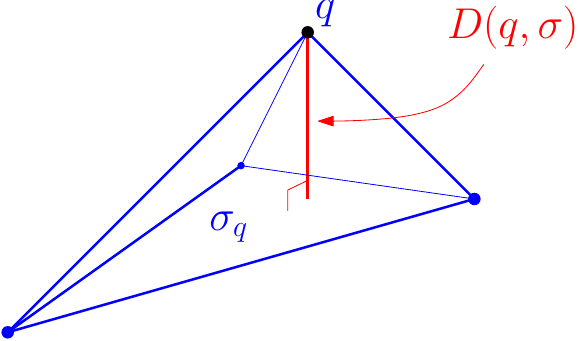}
  \end{center}
  \caption{Figure show altitude $D(q,\sigma)$ of the point $q$ in the simplex $\sigma$.}
  \label{fig-simplex-altitude-new}
\end{wrapfigure}

For a given vertex $p$
of $\sigma$, $\sigma_{p}$ denotes the subsimplex of $\sigma$ with the vertex set
$\{ p_{0}, \, \dots, \, p_{j}\} \setminus p$. If $\tau$ is a $j$-simplex, and $p$
is not a vertex of $\tau$, we can get a $(j+1)$-simplex $\sigma = p*\tau$,
called the {\em join} of $p$ and $\tau$. We will denote $\tau$ by $\sigma_{p}$
and will also write $\sigma = \sigma_{p} * p$.

The {\em altitude} of a vertex $p$ in $\sigma$ is $D(p, \sigma) =
\dist{p}{\aff(\sigma_{p})}$. See Figure \ref{fig-simplex-altitude-new}.
A poorly-shaped simplex can be characterized by the existence of a relatively small altitude.
The {\em thickness} of a $j$-simplex $\sigma$ of diameter $\Delta
(\sigma)$ is defined as
\begin{equation*}
   \thickness{\splxs} =
   \begin{cases}
     1& \text{if $j=0$} \\
     \min_{p \in \splxs} \frac{\splxalt{p}{\splxs}}{j
       \longedge{\splxs}}& \text{otherwise.}
   \end{cases}
\end{equation*}

Boissonnat, Dyer and Ghosh~\cite{boissonnat2012sdt} connected the geometric
properties of a simplex to the largest and smallest singular values
of the associated matrix:
\begin{lem}[Thickness and singular value~\cite{boissonnat2012sdt}]
  \label{lem:bound.skP}
  Let $\splxs = \simplex{p_0, \ldots, p_j}$ be a non-de\-generate
  $j$-simplex in $\rem$, with $j>0$, and let $P$ be the $m \times j$
  matrix whose $i^\text{th}$ column is $p_i - p_0$. Then
  \begin{enumerate}
    \item
        $\sing{1}{P}\leq \sqrt{j} \Delta(\sigma)$, and

    \item
        $\sing{j}{P} \geq \sqrt{j} \thickness{\splxs}\longedge{\splxs}$.
  \end{enumerate}
\end{lem}

A simplex that is not thick has a relatively small
altitude, but we want to characterize
bad simplices for which
\emph{all} the altitudes are relatively small.
This motivates the definition of  $\sliverbnd$-slivers.
\begin{definition}[$\sliverbnd$-good simplices and $\sliverbnd$-slivers]
  \label{def:slivers}
  \label{def:good.simplex}
 Let $\sliverbnd$ be a positive real
number smaller than one.
 A simplex $\splxs$ is \defn{$\sliverbnd$-good} if
  $\thickness{\splxs^j} \geq \sliverbnd^j$ for all $j$-simplices
  $\splxs^j \leq \splxs$. A simplex is \defn{$\sliverbnd$-bad} if it
  is not $\sliverbnd$-good. A \defn{$\sliverbnd$-sliver} is
  a $\sliverbnd$-bad simplex in which all the proper faces are
  $\sliverbnd$-good.
\end{definition}

\begin{remark}[On the good and bad simplex definitions]
\label{remark-good-bad-simplex}
    \begin{enumerate}
        \item
            Observe that in the definition of \defn{$\sliverbnd$-good} simplex
            the thickness bound goes down exponentially with dimensions. Ideally,
            one would like to have the thickness bound to be independent of the
            dimension of the simplex. We have defined it this way
            because with the current sliver removal
            technology we cannot guarantee
            the output triangulation to have
            thickness lower bound that is independent of the dimension
            of the simplices in the triangulation.

        \item
            Observe that a sliver must have dimension at least $2$, since
            $\thickness{\splxs^j} = 1$ for $j < 2$.  Observe also that our
            definition departs from the standard one since the slivers we
            consider have no upper bound on their circumradius, and in fact
            may be degenerate and not even have a circumradius. Also,
            observe that for a fixed $\sliverbnd$ we say a simplex $\splxs$
            is good if $\thickness{\splxs^j} \geq \sliverbnd^j$ for all
            $j$-simplices $\splxs^j \leq \splxs$.
    \end{enumerate}

\end{remark}

Ensuring that all simplices are $\sliverbnd$-good is the same as
ensuring that there are no slivers.
Indeed, if $\splxs$ is $\sliverbnd$-bad, then it has a $j$-face
$\splxs^j$
 that is not $\sliverbnd^j$-thick. By considering such a
face with minimal dimension we arrive at the following important
observation:
\begin{lem}
  \label{lem:bad.has.sliver}
  A simplex is $\sliverbnd$-bad if and only if it has a face that is a
  $\sliverbnd$-sliver.
\end{lem}


%

\subsection{Weighted points and weighted Delaunay complex}
\label{ssec-weighted-points-weighted-Delaunay-complex}

For a finite set of points $L$ in $\R^{d}$,  a weight assignment of $L$
is a non-negative real function from $L$ to $[0, \infty)$, i.e.,
${\varpi}: L \rightarrow [0, \infty)$.
A pair $(p, {\varpi}(p))$,  $p \in L$, is called a {\em weighted point}.
For simplicity, we  denote the weighted point $(p, {\varpi}(p))$ as $p^{{\varpi}}$.
The {\em relative amplitude} of ${\varpi}$ is defined as
\begin{equation}
\label{eq:relative.amplitude}
  \widetilde{{\varpi}} = \max_{p \in L}
  \max_{q\in L\setminus p} \frac{{\varpi}(p)}{\|p-q\|}.
\end{equation}

Given a point $x \in \R^{d}$, the {\em weighted distance} of $x$ from a weighted point
$(p, {\varpi}(p))$ is defined as
$$
  d(x, p^{{\varpi}}) = \|x-p\|^{2} - {\varpi}(p)^{2}.
$$

We say a sphere $S(c,r)$ is
{\em orthogonal} to $p^{{\varpi}} = (p, {\varpi}(p))$ if
$d(c,p^{{\varpi}}) = r^2$, i.e., if
$$
	\| p - c\|^{2} = {\varpi}(p)^{2} + r^{2}.
$$

For a simplex $\sigma = [p_{0}, \, \dots, \, p_{k}]$ with vertices in $L$
and ${\varpi} : L \rightarrow [0, \infty)$ a weight assignment,
we define the {\em ${\varpi}$-weighted normal space}, or just weighted normal space,
$N_{{\varpi}}(\sigma)$ of $\sigma$ as
$$
  N_{{\varpi}}(\sigma) = \left\{ x \in \reel^{d} :  \;
  d(x, p_{i}^{{\varpi}}) = d(x, p_{j}^{{\varpi}}),\,
  \forall \, p_{i},\, p_{j} \in \sigma \right\}.
$$
We call $S(c, r)$
a {\em ${\varpi}$-ortho sphere}, or just an {\em ortho sphere}, of $\sigma$ if it is orthogonal to the vertices of $\sigma$, i.e., if
for all $p_{i} \in \sigma$, we have
$r^{2} = d(c, p_{i}^{{\varpi}})$. Every $c \in N_\varpi(\sigma)$ is the center of an ortho sphere $S(c,r)$ with $r^2 = d(c,p_0^\varpi)$, and conversely, every ortho sphere is centered in $N_\varpi(\sigma)$.

We define the {\em ${\varpi}$-weighted} (or just weighted) center of $\sigma$
as
$$
  C_{{\varpi}}(\sigma) = \argmin_{x \in N_{{\varpi}}(\sigma)} d(x, p_{0}^{{\varpi}}).
$$
$N_{{\varpi}}(\sigma)$ is an
orthogonal compliment of $\aff(\sigma)$  intersecting $\aff(\sigma)$ at $C_{{\varpi}}(\sigma)$.

The {\em ${\varpi}$-weighted} (or just weighted) ortho-radius of $\sigma$ is defined
by
$$
  R_{{\varpi}}(\sigma)^{2} = d(C_{{\varpi}}(\sigma), p_{0}^{{\varpi}}).
$$
Note that weighted othro-radius $R_{\varpi}(\sigma)^{2}$ can be negative, i.e.,
$R_{\varpi}(\sigma)^{2} < 0$.

%
%
%


For a point $p \in L$ we define
the {\em weighted Voronoi cell} $\vor_{{\varpi}}(p)$ of $p$ as
$$
  \vor_{{\varpi}}(p) = \{ x \in \R^{d} :
  \forall~q\in L\setminus p,~d(x, p^{{\varpi}})\leq d(x, q^{{\varpi}})\}.
$$
\begin{wrapfigure}{r}{0.50\textwidth}
  \vspace{-80pt}
  \begin{center}
    \includegraphics[width=0.450\textwidth]{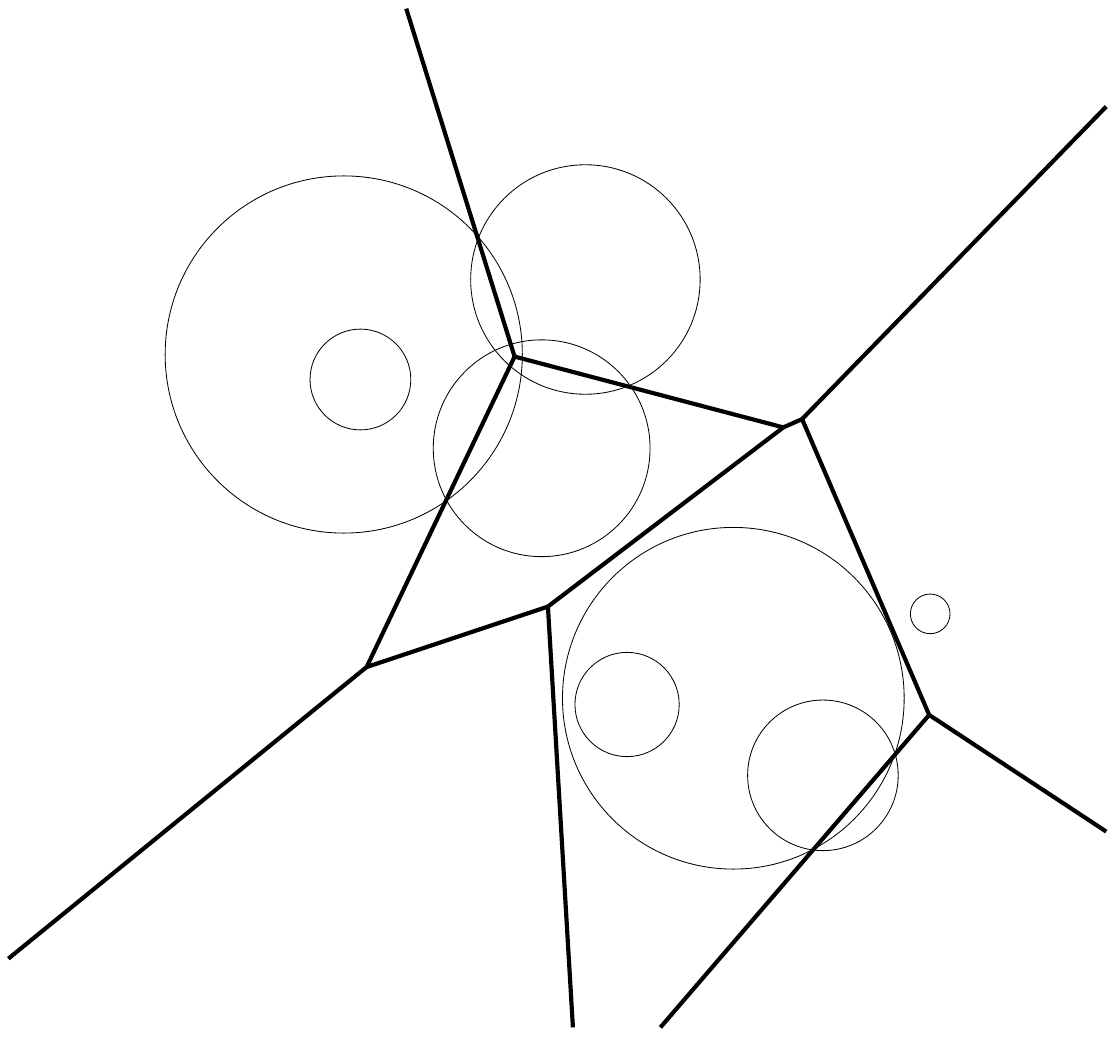}
  \end{center}
  \caption{The figure shows the weighted Voronoi diagram
  of weighted points, denoted by circles with centered at the points and
  radii equal to the weight of the points, in the plane.}
  \label{fig:powerdiagram-new}
\end{wrapfigure}

For a
simplex
$\sigma = [p_{0},\, \dots, \, p_{k}]$ with vertices in $L$,
the {\em weighted Voronoi face}
$\vor_{{\varpi}}(\sigma)$ of $\sigma$ is defined as
$$
  \vor_{{\varpi}}(\sigma) = \bigcap_{i=0}^{k} \vor_{{\varpi}}(p_{i}).
$$
Observe that the Voronoi faces are convex. We define
$\dim\vor_\varpi(\sigma)$ to be the dimension of
$\affhull{\vor_\varpi(\sigma)}$.

The weighted Voronoi cells give a decomposition of $\R^d$, denoted
$\vor_{{\varpi}}(L)$, called the {\em weighted Voronoi diagram} of $L$
corresponding to the weight assignment ${\varpi}$. 
See Figure \ref{fig:powerdiagram-new}.
Let $c \in \vor_{{\varpi}}(\sigma)$
and $r^{2} = d(c, p_{i}^{{\varpi}})$ where $p_{i} \in {\sigma}$. We
will call a $S(c, r)$ {\em ${\varpi}$-ortho Delaunay sphere}, or just
{\em Delaunay sphere}, of $\sigma$.

The {\em weighted Delaunay complex} $\del_{{\varpi}}(L)$ is defined as the
{\em nerve} of $\vor_{{\varpi}}(L)$, i.e.,
$$
  \sigma \in \del_{{\varpi}}(L) \; \; \mbox{iff} \; \; \vor_{{\varpi}}(\sigma) \neq \emptyset.
$$

\subsection{Manifolds and reach}
\label{ssec-manifolds-reach}


\begin{figure}[h]
  \begin{center}
    \includegraphics[width=2.750in]{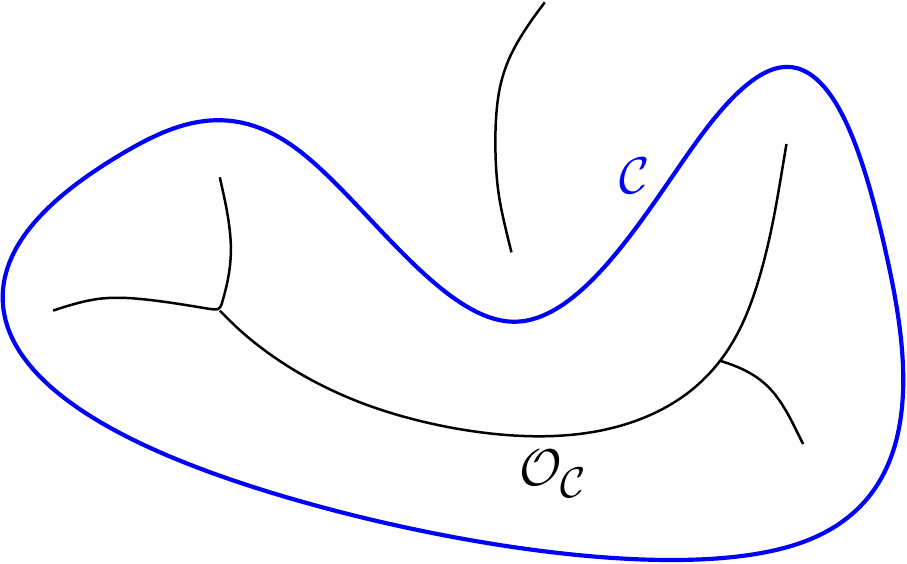}
  \end{center}
  \caption{The figure shows the medial axis ${\mathcal O}_{{\mathcal C}}$,
  drawn in black, of the blue curve ${\mathcal C}$.}
  \label{figure:manifold-medialaxis-new}
\end{figure}

For a given compact submanifold $\M$ of $\R^{d}$, the {\em medial axis} $\mathcal{O}_{\M}$ of
$\M$ is defined as the closure of the set of points in $\R^{d}$ that have
more than one closest points in $\M$. 
See Figure \ref{figure:manifold-medialaxis-new}.
The {\em reach} of
$\M$ is defined as
$$
  \reach = \inf_{x \in \M} \dist{x}{\mathcal{O}_{\M}}.
$$
Federer~\cite{federer} proved that $\reach$ is (strictly) positive
when $\M$ is of class $C^2$ or even $C^{1,1}$, i.e. the normal
bundle is defined everywhere on $\M$ and is Lipschitz continuous.
For simplicity, we are anyway assuming that $\M$ is a smooth
compact submanifold.

$T_{p}\M$ and $N_{p}\M$ denote
the tangent space and normal space at $p \in \M$.
%
We will use the following results
from~\cite{federer,giesen2004,boissonnat2012con.intrinsic}.
See~\cite[Lem.~6 \& 7]{giesen2004}
and~\cite[Lem.~B.3]{boissonnat2012con.intrinsic}.

\begin{figure}[h]
  \vspace{10pt}
  \begin{center}
    \includegraphics[width=4.50in]{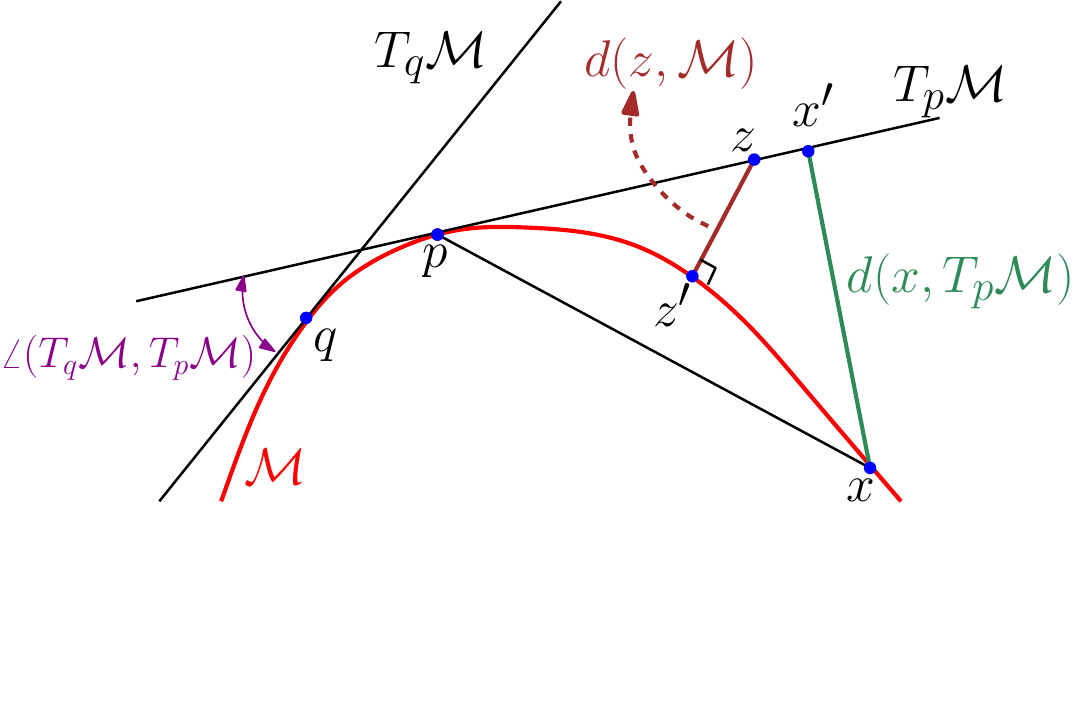}
  \end{center}
  \vspace{-40pt}
  \caption{Diagram for the Lemma~\ref{lem-sampling-property-manifold}.}
  \label{figure-manifold-properties-chord-angle-tangent-new}
\end{figure}

\begin{lem}\label{lem-sampling-property-manifold}
    Let $p$ be a point on the manifold $\M$.
    \begin{enumerate}
        \item
            If $x \in \M$ and $\|p-x\| < \reach$, then
            $\sin \angle (px,T_{p}\M) \leq \frac{\|p-x\|}{2\reach}$.
        \item
        		If $z \in T_{p}\M$ and $\|p-z\| < \frac{\reach}{4}$ then
            $\dist{z}{\M} \leq \frac{2\|p-z\|^{2}}{\reach}$.

        \item
        		If $q \in \M$ and $\|p-q\| < \frac{\reach}{4}$, then
            $\sin \angle(T_{p}\M, T_{q}\M) < \frac{6\|p-q\|}{\reach}$.
    \end{enumerate}
    See Figure \ref{figure-manifold-properties-chord-angle-tangent-new}.
\end{lem}

%

The following structural result is a restricted version\footnote{
Boissonnat, Guibas and Oudot~\cite[Lem.~4.4]{boissonnat2009}
proved a more general result
bounding the distance between $x$ and its
$(k+1)$-nearest weighted neighbor for all $k \leq d$. In
Lemma~\ref{lem-distance-point-on-M-with-sample}~(2) we only stated
the special case when $k \in \{0,\, 1 \}$.
}
of a result due to Boissonnat, Guibas and
Oudot~\cite[Lem.~4.3 \& 4.4]{boissonnat2009}.

\begin{lem}\label{lem-distance-point-on-M-with-sample}
    Let $L\subseteq \M$ be a $\epsilon$-sample of $\M$ with $\epsilon < \reach$,
    and ${\varpi} : L \rightarrow [0, \infty)$
    be a weight assignment with $\widetilde{{\varpi}} < \frac{1}{2}$.
    \begin{enumerate}
        \item
            For all $p \in L$, ${\varpi}(p) \leq 2 \widetilde{{\varpi}}\epsilon$.

        \item
            If
$\epsilon \leq \frac{\reach}{4}$,
	    then, for all $x \in \M$ and $k \in \{0,\, 1\}$,
            the Euclidean distance between $x$ and its $(k+1)$-nearest weighted
            neighbor in $L$ is at most $(1+2\widetilde{{\varpi}}+2k(1+3\widetilde{{\varpi}}))\epsilon$.

    \end{enumerate}
\end{lem}

The following result, due to~\cite{boissonnat2012csdt},
bounds the angle between the affine plane of a simplex with vertices on the
manifold $\M$ and the tangent planes to the manifold $\M$
at the vertices of the simplex.


\begin{cor}\label{cor-angle-bound-simplex-tangent-space}
    Let $\sigma$ be a $k$-simplex with $k \leq m$ and the vertices of $\sigma$
    are on the submanifold $\M$ of dimension $m$.
    If $\sigma$ is $\sliverbnd$-good and
    $\Delta(\sigma) <  \reach$, then for all $p \in \sigma$ we have
    $$
        \sin \angle (\aff(\sigma), T_{p}\M) \leq \frac{\Delta(\sigma)}{\Gamma^{m}_{0} \reach}.
    $$
\end{cor}

\subsection{Witness, cocone and tangential complex}

\begin{figure}
  \vspace{-100pt}
  \begin{minipage}[b]{0.5\linewidth}
    \centering
    \includegraphics[width=1.0\linewidth]{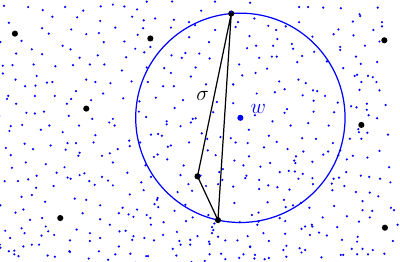}\\
    \vspace{10pt}
    {(a)}
  \end{minipage}
  \hspace{-30pt}
  \begin{minipage}[b]{0.5\linewidth}
    \centering
    \includegraphics[width=1.250\linewidth]{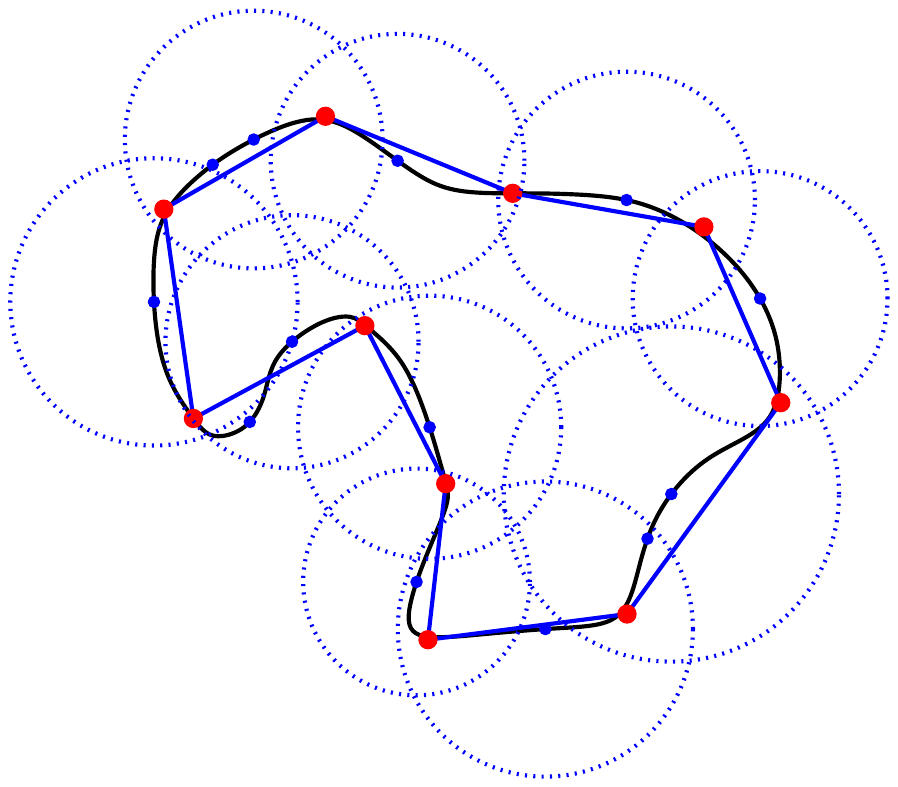}\\
    \vspace{-110pt}
    {(b)}
  \end{minipage}
  \caption{ (a) Blue points are the witnesses $W$ and the black points
    are the landmarks $L$.  All the landmarks in this example are
    assigned ``zero" weights.  The simplex $\sigma$ is witnessed by
    $w$. Note that the figure is taken from the
    paper~\cite{boissonnat2015esa}.\newline (b) In the figure witness
    set (blue points) and the landmark set (red) are sampled from the
    black curve. The landmarks are assigned ``zero" weight.  The
    one-dimensional complex (drawn in blue) is the witness complex
    approximating the black curve }
  \label{fig-witness-complex}
\end{figure}

We now recall the definition of the weighted {\em witness complex}
introduced by de Silva~\cite{deSilva2008}. Let $W \subset \R^{d}$,
and let $L \subseteq W$ be a finite set, and ${\varpi}: L \rightarrow [0, \infty)$
be a weight assignment of $L$. The points in the set $W$ are called {\em witnesses}
and the points in $L$ are called {\em landmarks}.
\begin{itemize}
    \item
	We say $w \in W$ is a {\em ${\varpi}$-witness} of a simplex
	$\sigma = [p_{0}, \, \dots, \, p_{k}]$ with vertices in $L$, if
	the $p_{0}, \, \dots,\, p_{k}$ are among the $k+1$ nearest neighbors
	of $w$ in the weighted distance,
	i.e., $p \in \sigma$, $q \in L \setminus \sigma$,
        $d(w, p^{{\varpi}}) \leq d(w, q^{{\varpi}})$. See
        Figure~\ref{fig-witness-complex}~(a).


    \item
        The {\em ${\varpi}$-witness complex} $\wit_{{\varpi}}(L, W)$
        is the maximum abstract simplicial
        complex with vertices in $L$, whose faces are ${\varpi}$-witnessed by points of $W$.
        When there is no ambiguity, we will call $\wit_{{\varpi}}(L, W)$ just witness complex
        for simplicity. See Figure~\ref{fig-witness-complex}~(b).

\end{itemize}


%
%
%
%


For any  point $p$ on a smooth submanifold $\M$ and
$\theta \in [0, \frac{\pi}{2}]$, we call the $\theta$-{\em cocone}
of $\M$ at $p$, or ${\rm K}^{\theta_{0}}(p)$ for short, the cocone of
semi-aperture $\theta$
around the tangent space $T_{p}\M$ of $\M$ at $p$:
$$
  {\rm K}^{\theta}(p) = \left\{ x \in \R^{d}: \angle (px, T_{p}\M) \leq \theta \right\}.
$$



Given an angle $\theta \in [0, \frac{\pi}{2}]$, a finite point set $\pts \subset \M$,
and a weight assignment ${\varpi} : \pts \rightarrow [0, \infty)$,
the {\em weighted $\theta$-cocone complex} of $\pts$, denoted by
${\rm K}^{\theta}_{{\varpi}}(\pts)$, is defined as
\begin{equation}
\label{eq:cocone.complex}
  {\rm K}^{\theta}_{{\varpi}}(\pts) =
  \left\{ \sigma\in \del_{{\varpi}}(\pts) : \vor_{{\varpi}}(\sigma)
  \cap \left( \bigcup_{p \in \sigma} {\rm K}^{\theta}(p) \right)
  \neq \emptyset\right\}.
\end{equation}
The cocone complex was first introduced by
Amenta, Choi, Dey and Leekha~\cite{amenta2002} in $\R^{3}$
for reconstructing surfaces and was later generalized by
Cheng, Dey and Ramos~\cite{cheng2005} for reconstructing submanifolds.

The {\em weighted tangential complex}, or just {tangential complex}, of $\pts$ is the
weighted $\theta$-cocone complex ${\rm K}^{\theta}_{{\varpi}}(\pts)$ with $\theta$
equal to ``zero" and will be denoted by $\del_{{\varpi}}(\pts, T\M)$.
The
Tangential complex
was first defined by
Boissonnat and Fl{\"{o}}totto~\cite{DBLP:journals/cad/BoissonnatF04}
for getting a coordinate
system from a point set sampled from a surface.
Boissonnat and Ghosh~\cite{boissonnat2011tancplx} later extended the definition to the
weighted setting and using the
weighted tangential complex they gave the first manifold reconstruction algorithm
whose time complexity depends linearly on the ambient dimension.

\begin{hyp}
  For the rest of this paper, we take
  \begin{eqnarray*}
    \theta_{0} &\stackrel{\rm def}{=}& \frac{\pi}{32}, \\
    {\rm K}(p) &\stackrel{\rm def}{=}& {\rm K}^{\theta_{0}}(p),\; \mbox{and}\\
    {\rm K}_{{\varpi}}(\pts) &\stackrel{\rm def}{=}& {\rm K}^{\theta_{0}}_{{\varpi}}(\pts).
  \end{eqnarray*}
\end{hyp}

%
%
%
%
%



\section{Power protection}
\label{ssec-power-protection}

\begin{figure}[h]
  \begin{center}
    \includegraphics[width=2.50in]{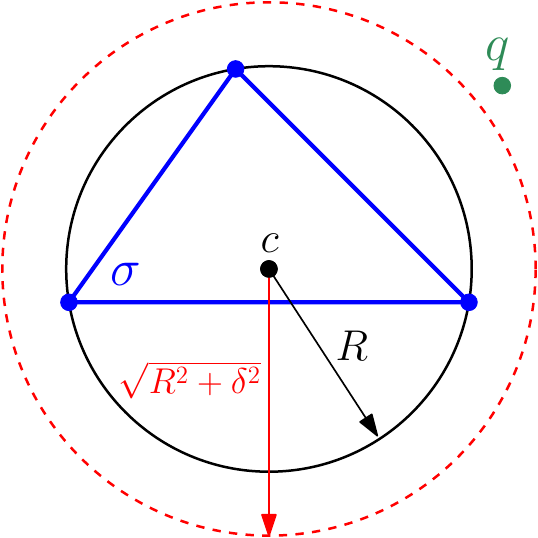}
  \end{center}
  \caption{Example of $\delta^{2}$-power protection when points
  are not weighted, i.e., the points have ``zero" weight.}
  \label{figure:shows-example-of-power-protection-new}
\end{figure}

Let $\pts \subset \R^{d}$ be a finite point sample.
A simplex $\sigma \in \del_{{\varpi}}(\pts)$ is {\em $\delta^{2}$-power
  protected}  at
$c \in \vor_{{\varpi}}(\sigma)$
if
$$
    \|q-c\|^{2} - {\varpi}(q)^{2} > \|p-c\|^{2}-{\varpi}(p)^{2} + \delta^{2}
\quad
\text{for all }q \in L \setminus \sigma \text{ and }p \in \sigma.
$$
For convenience,
we will say a simplex $\sigma \in \del_{{\varpi}}(\pts)$ is {\em $\delta^{2}$-power
protected} if $\exists\, c \in \vor_{{\varpi}}(\sigma)$ such that
$\sigma$ is {\em $\delta^{2}$-power protected}  at $c$.
See Figure \ref{figure:shows-example-of-power-protection-new}.

The following result shows that power protecting
$d$-simplices implies power protecting
lower dimensional subsimplices as well.
\begin{lem}\label{thm-protection-lower-dim-simplices}
    	Let $\pts \subset \R^{d}$ be a set of points, and let
	${\varpi}: \pts \rightarrow [0, \, \infty)$ be a weight
        assignment. In addition, let $p$ be a point of $\pts$  whose
        Voronoi cell
	$\vor_{{\varpi}}(p)$ is bounded. Then, if all the $d$-simplices
	incident to $p$ in $\del_{{\varpi}}(\pts)$
        are $\delta^{2}$-power
    	protected, with $\delta >0$, then
        any $j$-simplex in $\del_{{\varpi}}(\pts)$ incident to $p$ is
	$$
	  \frac{\delta^{2}}{d-j+1}\mbox{-power protected.}
	$$
\end{lem}

The proof of the Lemma~\ref{thm-protection-lower-dim-simplices}
is done in the lifted $\R^{d+1}$
space where power protection translates to {\em vertical distance} of
points from hyperplanes, see
Section~\ref{ssec-power-protection-space-of-spheres-framework}.

\begin{wrapfigure}{r}{0.50\textwidth}
  \begin{center}
    \includegraphics[width=0.450\textwidth]{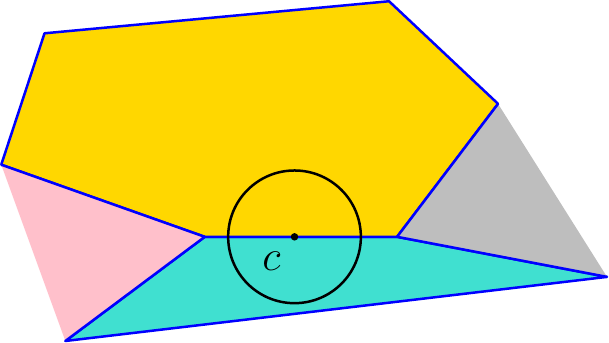}
  \end{center}
  \caption{The figure shows existence of a point on
  a Voronoi face that is far from other Voronoi faces.}
  \label{fig-protection-voronoi-faces-far}
\end{wrapfigure}

The above result in the unweighted case implies that if Voronoi vertices
are protected then any Voronoi face contains a point that is far
from any of the other Voronoi faces.
See Figure~\ref{fig-protection-voronoi-faces-far}.

\begin{figure}
  \centering
  \includegraphics[width=4.50in]{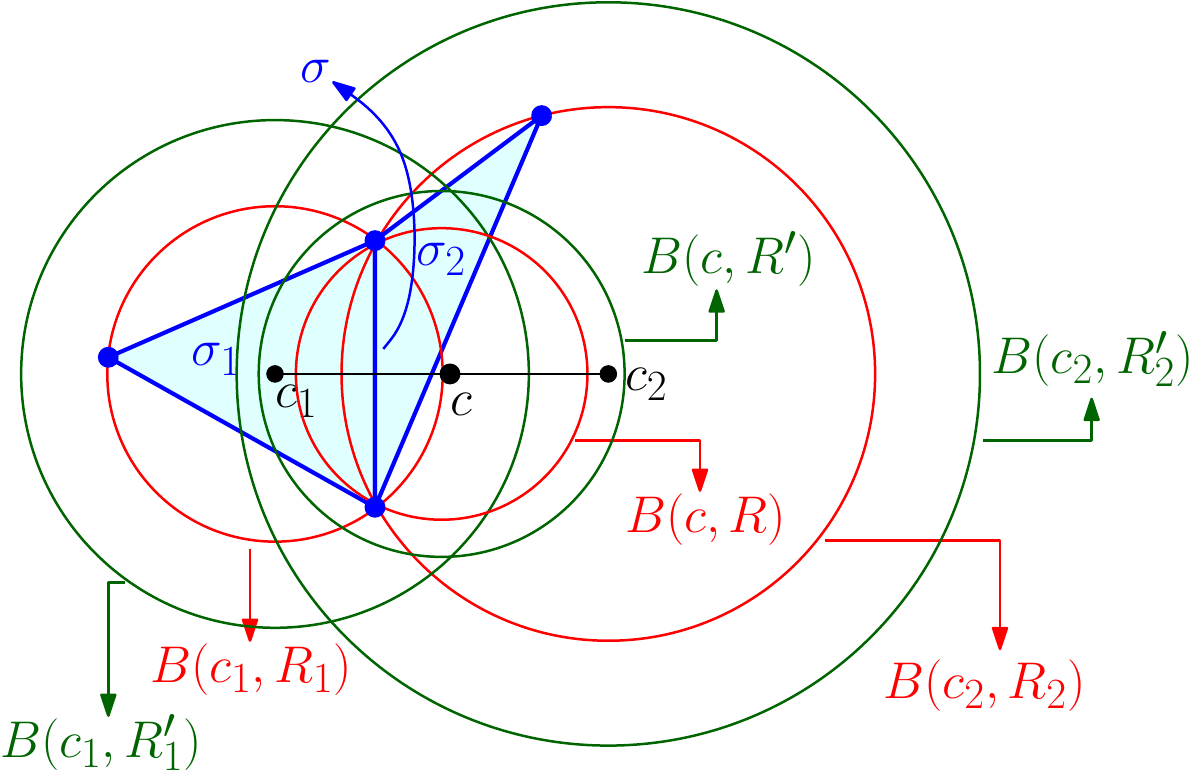}
  \caption{The two triangles $\sigma_{1}$ and $\sigma_{2}$ are
  $\delta^{2}$-power protected at $c_{1}$ and $c_{2}$ respectively. The
  figure shows that the edge $\sigma = \sigma_{1}\cap \sigma_{2}$ is
  $\frac{\delta^{2}}{2}$-power protected at
  $c = \frac{c_{1}+c_{2}}{2}$. Note that the vertices of the triangles have
  ``zero" weight and for $i \in \{1, \, 2\}$, $R'_{i} = \sqrt{R_{i}^{2}+\delta^{2}}$.
  See, Lemma~\ref{thm-protection-lower-dim-simplices} for a
  more general result.}
  \label{figure-showing-inheritance-of-protection-new}
\end{figure}

Lemma~\ref{thm-protection-lower-dim-simplices} also
implies that if $\aff \pts = \R^d$ and if all the $d$-simplices
in $\del_{{\varpi}}(\pts)$ are $\delta^{2}$-power protected then all the
simplices, not on the boundary of $\del_{{\varpi}}(\pts)$, are also power
protected. 
See Figure \ref{figure-showing-inheritance-of-protection-new}.
Another interesting aspect of this result is the fact that
the decay in power protection to lower dimensional simplices
goes down linearly with the dimension of the
ambient space.

To prove Lemma~\ref{thm-protection-lower-dim-simplices} we need the
following lemma on power protection.


\begin{lem}\label{lem:dim-voronoi-protected-simplex}
    Let $\pts \subset \R^{d}$ and $\varpi: \pts \rightarrow [0, \infty)$
    be a weight distribution. Let $p \in \pts$ such that $\vor_{\varpi}(p)$
    is bounded and all the $d$-simplices in $\del_{\varpi}(\pts)$
    incident to $p$
    are $\delta^{2}$-power for some $\delta > 0$.
    Then
    \begin{enumerate}

        \item
            the dimension of the maximal simplices in $\del_{\varpi}(\pts)$
	    incident to $p$
            is equal to $d$; and

        \item
            for all $j$-simplices $\sigma^{j} \in \del_{\varpi}(\pts)$
            incident to $p$,
	    $\dim \vor_{\varpi}(\sigma^{j}) = d-j$.
    \end{enumerate}
\end{lem}

\subsection{Proof of Lemma~\ref{lem:dim-voronoi-protected-simplex}}



\begin{remark}[On Lemma~\ref{lem:dim-voronoi-protected-simplex}]
	Observe that since
	$\vor_{\varpi}(p)$ is bounded, we have $\dim \aff (\pts) = d$.
\end{remark}

The following lemma is analogous to~\cite[Lem.~3.2]{boissonnat2012sdt}, and
the proof is exactly like the proof of that lemma.

\begin{lem}[Maximal simplices]\label{lem-maximal-simplices-incident-p}
    Every $\sigma \in \del_{\varpi}(\pts)$ incident to $p$
    is a face of a simplex $\sigma' \in \del_{\varpi}(\pts)$
    with $\dim\aff(\sigma')= d$.
\end{lem}
The following lemma is a direct consequence of the above result.
\begin{lem}[No degeneracies]\label{lem-no-degeneracies-incident-p}
    If every $d$-simplex in $\del_{\varpi}(\pts)$ incident to $p$
    is
    $\delta^{2}$-power protected for some $\delta > 0$, then there are no
    degenerate (see the definition of degenerate simplex given in
    Section~\ref{ssec-simplices-definition-background})
    simplices in $\del_{\varpi}(\pts)$ that are incident to $p$.
\end{lem}

Like in the case of Lemma~\ref{lem-maximal-simplices-incident-p},
following result is analogous to~\cite[Lem.~3.3]{boissonnat2012sdt}
and can be proved exactly along the same lines.
\begin{lem}[Separation]\label{lem-separation-result}
  If $\sigma^{j} \in \del_{\varpi}(\pts)$ is a $j$-simplex incident to $p$
with $\vor_\varpi(\sigma^j)$ bounded,
  and $q \in \pts \setminus \sigma^{j}$, then there is a $d$-simplex
  $\sigma^{d} \in \del_{\varpi}(\pts)$ incident to $p$ such that
  $\sigma^{j} \leq \sigma^{d}$ and $q \not\in \sigma^{d}$.
\end{lem}


\begin{proof}[of Lemma~\ref{lem:dim-voronoi-protected-simplex}]
The first assertion follows directly from Lemmas
\ref{lem-maximal-simplices-incident-p}
and
\ref{lem-no-degeneracies-incident-p}.

For the second assertion, we observe that
$\dim\vor_\varpi(\sigma^j) \leq d-j$ since
$\vor_\varpi(\sigma^j) \subseteq N_\varpi(\sigma^j)$, and
$\dim N_\varpi(\sigma^j) = d-j$ because $\sigma^j$ is nondegenerate. In
particular, if $j=d$, then we must have
$\dim\vor_\varpi(\sigma^j)=0$. We obtain the result for all $j$ by
showing, by induction on $i=d-j$,
 that if $\sigma^j \leq \sigma^{j+1}$, then
\begin{equation}
\label{eq:ind.vor.dim}
\dim \vor_{\varpi}(\sigma^{j}) > \dim \vor_{\varpi}(\sigma^{j+1}).
\end{equation}

Assume then that $\dim \vor_{\varpi}(\sigma^{j+1}) = d-(j+1)$.
We will show that for any facet $\sigma^j<\sigma^{j+1}$
there is a point $c \in \vor_\varpi(\sigma^j)$ such that 
$c \not\in N_\varpi(\sigma^{j+1})$. The claim
\eqref{eq:ind.vor.dim} then follows since 
$
\affhull{\vor_\varpi(\sigma^{j+1})}
\subseteq 
\affhull{\vor_\varpi(\sigma^{j})} 
$ 
and
Lemma~\ref{lem-no-degeneracies-incident-p} implies
$\dim N_\varpi(\sigma^{j+1}) = d-(j+1)$, and therefore
$N_\varpi(\sigma^{j+1}) = \affhull{\vor_\varpi(\sigma^{j+1})}$ by the
hypothesis on the dimension of $\vor_\varpi(\sigma^{j+1})$.


Let $q \in \sigma^{j+1}\setminus \sigma^{j}$.  
From Lemma~\ref{lem-separation-result}, there exists a $d$-simplex
$\sigma^d \in \del_{\varpi}(\pts)$ such that $\sigma^{j} < \sigma^d$
and $q \not\in \sigma^d$. Since all the $d$-simplices of
$\del_{\varpi}(\pts)$ incident to $p$ are $\delta^{2}$-protected, the
circumcentre $c \in \vor_{\varpi}(\sigma^d)$ satisfies
\begin{equation}\label{equation-very-useful}
  d(c, r^{\varpi}) < d(c, s^{\varpi}) -\delta^{2},
  \quad\text{for all } r \in \sigma^d 
~\mbox{and}~ s \in \pts \setminus \sigma^d.
\end{equation}
More specifically, this implies
\begin{equation}\label{equation-very-useful-1}
  d(c, r^{\varpi}) < d(c, q^{\varpi}) -\delta^{2},
  ~\forall~ r \in \sigma^{j},
\end{equation}
since $q \not\in \sigma^d$ and $\sigma^{j} < \sigma^d$.
Thus
$c \not \in N_{\varpi}(\sigma^{j+1})$,
and $c$ is the desired point
since 
$c \in \vor_{\varpi}(\sigma^d) \subseteq \vor_{\varpi}(\sigma^{j})$.
\end{proof}

\subsection{Lifting map, space of spheres and Voronoi diagram}


We are going to argue about the power protection of Delaunay simplices
in the ``space of spheres'' or ``lifting space''. For our purposes we
will be working primarily from the Voronoi perspective.
We will give a self-contained summary of the properties of the space
of spheres that we will use.
Full details can be found in~\cite[Chap.~17]{boissonnat1998}.


Since we will be dealing with ortho spheres, see the definition 
in Section \ref{ssec-weighted-points-weighted-Delaunay-complex}, 
in this section a sphere $S(c,r)$ can have $r^{2}<0$.

The {\em lifting map} $\glift$ takes a sphere
$\sphere{c}{r}$ in $\R^{d}$, with centre $c \in \R^{d}$ and radius $r$, to the
point $(c, \norm{c}^2 - r^2) \in \reel^{d+1}$, i.e.,
$$
	\phi(S(c, r)) = (c, \|c\|^{2} - r^{2}).
$$
We consider the points
in $\R^{d}$ to be spheres with $r=0$, and thus $\reel^{d}$ itself is
represented as a (hyper-) paraboloid in $\reel^{d+1}$, i.e, 
$x_{d+1} = \sum_{i=1}^{d} x^{2}_{i}$.

\begin{figure}
  \begin{center}
    \includegraphics[width=3.0in]{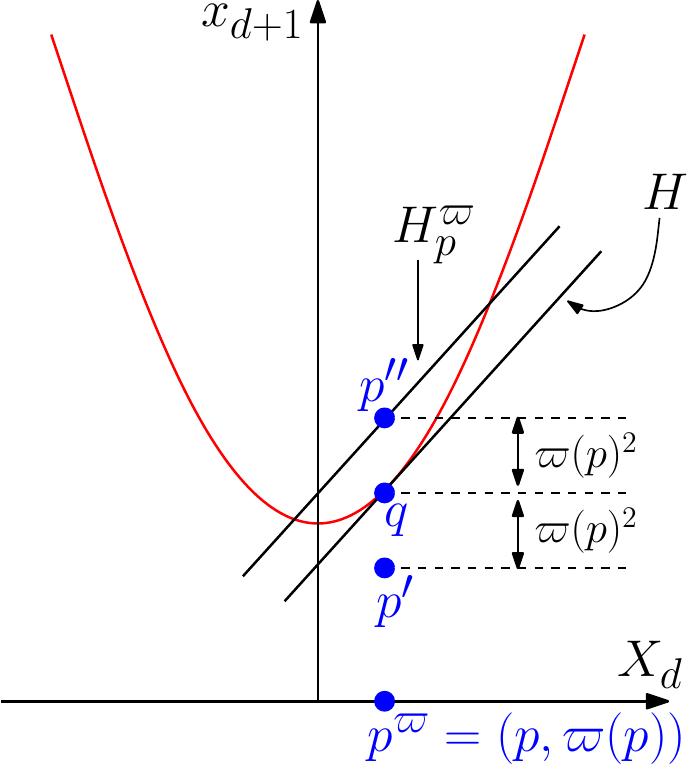}
  \end{center}
  \caption{In the figure $p' = \phi(S(p,\varpi(p)))$, $p'' =\phi(S(p,r))$
  where $r^{2} = -\varpi(p)^{2}$ and $q = \phi(S(p,0))$. Note that the hyperplanes
  $H$ and $H_{p}^{\varpi}$ are parallel and the hyperplane $H$ is tangential to the
  paraboloid (drawn in red), i.e., $x_{d+1} = \sum_{i=1}^{d} x^{2}_{i}$.
  Note that $X_{d} = (x_{1}, \, \dots, \, x_{d})$.}
  \label{fig-lifting-Voronoi-space-of-spheres}
\end{figure}

Let $\pts$ be a locally finite point set and
$\varpi: \pts \rightarrow [0, \infty)$ be a weight distribution.
The set of spheres that are orthogonal to point $p$, with weight $\varpi(p)$,
are represented by a hyperplane
$\hplane{\varpi}{p} \subset \reel^{d+1}$ that passes through
$\lift{\sphere{p}{r}}$ where $r^{2} = -\varpi(p)^{2}$. Indeed, for any sphere
$\sphere{c}{r}$ orthogonal to $p^{\varpi} = (p,\varpi(p))$
satisfies 
$$
  r^2  = \norm{c-p}^2 -  \varpi(p)^{2}.
$$
This implies
\begin{eqnarray*}
	\phi(S(c,r))
	&=& (c, \|c\|^{2}-r^{2})\\
	&=& (c, 2\langle c, \, p \rangle  + \varpi(p)^{2}-\norm{p}^2).
\end{eqnarray*}

So  the hyperplane is
\begin{equation*}
  \hplane{\varpi}{p}
=
\{ (c,h) \in \R^d\times \R \mid 
h = 2\langle c, p \rangle  + \varpi(p)^{2}-\norm{p}^2 \};
\end{equation*}
see Figure~\ref{fig-lifting-Voronoi-space-of-spheres}.

For any $p \in \pts \subset \R^{d}$, we represent its Voronoi cell
$\vor_{\varpi}(p)$ in the space of spheres by associating to each $c \in
\vor_{\varpi}(p)$ the unique sphere $\sphere{c}{r}$, where
$r^{2} = \|p-c\|^{2}- \varpi(p)^{2}$. Thus
$\lift{\vor_{\varpi}(p)} \subseteq \hplane{\varpi}{p}$.


For any Delaunay simplex $\splxs \in \del(\pts)$, its Voronoi cell
$\vor_{\varpi}(\splxs) = \bigcap_{p \in \splxs} \vor_{\varpi}(p)$
is mapped in the space of
spheres  to the intersection of the hyperplanes that support the
lifted Voronoi cells of its vertices:
$$
  \lift{\vor_{\varpi}(\splxs)}\subset\bigcap_{p \in \splxs}\hplane{\varpi}{p}.
$$
If $\pts$ is generic and $\splxs$ is a
$k$-simplex, then $\lift{\vor_{\varpi}(\splxs)}$ lies in a
$(d-k)$-dimensional affine space.

We can say more. The lifted Voronoi cell $\lift{\vor_{\varpi}(\splxs)}$ is
a convex polytope.
Any two points $z, z' \in
\lift{\vor_{\varpi}(\splxs)}$ have corresponding points $c,c' \in
\vor_{\varpi}(\splxs) \subset \R^{d}$, and a line segment between $c$ and
$c'$ gets lifted to a line segment between $z$ and $z'$ in
$\lift{\vor_{\varpi}(\splxs)}$.

\subsection{Proof of Lemma~\ref{thm-protection-lower-dim-simplices}:
power protection in the ``space of spheres" framework}
\label{ssec-power-protection-space-of-spheres-framework}

\begin{figure}[h]
  \begin{center}
    \includegraphics[width=2.250in]{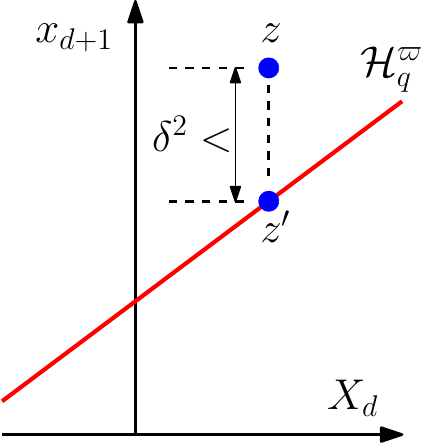}
  \end{center}
  \caption{Diagram showing connection between
  protection and lifting.}
  \label{fig-protection-lifting}
\end{figure}

We can talk about the power-protection at a point $c \in
\vor_{\varpi}(\splxs)$: it is the power-protection enjoyed by the Delaunay
sphere $\sphere{c}{r}$ centered at $c$. For a point $q \in \pts
\setminus \splxs$, we say that $c$ is
\defn{$\pdelta^2$-power-protected from $q$} if
$$
    \norm{q-c}^2  - \varpi(q)^{2} - r^2 > \pdelta^2.
$$
In the lifting space, if $z = \lift{\sphere{c}{r}}$, then the power
protection of $c$ from $q$ is given by the ``vertical'' distance of
$z$ above $\hplane{\varpi}{q}$, which we will refer to as the \defn{clearance} of $z$
above $\hplane{\varpi}{q}$. See, Figure~\ref{fig-protection-lifting}.

Thus for any $q \in \pts \setminus \splxs$ we have a function $f_q:
\lift{\vor_{\varpi}(\splxs)} \to \reel$ which associates to each $z \in
\lift{\vor_{\varpi}(\splxs)}$ the clearance of $z$ above $\hplane{\varpi}{q}$.
This is a
linear function of the sphere centres. Indeed, if $p \in \splxs$ and
$z = \lift{\sphere{c}{r}}$, then $r^2 + \varpi(p)^{2}= \norm{p-c}^2$, and
$$
  f_q(z) =
  2\langle c, \, p-q \rangle -(\norm{p}^2-\norm{q}^2)+(\varpi(p)^{2}-\varpi(q)^{2}).
$$


\begin{proof}[of Lemma~\ref{thm-protection-lower-dim-simplices}]
We wish to find a bound $h_{j}(\delta)$ such that if all the
$d$-simplices in $\del_{\varpi}(\pts)$ incident to $p$
are $\delta^2$-power-protected, then the
Delaunay $j$-simplices incident to $p$
will be $h_{j}(\delta)$-power-protected. 
Since $\vor_{\varpi}(\splxs^{j}) \subseteq \vor_{\varpi}(p)$ and
$\vor_{\varpi}(p)$ is bounded, we observe that for any $j$-simplex
$\splxs^{j}$ its Voronoi cell $\vor_{\varpi}(\splxs^{j})$ is the
convex hull of Voronoi vertices: the $\varpi$-weighted centres
of the Delaunay $d$-simplices that have $\splxs^{j}$ as a
face. It follows that $\lift{\vor_{\varpi}(\splxs^{j})}$ is the convex hull of
a finite set of points which correspond to these $d$-simplices.
Note that from Lemma~\ref{lem:dim-voronoi-protected-simplex} we have that
$\dim({\vor_{\varpi}(\splxs^{j})})$, which is
equal to $\dim(\lift{\vor_{\varpi}(\splxs^{j})})$, is $d-j$.
We
choose an affinely independent set $\{z_i\}$, $i \in \{0, \, 1,\, \ldots,\, k
\}$, of $k+1$ of these points, where $k= d-j$. 
See Figure~\ref{fig:splx.in.vor}.

\begin{figure}
  \begin{center}
    \includegraphics[width=2.50in]{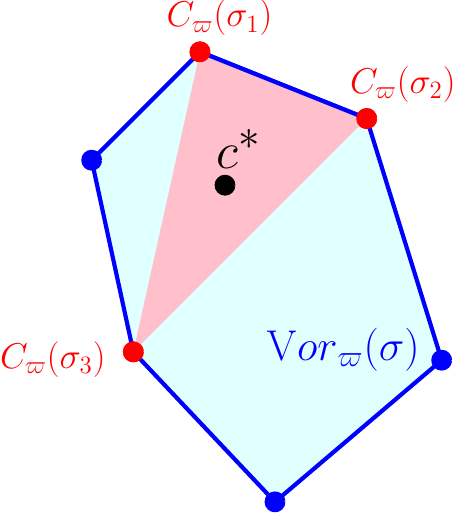}
  \end{center}
  \caption{Diagram for Lemma~\ref{thm-protection-lower-dim-simplices}. In the
  figure $\sigma$ is a $(d-2)$-dimensional simplex with $\sigma < \sigma_{i}$,
  with $i \in \{1, \, 2, \, 3\}$, where $\sigma_{i}$ is a $d$-dimensional
  simplex $\delta^{2}$-power protected at $C_{\varpi}(\sigma_{i})$.
  From the proof of Lemma~\ref{thm-protection-lower-dim-simplices} we get that
  $\sigma$ is $\frac{\delta^{2}}{3}$-power protected at
  $c^{*} = \frac{1}{3} \left( \sum_{i=1}^{3} C_{\varpi}(\sigma_{i}) \right)$.
\label{fig:splx.in.vor}
  }
\end{figure}
Let
$$
  z^* = \frac{1}{k+1} \sum_{i= 0}^{k} z_i
$$
be the barycenter of these lifted
Delaunay spheres, and consider the clearance, $f_q(z^*)$, of $z^*$
above $\hplane{\varpi}{q}$, where $q \in \pts \setminus \splxs^{j}$.
Observe that $z^{*}$ is an interior point of
$\lift{\vor_{\varpi}(\splxs^{j})}$.
Let $\splxs_i$ be
the Delaunay $d$-simplex corresponding to $z_i$. There must be a
$\splxs_{\ell}$, $l \in \{0, \, \dots, \, k\}$,
which does not contain $q$, since otherwise the $k$-simplex
defined by the set $\{z_i\}_{i\in \{0,\ldots,k\}}$ would lie in
$\lift{\vor_{\varpi}(\splxjoin{q}{\splxs^{j}})}$,
implying
$\dim \left( \lift{\vor_{\varpi}(\splxjoin{q}{\splxs^{j}})}\right)
\geq k$,
which contradicts
Lemma~\ref{lem:dim-voronoi-protected-simplex}. 
Since $\splxs_{\ell}$ is
$\delta^2$-power-protected, we have $f_q(z_{\ell}) > \delta^2$, and by
the linearity of $f_q$ we get a bound on the clearance of $z^*$ above
$\hplane{\varpi}{q}$:
\begin{eqnarray*}
  f_q(z^*) = \frac{1}{k+1}\sum_{i=0}^{k} f_q(z_i)
  \geq \frac{f_q(z_{\ell})}{k+1}
  > \frac{\delta^2}{k+1}.
\end{eqnarray*}
Since $q$ was chosen arbitrarily from $\pts \setminus \splxs^{j}$, this
provides a lower bound on the power protection at $c^* \in
\vor_{\varpi}(\splxs^{j})$, where $z^* = \lift{\sphere{c^*}{r^*}}$, and hence a
lower bound on the power protection of $\splxs^{j}$.
\end{proof}

\begin{remark}
If we
could find two lifted Voronoi vertices $z_1$ and $z_2$ such that the
line segment between them lies in the relative interior of
$\lift{\vor_{\varpi}(\splxs^{j})}$, then the midpoint of that segment would
have a power protection of $\frac{\delta^2}{2}$. However, this isn't
possible in general, for $\vor_{\varpi}(\splxs^{j})$ could be a 
$(d-j)$-simplex,
when $\splxs^{j}$ is not a maximal shared face of any two Delaunay
$d$-simplices.
\end{remark}






\section{Stability, protection, and the witness complex}
\label{ssec-stability-protection-wit-complex}

Our main structural result,
Theorem~\ref{thm-output-weighted-rDT-eqal-wit} below, gives conditions
that guarantee that $\del_{{\varpi}}(L, T\M) = \wit_{{\varpi}}(L, W)$.
Since the proof of Theorem~\ref{thm-output-weighted-rDT-eqal-wit} is
quite long and technical, it goes through multiple stages which we
will only outline in this section. For full details refer to
Appendix~\ref{ssec-faces-inherit-protection}. We begin by introducing some parameters and terminology employed in the statement of the theorem.

Let $\M$ be a $m$-dimensional submanifold of $\reel^{d}$,
 $\alpha_{0} < \frac{1}{2}$  an absolute constant%
\footnote{By absolute constant we mean that $\alpha_{0}$
is independent of the dimension of the manifold or other parameters
of the algorithm. For simplicity, the reader can take $\alpha_{0} = \frac{1}{3}$
for the rest of this paper.}, and
$\Gamma_{0} < 1$ and $\delta_{0} < \alpha_{0}$ parameters to the
algorithm satisfying
Inequality~\eqref{eqn-ineq-gamma-delta-alpha}; see
Lemma~\ref{lem-existence-of-good-weights} from
Section~\ref{sec:reconstruction-algorithm}.
We define
$$
  \tilde{\alpha}_{0} \stackrel{\rm def}{=} \sqrt{\alpha^{2}_{0} - \delta^{2}_{0}}.
$$

\begin{definition}[Elementary weight perturbations and stable weight assignments]
  Let $W \subset \M$ be an $\e$-sample of $\M$, $L \subset W$ a
  $\lambda$-net of $W$ with $\e \leq \lambda$, and
  ${\varpi}: L \rightarrow [0, \infty)$ a weight assignment with
  relative amplitude \eqref{eq:relative.amplitude} satisfying
  $\widetilde{{\varpi}} \leq \tilde{\alpha}_{0}$.
  A weight assignment $\xi: L \rightarrow [0, \infty)$ will be called
  an {\em elementary weight perturbation} of ${\varpi}$ ({\em ewp} for
  short) if 
$$
	\exists~p \in L, ~ \xi(p) \in \left[ {\varpi}(p),\,
  	\sqrt{{\varpi}(p)^{2}+ \delta^{2}_{0}\lambda^{2}} \right]
  	\; \;  \mbox{and}  \; \;
	\xi(q) = {\varpi}(q) \;\; \mbox{if} \;\; q \in L \setminus p.
$$
We call the weight assignment ${\varpi}: L \rightarrow [0, \infty)$
{\em stable} (resp., {\em locally stable at $p\in L$}) if for all ewp
$\xi$ of ${\varpi}$, ${\rm K}_{\xi}(L)$ contains no
$\Gamma_{0}$-slivers of dimension $\leq m+1$ (resp., no such slivers
incident to $p$).
\end{definition}

\begin{thm}\label{thm-output-weighted-rDT-eqal-wit}
	Let $\M$ be a $m$-dimensional submanifold of $\reel^{d}$,
    $W \subset \M$  an $\e$-sample of $\M$, $L \subset W$ 
    a $\lambda$-net of $W$ with $\e \leq \lambda$, and
    ${\varpi}: L \rightarrow [0, \infty)$ a stable weight assignment
    with $\widetilde{\varpi} \leq \tilde{\alpha}_{0}$.
%
%
%
    If
    \begin{equation}\label{equation-main-structural-theorem-1}	  
    		\lambda < \frac{\reach}{2^{15}(m+1)} \min\Big\{ \Gamma^{2m+1}_{0},\, \delta^{2}_{0} \Big\}
    \end{equation}
%
    and
    \begin{equation}\label{equation-main-structural-theorem-2}
      \e < \frac{\lambda}{24}\left( \frac{\delta^{2}_{0}}{m+1} - \frac{2^{15}\lambda}{\reach}\right)
    \end{equation}
%
    then,
    $$
      \del_{{\varpi}}(L, T\M) = \wit_{{\varpi}}(L, W).
    $$
    In addition, if $\lambda$ is sufficiently small, then
    $\wit_{{\varpi}}(L, W)$ is homeomorphic to, and a close geometric
    approximation of, $\M$.
  \end{thm}


    Since ${\varpi}$ is a stable weight assignment,
    ${\rm K}_{{\varpi}}(L)$ contains no $\Gamma_{0}$-slivers
    of dimension $\leq m+1$. 
    The proof of Theorem \ref{thm-output-weighted-rDT-eqal-wit} relies on the following 
    three properties {\bf P1}, {\bf P2} and {\bf P3}, which hold for 
    $\lambda$ sufficiently small. 
    \begin{description}
     \item[P1]
	 For all $\sigma \in {\rm K}_{{\varpi}}(L)$,
	 $\angle (\aff\sigma, T_{p}\M) = O(\lambda)$. 

     \item[P2]
	The  simplices of $\del_{{\varpi}}(L, T\M)$
	have dimension at most $m$, and the maximal
	dimension of simplices in ${\rm K}_{{\varpi}}(L)$
	is $\leq m$.

    \item[P3]
	For all  $\sigma \in \del_{{\varpi}}(L, T\M)$ and $p \in \sigma$,
	$\vor_{{\varpi}}(\sigma) \cap T_{p}\M\neq\emptyset$.


      \end{description}
      The above properties are direct consequence of results
      from~\cite{cheng2005,boissonnat2009,boissonnat2011tancplx}.
	For the full details see Lemma~\ref{lem-property-cocone-complex}
	in Appendix~\ref{ssec-faces-inherit-protection}.

	Using properties
      {\bf P1}, {\bf P2} and {\bf P3}, we will give the outline of the proof of 
      $$
      	\wit_{{\varpi}}(L, W) = \del_{{\varpi}}(L, T\M).
      $$
	The part about homeomorphism and close geometric
	aspect of Theorem~\ref{thm-output-weighted-rDT-eqal-wit}	
	will directly follow from a result of Boissonnat and
	Ghosh~\cite{boissonnat2011tancplx}.


We will now give an outline of the proof of
$\wit_{{\varpi}}(L, W) = \del_{{\varpi}}(L, T\M)$.


\paragraph{Step 1: $\wit_{{\varpi}}(L, W) \subseteq \del_{{\varpi}}(L, T\M)$.}
This step is proved by contradiction in 
Lemma~\ref{lem-conditions-weighted-wit-subseteq-tancomplex}.
Let $\sigma^{k} \in\wit_{{\varpi}}(L,W)$ be a $k$-simplex with 
$\sigma^{k} \not\in \del_{{\varpi}}(L, T\M)$ and $p$ a vertex of $\sigma^{k}$.  
We will show that if this is the case then there exists $\sigma^{m+1}$ with 
$\sigma^{k} \leq \sigma^{m+1}$ and $\sigma^{m+1} \in {\rm K}_{\varpi}(L)$. 
We will 
reach a contradiction via Property P2.  
  
  Using the sampling
  assumptions on $L$ and $W$, we can show that for any
  $w \in W$ that is a
  ${\varpi}$-witness of $\sigma^{k}$ or of its subfaces,
  $\|p-w\| = O(\lambda)$~\cite[Lem.~4.4]{boissonnat2009}.
  This implies,
  from~\cite[Lem.~6]{giesen2004},
  $$
    d(w, T_{p}\M) = O\left(\frac{\lambda^{2}}{\reach}\right).
  $$
  From~\cite[Thm.~4.1]{deSilva2008},
  we know that $\vor_{{\varpi}}(\sigma^{k})$ intersects the convex hull
  of the ${\varpi}$-witnesses of $\sigma^{k}$ and its
  subfaces. Let
  $c_{k} \in\vor_{{\varpi}}(\sigma^{k})$ be a point in this
  intersection.  We have
  $$
    d(c_{k}, T_{p}\M) = O\left(\frac{\lambda^{2}}{\reach}\right)
  $$
  and, since $L$ is 
$\lambda$-sparse and $\widetilde{\varpi} < \frac{1}{2}$,
$$
    d(p, c_{k}) \geq d(p, C_{\varpi}(\sigma^{k})) \geq \frac{3\lambda}{8}.
    \footnote{Let $p$ and $q$ be distinct vertices of $\sigma^{k}$, 
    and let $x$ be an orthogonal projection of $C_{\varpi}(\sigma^{k})$ 
    on the line through
$p$ and $q$. Since $\widetilde{\varpi} < \frac{1}{2}$,
    $x \in pq$. 
Using the facts that 
    $d(x, p^{\varpi}) = d(x, q^{\varpi})$, $\|p-q\| \geq \lambda$
    and $\varpi(q) < \frac{\|p-q\|}{2}$ (as $\widetilde{\varpi} < 1/2$), we get 
    \begin{eqnarray}
    		\|p-x\| &=& \frac{\|p-q\|}{2} \left( 1+ \frac{\varpi(p)^{2} - \varpi(q)^{2}}{\| p - q \|^{2}} \right)\nonumber\\ 
    		&\geq& \frac{\|p-q\|}{2} \left( 1 - \frac{\varpi(q)^{2}}{\|p-q\|^{2}}\right) \nonumber\\
    		&\geq& \frac{3\lambda}{8}
    \end{eqnarray}
    The bound on $d(p, C_{\varpi(\sigma^{k})})$ follows from the fact that 
    $d(p, x) \geq d(p, C_{\varpi}(\sigma^{k}))$.
    }
$$
  Therefore, using the sampling assumption on $\lambda$, we get
  $$
    \frac{d(c_{k}, T_{p}\M)}{d(c_{k}, p)} =
    O\left(\frac{\lambda}{\reach}\right) < \sin \theta_{0}.
  $$
  By the definition of the cocone complex, 
  see Equation~\eqref{eq:cocone.complex},
  this implies
  that $\sigma^{k} \in {\rm K}_{{\varpi}}(L)$.
  Since $\angle (\aff\sigma^{k}, T_{p}\M)$ is small (property P1),
  there exists $c_{k}'\in T_{p}\M$ such that the line segment $[c_{k},
  c_{k}']$ is orthogonal to $\affhull{\sigma^{k}}$ and
  $$
    d(c_{k}, c_{k}') = O\left(\frac{\lambda^{2}}{\reach}\right).
  $$
  Again, as
  $\lambda$ is small, the line segment
  $[c_{k}, c_{k}']$ is contained in ${\rm K}(p)$. Since
  $\sigma^{k} \not \in \del_{{\varpi}}(L, T\M)$, $\exists~ c_{k+1} \in
  [c_{k}, c_{k}']$ and a Delaunay $(k+1)$-simplex $\sigma^{k+1}$ such that
  $$
    c_{k+1} \in \vor_{{\varpi}}(\sigma^{k+1})
    \quad \mbox{with} \quad
    \sigma^{k} < \sigma^{k+1}.
  $$
  Therefore, $\sigma^{k+1} \in {\rm K}_{{\varpi}}(L)$.
  If  $k = m$,  we have reached a
  contradiction with property P2.
  Otherwise, using the facts that $\angle(\sigma^{k+1}, T_{p}\M)$
  is small, $d(c_{k+1}, T_{p}\M) = O\left(\frac{\lambda^{2}}{\reach}\right)$
  and $d(p, c_{k+1}) = {\Omega}(\lambda)$,
  we will find a $c'_{k+1} \in T_{p}\M$ such that
  $[c_{k+1},c'_{k+1}] \in {\rm K}(p)$. Since
  $\sigma^{k+1} \in {\rm K}_{{\varpi}}(L)$, $\exists~c_{k+2} \in [c_{k+1},c'_{k+1}]$
  and $(k+2)$-simplex $\sigma^{k+2} \in {\rm K}_{{\varpi}}(L)$
  such that
  $$
    c_{k+2} \in \vor_{{\varpi}}(\sigma^{k+2})
    \quad \mbox{and} \quad
    \sigma^{k+1}< \sigma^{k+2}.
  $$
  Continuing this procedure of walking
  on the Voronoi cell of the simplex from a point, like $c_{k+1}$,
  in the intersection the Voronoi cell of the simplex and ${\rm K}(p)$
  towards $T_{p}\M$, we will get a sequence of points
  $$
    c_{k}, \dots, c_{m+1}
  $$
  and simplies
  $$
    \sigma^{k} < \dots  < \sigma^{m+1}
  $$
  with
  $$
    c_{j} \in \vor_{{\varpi}}(\sigma^{j})\cap {\rm K}(p)
    \quad\mbox{and}\quad \sigma^{j} \in {\rm K}_{{\varpi}}(L).
  $$
  We have now reached a contradiction
  via property P2. This concludes the proof of Step~1.
%

\paragraph{Step 2: $\del_{{\varpi}}(L, T\M) \subseteq \wit_{{\varpi}}(L, W)$.}
%
%

We say that a simplex $\sigma \in \del_\varpi(L,T\M)$ is \defn{$\delta^2$-power protected on $T_p\M$} if it is $\delta^2$-power protected at a point $c\in \vor_\varpi(\sigma) \cap T_p\M$.
In Lemma~\ref{lem-props-we-want-omega-to-satisfy} we show that the
stable weight assignment implies that all $m$-simplices in
$\del_{{\varpi}}(L, T\M)$ are $\delta^{2}$-power protected on
$T_{p}\M$, where $\delta = \delta_{0}\lambda$.  
To reach a
contradiction, let us assume that there exists a $m$-simplex
$\sigma \in \del_{{\varpi}}(L, T\M)$ that is not power protected on
$T_{p}\M$ for some $p \in \sigma$. Then for any
$c \in \vor_{{\varpi}}(\sigma)\cap T_{p}\M$ there exists
$q \in L \setminus \sigma$ such that
    $$
	d(c,p^{{\varpi}}) \geq d(c, q^{{\varpi}}) - \delta^{2}.
    $$
    Consider now the following
    weight assignment:
    $$
          \xi(x) = \left\{
        \begin{array}{l l}
            {\varpi}(x) & \quad \text{if $x\neq q$}\\
            \sqrt{{\varpi}(q)^{2}+\beta^{2}} & \quad \text{if $x=q$}\\
        \end{array} \right.
    $$
    where
    $$
      \beta^{2} = d(c,q^{{\varpi}}) - d(c, p^{{\varpi}}).
    $$
    It is easy to
    see that $\xi$ is an ewp of ${\varpi}$ and $\tilde{\xi} < 1/2$.
    Observe that the $(m+1)$-simplex $\sigma' = q * \sigma$ is 
    in ${\rm K}_{\xi}(L)$.     
    Since $\lambda$ is sufficiently small,
    $\sigma'$ is a $\Gamma_{0}$-bad $(m+1)$-simplex;
    broadly, the idea 
(see also the proofs of~\cite[Lem.~13]{cheng2005}
    and \cite[Lem.~4.9]{boissonnat2011tancplx})
is that
the thickness of any
    $(m+1)$-simplex embedded in $\R^{m}$ is zero, and here  $\sigma '$ is
    a $(m+1)$-simplex embedded in $\R^d$, but whose vertices belong to a
    small neighborhood of a $m$-dimensional submanifold
    $\M$ of $\R^d$ so we can show that its
    thickness  is 
small.
By proving that $\sigma'$ is $\Gamma_{0}$-bad, we arrive at a
contradiction with the fact that ${\varpi}$ is a stable weight
assignment.

	In the first part of the proof of Lemma~\ref{lem-rDt-protected-M}
    we prove that all simplices (of all dimensions) in $\del_{{\varpi}}(L, T\M)$ are
    $\frac{\delta^{2}}{m+1}$-power protected on $T_{p}\M$ for all $p \in
    \sigma$. To establish this result, we want to use
    Lemma~\ref{thm-protection-lower-dim-simplices} but we cannot use
    the lemma directly since it only holds
    for $d$-simplices of $\R^d$. To overcome this issue, we resort to
    Lemma~2.2 of~\cite{boissonnat2011tancplx} which states that
    $\vor_{{\varpi}}(L)\cap T_{p}\M$ is identical to
    a weighted Voronoi diagram $\vor_{\psi}(L')$  where
    $L'$ is the orthogonal projection of $L$ onto $T_{p}\M$, i.e.,
    $\vor_{{\varpi}}(\sigma) \cap T_{p}\M = \vor_{{\psi}}(\sigma')$
    where $\sigma'$ is the projection of $\sigma$ onto $T_{p}\M$.
    Also we can prove, using P1, that $\delta^{2}$-power protection
    of a simplex $\sigma \in \del_{{\varpi}}(L, T\M)$ incident to $p$
    on $T_{p}\M$
    implies
    $\delta^{2}$-power protection of $\sigma' \in \del_{\psi}(L')$.
    Using this correspondance, we can show
    that all $m$-simplices incident to $p$ in $\del_{\psi}(L')$ are
    $\delta^{2}$-power protected since all the $m$-simplices incident to
    $p$ in $\del_{{\varpi}}(L, T\M)$ are $\delta^{2}$-power protected on
    $T_{p}\M$.
    We can now use
    Lemma~\ref{thm-protection-lower-dim-simplices}.  Using the
    bound on $\lambda$,  
    we can show that $\vor_{\psi}(p) =
    \vor_{{\varpi}}(p)\cap T_{p}\M$ is bounded,
    see~\cite[Lem.~4.4]{boissonnat2011tancplx}.
    From
    Lemma~\ref{thm-protection-lower-dim-simplices}, we then get that all
    $j$-simplices in $\del_{\psi}(L')$
    incident to $p'$ are $\frac{\delta^{2}}{m+1}$-power
    protected. This result, together with the correspondence
    we have established between the power protection of
    simplices incident to $p$ in $\del_{\psi}(L')$ and the
    power protection on $T_{p}\M$ of
    simplices incident to $p$ in $\del_{{\varpi}}(L, T\M)$,
    we deduce that all $j$-simplices  incident to $p$ in
    $\del_{{\varpi}}(L, T\M)$ are $\frac{\delta^{2}}{m+1}$-power protected
    on $T_{p}\M$.

    Let $\sigma$ be $\frac{\delta^{2}}{m+1}$-power protected
    at $c \in \vor_{{\varpi}}(\sigma) \cap T_{p}\M$, where $p \in \sigma$.
    We can show that there exists $c' \in \M$, such that $\|c-c'\|$ is
    small compared to $\frac{\delta^{2}}{m+1}$ and the line passing
    through $c$ and $c'$ is orthogonal to $\aff (\sigma )$. Using simple
    triangle inequalities, we can prove that $\sigma$ is
    ${\Omega}\left(\frac{\delta^{2}}{m}\right)$-power protected at $c'$.
	See Lemma~\ref{lem-rDt-protected-M}.

    As $W$ is an $\e$-sample of $\M$, we can find a $w \in W$
    such that $\|w-c'\| < \e$. Using the facts that $\e$ is much
    smaller than $\delta^{2}= \delta^{2}_{0}\lambda^{2}$ and
    $\sigma$ is ${\Omega}\left(\frac{\delta^{2}}{m}\right)$-power protected at $c'$,
    we get $w$ to be a ${\varpi}$-witness of $\sigma$. Since $\sigma$ is an
    arbitrary simplex of $\del_{{\varpi}}(L, T\M)$, we have proved that
    $\del_{{\varpi}}(L, T\M) \subseteq \wit_{{\varpi}}(L, W)$.
	See Lemma~\ref{lem-conditions-rDt-subseteq-witness}.    
    

This ends the proof of Theorem~\ref{thm-output-weighted-rDT-eqal-wit}.

\section{Reconstruction algorithm}
\label{sec:reconstruction-algorithm}

Let $\man$ be a smooth submanifold with known dimension $m$,
let $W \subset \man$ be an $\e$-sample of $\man$, and let
$L \subset W$ be a $\lambda$-net of $W$ for some known $\lambda$.
We will also assume that $\e < \lambda$, which implies that $L$
is a $(\lambda, 2\lambda)$-net of $\man$. We will discuss the
reasonability of these assumptions in
Section~\ref{ssec-regarding-assumptions}.

The primary task of the algorithm is to find a
stable weight assignment
${\varpi} : L \rightarrow [0, \, \infty)$. We will prove that this is
possible if $\Gamma_{0}$, $\delta_{0}$, and the absolute constant
$\alpha_{0} < \frac{1}{2}$ satisfy
Inequality~\eqref{eqn-ineq-gamma-delta-alpha}
(Lemma~\ref{lem-existence-of-good-weights}).

Once we have calculated a stable weight assignment ${\varpi}$,
we can just output the witness complex $\wit_{{\varpi}}(L, W)$, which is
a faithful reconstruction of $\man$ by Theorem~\ref{thm-output-weighted-rDT-eqal-wit}.


%


\subsection{Outline of the algorithm}
\label{ssec-outline-algorithm}

We initialize all weights by setting  ${\varpi}_{0}(q) = 0$ for all
$q \in L$.  We then process each point $p_i\in L$, $i \in \{1,\, \dots,
\, n \}$. At step $i$, we compute a new weight assignment ${\varpi}_{i}$
satisfying the following properties:
\begin{description}
    \item[C1.]
        $\widetilde{{\varpi}}_{i} \leq \tilde{\alpha}_{0}$, and
	$\forall~q \in L \setminus \{ p_{i} \}$, ${\varpi}_{i}(q) = {\varpi}_{i-1}(q)$.



    \item[C2.] ${\varpi}_{i}$ is locally stable at $p_{i}$.


%
\end{description}


Once we have assigned weights to all the points of $L$  in the above
manner,
the algorithm outputs $\wit_{{\varpi}}(L, W)$ where ${\varpi} =
{\varpi}_{n}$ is
the final weight assignment ${\varpi}_{n}: L \rightarrow [0, \,
\infty)$.

The crux of our approach is that weight assignments will be done
without computing the cocone complex
or any other sort of Voronoi/Delaunay subdivision. Rather, we just look at
local neighborhoods
\begin{eqnarray*}
  N(p_{i}) \stackrel{\rm def}{=}
  \Big\{x \in L : \#(\overline{B}(p_{i}, d(p_{i},x))\cap L) \leq N_{1}\Big\},
\end{eqnarray*}
where
$N_{1}$ is defined in Lemma~\ref{lem-bound-on-neighbors}. The main idea is the
following. We define the {\em candidate simplices} of $p_{i}$ as the
$\Gamma_{0}$-slivers $\sigma$ of dimension $\leq m+1$,
with vertices in $N(p_{i})$, $p_{i} \in \sigma$, and
of diameter $\Delta(\sigma) \leq 16\lambda$. For such a candidate
simplex $\sigma$, we compute
a {\em forbidden interval} $I_{{\varpi}_{i-1}}(\sigma, p_{i})$ 
(to be defined in Section~\ref{ssec-algorithm-analysis}).
We then select a weight for $p_{i}$ that is outside all the forbidden intervals
of the candidate simplices of $p_{i}$.

We will denote by $S(p_{i})$ the set of candidate simplices of $p_{i}$.
For a point $p$ in $L$, we write
$$
  nn(p) \stackrel{\rm def}{=} \min_{q \in L\setminus p} \|p-q\|.
$$

\begin{algorithm}
  \caption{Pseudocode of the algorithm}
  \begin{algorithmic}
    \STATE{\bf Input:} $L$, $W$, $\Gamma_{0}$, $\delta_{0}$ and $m$
    \STATE // let $L = \{p_{1}, \, \dots, \, p_{n}\}$
    \STATE // parameters $\Gamma_{0}$, $\delta_{0}$ and $m$
    satisfy Eq.~\eqref{eqn-ineq-gamma-delta-alpha}
    \STATE{\bf Initialization:} ${\varpi}_{0}: L \rightarrow [0, \infty)$ with
    ${\varpi}_{0}(p) = 0$, $\forall ~p \in L$;
    \STATE{\bf Compute:} $nn(p)$, $N(p)$ for all $p \in L$
    \FOR{$i = 1$ to $n$}
        \STATE{\bf Compute:} {\em candidate simplices} $S(p_{i})$;
        \STATE $I \gets \bigcup_{\sigma \in S(p_{i})} I_{{\varpi}_{i-1}}(\sigma, p_{i})$;
        \STATE ${\varpi}_{i}(q) \gets {\varpi}_{i-1}(q)$ for all $q \in L\setminus\{p_{i}\}$;
        \STATE $x \gets$ a point from $[0, \tilde{\alpha}_{0}^{2}\, nn(p_{i})^{2}] \setminus I$;
        \STATE ${\varpi}_{i}(p_{i}) \gets \sqrt{x}$;
    \ENDFOR
    \STATE{\bf Output:} $\wit_{{\varpi}_{n}}(L, W)$;
  \end{algorithmic}
  \label{algorithm-1}
\end{algorithm}


\subsection{Analysis}
\label{ssec-algorithm-analysis}

\subsubsection{Correctness of the algorithm}
\label{ssec-correctness-algorithm}

Forbidden intervals and elementary weight perturbations are closely
related (see Lemma~\ref{lem-length-forbidden-interval} below) and we
will prove in Lemma~\ref{lem-existence-of-good-weights} that, if
Inequality~\eqref{eqn-ineq-gamma-delta-alpha} is satisfied, we can find
a locally stable weight assignment ${\varpi}_{i}$ at each iteration of
the algorithm.  Moreover, we will prove that if all ${\varpi}_{i}$ are
locally stable, then we will end up with a stable weight assignment
${\varpi} = {\varpi}_{n}$ for which
Theorem~\ref{thm-output-weighted-rDT-eqal-wit} applies. In this
respect our algorithm is in the same vein as the seminal work of Cheng
{\em et al.}~\cite{cheng2000}.  See
also~\cite{cheng2005,boissonnat2009,boissonnat2011tancplx}.


For a given weight assignment ${\varpi}: L \rightarrow [0, \, \infty)$,
and a simplex $\sigma$ with vertices in $L$, we define
\begin{equation}\label{eqn-forbidden-interval-1}
    F_{{\varpi}}(p, \sigma) \stackrel{{\rm def}}{=} D(p, \sigma)^{2} +
    d(p, N_{{\varpi}}(\sigma_{p}))^{2} - R_{{\varpi}}(\sigma_{p})^{2}.
\end{equation}
Note that $F_{{\varpi}}(p, \sigma)$ depends on the weights of the
vertices of $\sigma_{p}$ and not on the weight of $p$. This
crucial fact will be used in the  
analysis of the algorithm.

If $\sigma$
is a candidate simplex of $p$,
the {\em forbidden interval} of $\sigma$ with respect to $p$ is
\begin{equation}\label{eqn-forbidden-interval-2}
    I_{{\varpi}}(\sigma,p) \stackrel{{\rm def}}{=}
    \left[ F_{{\varpi}}(p,\sigma)- \frac{\eta}{2}, \,
    F_{{\varpi}}(p,\sigma)+ \frac{\eta}{2} \right],
\end{equation}
where
\begin{equation}\label{eqn-forbidden-interval-3}
    \eta \stackrel{{\rm def}}{=}
    2^{14} \left( \Gamma_{0} + \frac{\delta^{2}_{0}}{\Gamma^{m}_{0}}
    \right)\lambda^{2}.
\end{equation}

The following  result relates candidate simplices,
forbidden intervals and stable weight
assignments. The proof is included in
Appendix~\ref{appendix-length-forbidden-interval}.
\begin{lem}
    \label{lem-length-forbidden-interval}
    Let $L\subset \M$ be a $(\lambda, 2\lambda)$-net of $\M$
    with $ \lambda < \frac{1}{18} (1 - \sin\theta_{0})^{2} \reach$ and
    ${\varpi} : L \rightarrow [0, \infty)$ be a weight assignment with
    $\widetilde{{\varpi}} \leq \tilde{\alpha}_{0}$. Let, in addition, $p$ be
    a point of $L$, and $\sigma$ a candidate simplex of $p$.
    If there exists an ewp  ${\varpi}_{1}$ of ${\varpi}$ satisfying
    $\widetilde{{\varpi}}_{1}\leq \alpha_{0}$ and $\sigma \in {\rm K}_{{\varpi}_{1}}(L)$,
    then ${\varpi}(p)^{2} \in I_{{\varpi}}(p, \sigma)$.
\end{lem}
Lemma~\ref{lem-length-forbidden-interval} shows that the emergence of
a candidate simplex $\sigma$ incident to $p$ in the weighted cocone
complex under an ewp implies that the original weight of $p$ had to be
in the forbidden interval, i.e., $I_{\varpi}(p,\sigma)$.

The following lemma shows that {\em good weights}, i.e., weights
which do not lie in any forbidden intervals,
exist, which ensures that
the algorithm will terminate.

\begin{lem}[Existence of good weights]\label{lem-existence-of-good-weights}
    Assume that $\lambda \leq \frac{\reach}{512}$,  and
    $\Gamma_{0}$, $\delta_{0}$ and $\tilde{\alpha}_{0}$
    $(= \sqrt{\alpha^{2}_{0} - \delta^{2}_{0}})$
    satisfy 
    \begin{equation}\label{eqn-ineq-gamma-delta-alpha}
        \Gamma_{0}+\frac{\delta^{2}_{0}}{\Gamma^{m}_{0}} <
	\frac{\tilde{\alpha}_{0}^{2}}{2^{14} N},
    \end{equation}
    where $N = 2^{O(m^{2})}$ and will be defined explicitly in the proof.
    Then, at the $i^{th}$ step, one can find a weight
    ${\varpi}_{i}(p_{i}) \in [0, \tilde{\alpha}_{0}\, nn(p_{i}) ]$ outside the
    forbidden intervals of the candidate simplices of $S(p_{i})$. Moreover, ${\varpi}_{i}$
    satisfies properties {\bf C1} and {\bf C2}.
%
\end{lem}



Using simple packing arguments and Lemma~\ref{lem-sampling-property-manifold}~(1), we get the
following bound (similar arguments were used, for example, in~\cite[Lem.~9]{giesen2004}
and \cite[Lem.~4.12]{boissonnat2011tancplx}).
\begin{lem}\label{lem-bound-on-neighbors}
    If $\lambda \leq \frac{\reach}{512}$, then for any $p\in L$,
    $\# (B(p, 16\lambda) \cap L ) \leq 66^{m} \stackrel{\rm def}{=} N_{1}$.
\end{lem}

\begin{proof}[of Lemma~\ref{lem-existence-of-good-weights}]
%
    Write    $S(p_{i})$ for the set of candidate simplices of $p_{i}$.
    We have
    $$
      \# S(p_{i}) \leq N \stackrel{\rm def}{=} \sum_{j=2}^{m+1} N_{1}^{j}.
    $$

    For all ${\varpi} : L \rightarrow [0, \infty)$ with $\widetilde{{\varpi}}\leq {\alpha}_{0}$,
    we get from Lemmas~\ref{lem-property-cocone-complex}~(2) and~\ref{lem-bound-on-neighbors}
    that 
    the set of $\Gamma_{0}$-slivers of dimension $\leq m+1$
    in ${\rm K}_{{\varpi}}(L)$ that are incident to $p_{i}$ is a
    subset of $S(p_{i})$.

	Recall, from Section~\ref{ssec-general-notations},
	that for $X \subseteq \mathbb{R}$, $\mu(X)$ denotes the 
	standard Lebesgue measure of $X$. 
    Since
    \begin{eqnarray*}
	\mu \left(\bigcup_{\sigma \in S(p_{i})} I_{{\varpi}_{i-1}}(\sigma, p_{i})\right)
	&\leq&
        \sum_{\sigma \in S(p_{i})} \mu (I_{{\varpi}_{i-1}}(\sigma, p_{i}))\\
	&\leq& N \eta \\
	&<& \tilde{\alpha}_{0}^{2}\lambda^{2} \\
	&\leq& \tilde{\alpha}_{0}^{2}\, nn(p_{i})^{2},
    \end{eqnarray*}
    we can select ${\varpi}(p_i) \in [ 0, \tilde{\alpha}_{0}\, nn(p_{i})]$
    such that ${\varpi}(p_{i})^{2}$ is
    outside the forbidden intervals of the candidate simplices of $p_{i}$,
    i.e.,
    $$
      {\varpi}(p_{i})^{2} \not\in
      \bigcup_{\sigma \in S(p_{i})} I_{{\varpi}_{i-1}}(\sigma, p_{i}).
    $$
    By Lemma~\ref{lem-length-forbidden-interval},
    the weight assignment ${\varpi}_{i}$ we obtain is a locally stable weight assignment for $p_{i}$.
%
%
%
%
%
\end{proof}
%
%
%
%
%
The following lemma shows that getting a locally stable weight
assignment ${\varpi}_{i}$ at each iteration of the algorithm gives
a globally stable weight assignment ${\varpi}_{n}$ at the end of
the algorithm.

\begin{lem}
    The weight assignment ${\varpi}_{n} : L \rightarrow [0, \infty)$
    is stable.
\end{lem}
\begin{proof}
    It is easy to see that $\widetilde{{\varpi}}_{n} \leq \tilde{\alpha}_{0}$,
    since for all $p \in L$, the weights were chosen from the
    interval $[0, \, \tilde{\alpha}_{0} \, nn(p)]$.

	
    We will prove the stability of ${\varpi}_{n}$ by contradiction.
    Let $\xi: L \rightarrow [0, \infty)$ be an ewp of ${\varpi}_{n}$
    that modifies the weight of $q \in L$, and
    assume that there exists a $\Gamma_{0}$-sliver
    $\sigma = [p_{i_{0}}, \, \dots, \, p_{i_{k}}] \in {\rm K}_{\xi}(L)$.
    Note that $\tilde{\xi} \leq \alpha_{0}$, and that for any $p \in {\sigma}$,
    $\sigma \in S(p)$
    (from the definition of $S(p)$ and Lemma~\ref{lem-bound-on-neighbors}).
    Without loss of generality assume that
    $$
      i_{0}< \dots < i_{k}.
    $$
    We will have to consider the following two cases:

    \paragraph{Case 1.} $q$ is not a vertex of $\sigma$.
    This implies that
    $\sigma \in {\rm K}_{{\varpi}_{n}}(L)$ since $\xi(x) = {\varpi}_{n}(x)$ for all
    $x \in L\setminus \{q\}$, and $\xi(q) \geq {\varpi}_{n}(q)$. Using the same arguments, we can
    show that $\sigma \in {\rm K}_{{\varpi}_{i_{k}}}(L)$.
    From Lemma~\ref{lem-existence-of-good-weights}
    and the fact that ${\varpi}_{i_k}$ is an ewp of itself,
    we have reached a contradiction as ${\varpi}_{i_k}$
    is a locally stable weight assignment for $p_{i_k}$.


    \paragraph{Case 2.} $q$ is a vertex of $\sigma$. Using the same arguments
    as in  Case~1 we can show that
    $\sigma \in {\rm K}_{\xi_{1}}(L)$ where $\xi_{1}: L \rightarrow [0, \infty)$
    is a weight assignment satisfying: $\xi_{1}(q) = \xi(q) $
    and $\xi_{1}(x) = {\varpi}_{i_{k}}(x)$ for all $x \in L \setminus \{q\}$.
    Observe that $\xi_{1}$ is an ewp of ${\varpi}_{i_{k}}$.
    As in Case~1, we have reached a contradiction since ${\varpi}_{i_{k}}$
    is a locally stable weight assignment for $p_{i_k}$.
%
\end{proof}

\subsubsection{Complexity of the algorithm}
\label{ssec-complexity-algorithm}

The following theorem easily follows from the algorithm and the previous analysis.
\begin{thm}\label{lem-complexity}
	Let $\M$ be a $m$-dimensional submanifold of $\R^{d}$, and 
	let $W \subset \M$ be an $\e$-sample of $\M$ and $L \subseteq W$ be a 
	$\lambda$-net of $W$ with $\e \leq \lambda$. Also, assume that the
	parameters $\alpha_{0}$, $\delta_{0}$ and $\Gamma_{0}$ satisfy 
	Equation~\eqref{eqn-ineq-gamma-delta-alpha} from 
	Lemma~\ref{lem-existence-of-good-weights}, and $\e$ and $\lambda$ 
	satisfy 
	Equations~\eqref{equation-main-structural-theorem-1} and 
	\eqref{equation-main-structural-theorem-2} from 
	Theorem~\ref{thm-output-weighted-rDT-eqal-wit}.
  The time and space complexity of
  Algorithm~\ref{algorithm-1}
  are
  $$
    d \# L \left(2^{O(m)} \# L + 2^{O(m^{2})} + O(m) \# W \right) + O(m^{3}\# W)
  $$
  and
  $$
    d\# W+ \# L \left( 2^{O(m^{2})}+d \right) + O(m\#L \times \#W)
  $$
  respectively.
\end{thm}
\begin{proof}
    In the initialization phase of the algorithm one needs to
    compute $nn(p)$ and $N(p)$ for all $p \in L$. Time time complexity
    for this part of the procedure will be $d 2^{O(m)} (\# L)^{2}$.

    Inside the for-loop for each $p_{i} \in L$, one needs to do the following:
    \begin{enumerate}
        \item
            compute candidate simplices $S(p_{i})$,

        \item
            compute $I = \bigcup_{\sigma \in S(p_{i})} I_{\varpi_{i-1}}(\sigma, p_{i})$,

        \item
            find a point $x$ in
            $\left[ 0, \tilde{\alpha}_{0}^{2}\, nn(p_{i})^{2}\right] \setminus I$.
    \end{enumerate}
    Observe that for all $\sigma^{j} \subseteq N(p_{i})$ with $p_{i} \in \sigma$
    and $j \leq m+1$, we need to check if $\sigma^{j}$ is in $S(p_{i})$ and the
    time complexity for this procedure for a given $\sigma^{j}$ is $d\, 2^{O(j)}$.    
	The bound follows from the facts that the number of 
	faces of $\sigma^{j}$ is $2^{j}$, and for a given face $\sigma^{k} \leq \sigma^{j}$ 
	and a vertex $q \in \sigma^{k}$, we can compute $D(p,\sigma^{k})$ in time 
	complexity $O(d\, {\rm poly}(k))$
	\footnote{Note that ${\rm poly}(k)$ denotes a polynomial in $k$ of degree $O(1)$.
	}.
    If $\sigma^{j} \in S(p_{i})$, then the time complexity of computing
    $I_{\varpi_{i-1}}(\sigma^{j}, p_{i})$ is $O(d \, {\rm poly}(j))$. Time complexity of
    computing $I$ and finding a point
    $x \in \left[ 0, \tilde{\alpha}_{0}^{2}\, nn(p_{i})^{2}\right] \setminus I$
    will be at most $O((\# S(p_{i}))^{2})$. Since $\# S(p_{i}) = 2^{O(m^{2})}$,
    therefore the overall time complexity for one execution of the for-loop will be
    $d 2^{O(m^{2})}$.

    The above discussion implies the overall time complexity of computing a stable
    weight assignment $\varpi : L \rightarrow [0, \infty)$ will be
    $d\, 2^{O(m^{2})} (\# L)$.

    Once a stable weight assignment has been computed then
    we can use Boissonnat and Maria's simplex-tree based
    witness complex computation algorithm~\cite[Sec.~3.2]{DBLP:journals/algorithmica/BoissonnatM14}
    for computing $\wit_{\varpi}(L,W)$.
    Observe that since $\varpi: L \rightarrow [0, \infty)$ is a stable
    weight assignment the $m$-skeleton\footnote{
    For a simplicial complex $\mathcal{K}$, the $m$-skeleton of ${\mathcal K}$
    is set of simplices of ${\mathcal K}$ of dimension at most $m$.}
    of $\wit_{\varpi}(L,W)$
    is equal to $\wit_{\varpi}(L, W)$,
    see Theorem~\ref{thm-output-weighted-rDT-eqal-wit}, and therefore we will only
    compute in the algorithm the $m$-skeleton of $\wit_{\varpi}(L,W)$.
    Note that for constructing $m$-skeleton of $\wit_{\varpi}(L, W)$,
    the simplex-tree based algorithm of Boissonnat and
    Maria~\cite{DBLP:journals/algorithmica/BoissonnatM14} will need access to
    $(m+1)$-nearest $\varpi$-weighted neighbors in $L$ for
    each $w \in W$. The time complexity for computing
    this $(m+1)$-nearest $\varpi$-weighted neighbors in $L$
    for all witness in $W$ will be $O(m \# L \times \# W)$. The time complexity of
    Boissonat and Maria's witness complex construction algorithm
    will be
    $$
        O\left( m^{3}\left(\#\wit_{\varpi}(L,W) + \# W \right)\right)
        = O\left( 2^{O(m^{2})} \# L + m^{3} \# W\right).
    $$
    The last bound follows from the fact that
    $\#\wit_{\varpi}(L,W) = 2^{O(m^{2})}\# L$. Note that the upper bound on
    $\#\wit_{\varpi}(L,W)$ follows from the facts that the dimensions
    of simplices in $\wit_{\varpi}(L,W)$ are at most $m$, using $\e \leq \lambda$
    and from triangle inequality we have
    $L$ is a $(\lambda, 2\lambda)$-net
    of ${\mathcal M}$,
    for all
    $\sigma \in \wit_{\varpi}(L,W)$ we have $\Delta(\sigma)\leq 16 \lambda$
    (see Lemma~\ref{lem-property-cocone-complex-basic}),
    and for all $p \in L$ we have
    $\# \left(B(p, 16\lambda)\cap L\right) = 2^{O(m)}$
    (see Lemma~\ref{lem-bound-on-neighbors}).

    Combining everything, we get that the time complexity of the
    algorithm is
    $$
        d\, \# L \left( 2^{O(m)} \# L + 2^{O(m^{2})} + O(m) \#W\right)\#L + O(m^{3}\# W).
    $$

    The space complexity of storing $\wit_{\varpi}(L,W)$ in the simplex-tree
    data structure is bounded by $O\left( \# \wit_{\varpi}(L,W)\right)$. Hence the
    overall space complexity of the
    algorithm is
    $$
        d\# W+ (2^{O(m^{2})}+d)\#L + O(m\#L \times \#W).
    $$
    Note that the $2^{O(m^{2})}\# L$ term bounds the space complexity of
    storing $N(p)$ and $S(p)$ for all $p \in L$ and $\wit_{\varpi}(L,W)$, 
    the $O(m\#L \times \#W)$ term comes
    from storing the $\varpi$-weighted $(m+1)$-nearest neighbors in $L$
    for each witness $w$ in $W$, and finally the terms $d \# L$ 
    and $d \# W$ come from storing the coordinates 
    of the points in $L$ and $W$ respectively.
\end{proof}

\subsection{Regarding the assumptions}
\label{ssec-regarding-assumptions}

We have assumed that we know the dimension of the
manifold $m$, and the value of $\lambda$ (having an upper
bound would have been good enough) where $L$ is a $\lambda$-net
of $W$.


We will address the second question first. Given a point sample $W$,
and beginning with an arbitrary point from $W$, it is simple to show that
a furthest point sampling~\cite{DBLP:journals/tcs/Gonzalez85}
from $W$ will generate a
$\lambda$-net of $W$, for some $\lambda >0$, and it is possible
to keep track of the value of $\lambda$. For an analysis of this
procedure, refer to~\cite[Lem.~5.1]{boissonnat2009}.

Let $\pts \subset \M$ be an  $(\nu, \epsilon)$-net of $\M$.
If $\frac{\nu}{\epsilon} = O(1)$ and if we
know an upper bound on this quantity
and if $\epsilon \leq \epsilon_{0}$, where $\epsilon_{0}$
depends only on the reach and the dimension of $\M$, 
then we can learn the local dimension of the manifold at each sample
point with time and space complexity $2^{O(m)}(\#\pts)^2$ and $2^{O(m)}\#\pts$
respectively,
see~\cite{DBLP:journals/ijcga/ChengWW08,cheng2009dimension,giesen2004}.
Note that, in these papers, the dimension estimation is done
locally around each sample point and
therefore is exactly in the spirit of this paper.
For a more detailed discussions on these things refer to
Section~\ref{sec:conclusion-only-distances-required}.


\section{Conclusion: only distances required}
\label{sec:conclusion-only-distances-required}

The algorithm we have outlined can be simply adapted to
work in the setting where the
input is just a {\em distance matrix} corresponding to a dense point sample
on the submanifold $\M$. Rather than giving explicit
coordinates of the points, we will be given a distance matrix
$M= (a_{ij})$ where $a_{ij} = \|p_{i} - p_{j}\|$ and
$p_{i}, \, p_{j} \in W$.

In our
reconstruction algorithm, we have to compute things like
local neighbors $N(p)$, candidate simplices $S(p)$,
forbidden intervals $I_{\varpi}(\sigma,p)$, and the
witness complex. We will end the section with a discussion
on the sampling conditions, extension to noisy
distance matrix and comparisons with other methods.

\paragraph{$\lambda$-net $L$ of $W$.}
As already discussed in Section~\ref{ssec-complexity-algorithm},
the distance matrix can be used to generate a $\lambda$-net
$L$ of $W$ by repeatedly inserting a farthest point.


\paragraph{Computing $N(p)$ and $S(p)$.}
Computing
$N(p)$ is simple. For computing $S(p)$, we need to compute
the altitude of simplices. This reduces to
computing volume of simplices, 
$$
  D(p, \sigma^{j}) = \frac{j \vol(\sigma^{j})}{\vol(\sigma^{j}_{p})},
$$
which can be done from the knowledge of the lengths of its edges.
To see this,
observe that for a simplex $\sigma = [p_{0},\, \dots , \, p_{k} ]$
$$
  \vol(\sigma) = \frac{1}{k!} \sqrt{\mid \det M(\sigma) \mid},
$$
where $M(\sigma) \stackrel{\rm def}{=} \left(b_{ij}\right)_{ 1\leq i,\, j \leq k}$ with
$$
  b_{ij} = \langle p_{i} - p_{0}, \, p_{j} - p_{0}\rangle
  = \frac{\| p_{i} - p_{0}\|^{2} + \| p_{j} - p_{0}\|^{2} - \| p_{i} - p_{j} \|^{2}}{2}.
$$

The above discussion shows that $N(p)$ and $S(p)$ can be computed directly from
the distance matrix.

\paragraph{Computing forbidden intervals $I_{\varpi}(\sigma,p)$.}
Assume $\sigma$ is a $k$-simplex.
Recall that computing $I_{\varpi}(\sigma,p)$ will boil down to
computing $D(p,\sigma)$, $d(p, N_{\varpi}(\sigma_{p}))$ and $R_{\varpi}(\sigma_{p})$,
see Equations~\eqref{eqn-forbidden-interval-1}, \eqref{eqn-forbidden-interval-2} and \eqref{eqn-forbidden-interval-3}.
We have already discussed how to compute $D(p,\sigma)$, but observe that
$d(p, N_{\varpi}(\sigma_{p}))$ and $R_{\varpi}(\sigma_{p})$ can be computed
if we can find a distance preserving embedding of $\sigma$ into an Euclidean space.
Since we know the pairwise distance between vertices of the simplex, a distance preserving embedding
of $\sigma$ can be computed in $O(k^{3})$, where $\sigma$ is a $k$-simplex.
See~\cite{discrete-geometry-Matousek,metric-embeddings-Matousek}.

\paragraph{Dimension estimation from distance matrix.}
We will now show that using known algorithms, such as, for
example,~\cite{giesen2004,cheng2009dimension,DBLP:journals/ijcga/ChengWW08},
on dimension estimation of submanifolds, one can estimate the
dimensions of submanifolds from distance matrices.  We will be
adapting Cheng, Wang and Wu's
approach~\cite{DBLP:journals/ijcga/ChengWW08}.  Let $L \subset \M$ be
a $(\lambda, 2\lambda)$-net of the manifold $\M$ and $p \in L$.  For
the time being we will assume that we have explicit coordinates of the
points in the $\R^{d}$, the ambient space, and $\reach =1$.  We want
to estimate the unknown dimension $m$ of the manifold at $p$.
Cheng et al.~\cite{DBLP:journals/ijcga/ChengWW08}
showed that if $\lambda$ is less than some $\lambda_{0}$, where
$\lambda_{0}$ depends only on $m$, then
there
exist an absolute constant $C$ such that the
{\em covariance matrix}
\footnote{
Covariance matrix $Cov(V)$ of the set of vectors $V =\{ v_{1}, \, \dots, \, v_{l}\}$
in $\reel^{d}$ is defined as
$$
	Cov(V) = \sum_{i=1}^{l} (v_{i1}, \, v_{i2}, \, \dots, \, v_{id})^{T}
	(v_{i1}, \, v_{i2}, \, \dots, \, v_{id}),
$$
where $v_{i} = (v_{i1}, \, v_{i2}, \, \dots, \, v_{id})$. Observe that
$Cov(V)$ is a $d\times d$ matrix.
}
of the
set of vectors $\left\{ q - p \mid q \in X(p)\right\}$, where
$X(p) = \{ q \in L \mid \|p-q\| \leq C\lambda\}$, has a sharp gap between the
top $m$ eigenvalues of the covariance matrix and the rest of the eigenvalues.
More explicitly, if the $\beta_{1} \geq \beta_{2} \geq \dots \geq \beta_{d}$
are the eigenvalues of the covariance matrix then
\begin{eqnarray}
    \frac{\beta_{k}}{\beta_{i}} = \Theta(1) \quad \mbox{and} \quad
	\frac{\beta_{j}}{\beta_{i}} = O(\lambda^{2}),
	\quad \quad \mbox{where $1 \leq i, \, k \leq m$ and $m+1 \leq j \leq d$}
\end{eqnarray}
Using this gap Cheng et al.~\cite{DBLP:journals/ijcga/ChengWW08}
estimated the dimension of $\M$ at $p$.
If the set $X(p)$ is given, then the running time of this
algorithm will be $O(d 2^{O(m)})$.
First note that using the sparsity condition on $L$, as in
Lemma~\ref{lem-bound-on-neighbors},
we can show that $\# X(p) = 2^{O(m)}$, and secondly the set $X(p)$
can be computed directly from the distance matrix.
The part about
covariance matrix can be done by first computing a distance
preserving Euclidean embedding of the point set $X(p)$ and then
checking the gap in the eigenvalues of the covariance matrix
constructed using the coordinate of the embedded points.
Note that the dimension of the Euclidean space where the points in $X(p)$
will be embedded is bounded by $\# X(p) = 2^{O(m)}$.
Therefore given the set $X(p)$, the
time complexity of the dimension procedure will be
$2^{O(m)}$.

\paragraph{Computing witness complex.}
By its very definition, the witness
complex can be built from an interpoint distance matrix. So, we can
easily adapt our algorithm, without increasing its complexity, to the
setting of interpoint distance matrices, which was not possible  with the other
reconstruction algorithms that explicitly need coordinates of the
points~\cite{cheng2005,boissonnat2009,boissonnat2011tancplx}.

\paragraph{Noisy distances and geodesic distances.}
The algorithm given in this paper is quite robust to noise and other
distortions of the Euclidean distances. For example, as we discuss
below, it could accommodate a distance matrix defined via
\emph{geodesic distances} on the manifold.
Recalling the input to the Algorithm~\ref{algorithm-1},
let $W \subset \M$ be an $\e$-sample of $\M$, and $L\subset W$
be a $\lambda$-net of $W$
with $\e \leq \lambda$. This implies that $L$ is a
$(\lambda, 2\lambda)$-net of $\M$. Without loss of generality we will
assume that $\reach = 1$.

We will have access to a {\em noisy distance matrix}
$\tilde{d}(\cdot , \cdot)$ of $W$ satisfying
the following inequality:
$$
\left| \frac{\tilde{d}(p,q)}{\|p-q\|} - 1 \right| \leq \gamma.
$$
We will call $\gamma$ the {\em error fraction} of the noisy distances.
We will assume that $\gamma = O(\lambda^{\rho})$ for some $\rho >0$.
This is a standard noise
model~\cite{DBLP:journals/siamcomp/NiyogiSW11, DBLP:journals/ijcga/ChengWW08,
DBLP:journals/comgeo/ChengFGKPR05,DBLP:journals/comgeo/DeyG06}, and 
we will show how the algorithm given in this paper can be adapted to 
handle this amount of noise, i.e., the case when the error fraction 
$\gamma = O(\lambda^{\rho})$ for some constant $\rho > 0$. 

One of the central objects in the paper is thickness, but for
the discussion on noisy distances we will be using the notion of
{\em fatness}. The fatness of a $j$-simplex $\sigma$ is defined as
$$
	\Theta(\sigma) = \frac{\vol(\sigma)}{\Delta(\sigma)^{j}}.
$$
Like in the case of thickness,
using the definition of fatness one can characterize bad simplices.
\begin{definition}[$\Theta_{0}$-good simplices and $\Theta_{0}$-slivers]
  \label{def:slivers-new}
  \label{def:good.simplex-new}
 Let $\Theta_{0}$ be a positive real
number smaller than one.
 A simplex $\splxs$ is \defn{$\Theta_{0}$-good} if
  $\Theta(\splxs^j) \geq \Theta_{0}^j$ for all $j$-simplices
  $\splxs^j \leq \splxs$. A simplex is \defn{$\Theta_{0}$-bad} if it
  is not $\Theta_{0}$-good. A \defn{$\Theta_{0}$-sliver} is
  a $\Theta_{0}$-bad simplex in which all the proper faces are
  $\Theta_{0}$-good.
\end{definition}
Thickness and fatness are analogous concepts and the calculations done in this
paper can easily be done using fatness. For more details
on fatness refer to the manifold reconstruction
paper of Boissonnat and Ghosh~\cite{boissonnat2011tancplx}.

Let $\sigma$ be a $j$-simplex with vertices from $L$, and let
$\widetilde{\vol}(\sigma)$ and $\widetilde{\Delta}(\sigma)$ denote
{\em noisy volume} and {\em noisy diameter} of $\sigma$
obtained by using noisy distances. Expressions for
$\widetilde{\vol}(\sigma)$ and $\widetilde{\Delta}(\sigma)$
for the simplex $\sigma = [p_{0}, \, \dots ,  \, p_{j}]$
are the
following:
$$
    \widetilde{\Delta}(\sigma) \stackrel{{\rm def}}{=}
    \max_{p_{i}, p_{k} \in \sigma} \tilde{d}(p_{i},p_{k}),
$$
and
$$
    \widetilde{\vol}(\sigma) \stackrel{{\rm def}}{=}
    \sqrt{\det (\widetilde{M}(\sigma))},
$$
where $\widetilde{M}(\sigma)
\stackrel{\rm def}{=} \left(a_{ik}\right)_{ 1\leq i,\, k \leq j}$ with
$$
  a_{ik} = \frac{\tilde{d}(p_{i}, p_{0})^{2} + \tilde{d}(p_{k}, p_{0})^{2} -
  \tilde{d}(p_{i},p_{k})^{2}}{2}.
$$
We came up with this definition of $\widetilde{M}(\sigma)$ to mimic the
definition of $M(\sigma)$ given earlier in the section.

We will call
$\widetilde{\Theta}(\sigma) :=
\frac{\widetilde{\vol}(\sigma)}{\widetilde{\Delta}(\sigma)^{j}}$
the {\em noisy fatness} of the $j$-simplex $\sigma$.
Using~\cite[Lem.~4.3.6]{ghosh:hal-01095861} we can show that noisy fatness and
actual fatness are closely related:
\begin{equation}
	\left| \frac{\widetilde{\Theta}(\sigma)^{2}}{\Theta(\sigma)^{2}} -1 \right| \leq O(j \lambda^{\rho}).
\end{equation}
This shows that we can modify Algorithm~\ref{algorithm-1} in terms of fatness and
the candidate
simplices can be detected using noisy fatness. Note that candidate simplices in
the case of a noisy distance matrix will be defined using $\Theta_{0}$-slivers
(defined in terms of fatness) rather than $\Gamma_{0}$-slivers (defined in
terms of thickness).

The calculation for computing forbidden intervals in
Section~\ref{appendix-length-forbidden-interval} can be easily extended to the
case of noisy distances. Also the dimension estimation
algorithm of Cheng et al.~\cite{DBLP:journals/ijcga/ChengWW08}
works for noisy point samples and also
can be directly extended to the case of noisy distance matrix.
This shows that the algorithm given in this paper can be adapted to 
handle this amount of noise, i.e., the case when the error fraction 
$\gamma = O(\lambda^{\rho})$ for some constant $\rho > 0$.

Another interesting problem to consider is the case when the
entries to the distance matrix are
{\em geodesic distances}, i.e., the entries in the distance matrix are
geodesic distances $d_{\M}(p,q)$ between the points $p$ and $q$ on
the manifold $\M$. For the case of submanifolds,
Niyogi, Smale and Weinberger~\cite[Prop.~6.3]{niyogi2008}
proved the following result
connecting geodesic distance and Euclidean distances in a small neighborhood
of the submanifold $\M$.
\begin{lem}
	Let $p$ and $q$ be points on the submanifold $\M$ of $\reel^{d}$ with $\reach =1$ and
	$\|p-q\| \leq \frac{1}{2}$. Then
	$$
		\left| \frac{d_{\M}(p,q)}{\|p-q\|} - 1 \right| \leq O\left(\|p-q\|\right).
	$$
\end{lem}
Therefore for the case of geodesic distances we are again back to the noisy
distance framework with the error fraction $\gamma = O(\lambda)$.

\paragraph{Sampling conditions.}
The sampling condition,
i.e., the bound on $\lambda$ in Theorem~\ref{thm-output-weighted-rDT-eqal-wit}
is quite pessimistic
with respect to the reach of the manifold.
Naturally this makes the results in this paper to be of more theoretical
interest and less relevant for applications. But we feel that the
techniques introduced in this paper could be used to get
better reconstruction algorithms, and or design new and better
reconstruction heuristics. Also, note that the only assumption
in this paper was that $\M$ is a smooth submanifold with positive reach.
If one restrict to a more narrow class of manifolds
such as compact flat manifolds
then one could easily improve on the
sampling conditions. For more details on triangulating
closed Euclidean Orbifolds refer to a recent paper of
Caroli and Teillaud~\cite{DBLP:journals/dcg/CaroliT16}.

\paragraph{Comparison with previous works.}
It is a natural question to ask how other manifold reconstruction
algorithms~\cite{cheng2005,boissonnat2009,boissonnat2011tancplx}
will fair when given only a distance matrix to work with.
We will assume that input is a distance matrix corresponding
to a dense point sample $\pts$ of $\M$.
Since all the
previous algorithms work with explicit coordinates of the points
in the point sample one needs to start by first getting a
distance preserving embedding into an Euclidean space.
Note that this can be done in time $O((\# \pts)^3)$ via
{\em Cholesky factorization} of positive definite matrix.
For more details on Cholesky factorization refer
to~\cite{matrix-computations, trefethen1997}, and
for details on distance preserving embedding into Euclidean space
see~\cite{discrete-geometry-Matousek,metric-embeddings-Matousek}.
Once we have the embedding then
we can apply the previous manifold reconstruction
algorithms~\cite{cheng2005,boissonnat2009,boissonnat2011tancplx}.

We will now outline some of the issues with this approach:
\begin{enumerate}


    \item {\bf Time complexity of getting a distance
    preserving embedding.} The distance
    preserving embedding will be computed via
    Cholesky factorization with the time
    complexity $O((\# \pts)^{3})$.

    \item {\bf Noisy distance matrix and geodesic distances.}
    These approaches won't work if the distance matrix given is
    noisy. Since the entries in the noisy distance matrix
    may not be Euclidean distance like for example when the
    entries are geodesic distance on the manifold $\M$. 
    Since in these cases
    the entries to the
    noisy distance matrix are not Euclidean distances, the
     approach via distance preserving Euclidean embedding
     won't work.

    \item {\bf Large dimension of the embedding space.}
    Observe that the dimension $d$ of the Euclidean space we can get
    via the above distance preserving embedding can be as large as
    $d = \Omega(\# \pts)$. This will make time complexity of
    the manifold reconstruction algorithms given
    in~\cite{cheng2005,boissonnat2009}
    exponential in $\#\pts$, since both these approaches compute
    Voronoi diagrams in the ambient space using the whole
    point sample or a subset of the sample.

    \item {\bf Problems with tangential Delaunay complex.}
        Once we have the distance preserving embedding of the point sample
        $\pts$ we can use the tangential Delaunay complex (TDC)
        for reconstruction~\cite{boissonnat2011tancplx}.
        As already mentioned, the dimension of the Euclidean space obtained
        from a distance preserving embedding of $\pts$ can be as large as
        $\Omega(\# \pts)$. Note that TDC
        construction will compute the following structures:
        \begin{enumerate}
            \item The
                approximate tangent space at each point of the point sample.
                This if the dimension of the embedding space is $\Omega(\# \pts)$
                will incur an $\Omega(\# \pts)$ time complexity.

            \item The
                TDC construction will need to compute
                Delaunay triangulations restricted to the approximate tangent
                spaces. So the problem of using complicated and highly
                structured data structure still stays.

            \item
                In the TDC construction one needs to compute $(m+1)$-dimensional
                simplices corresponding to each {\em inconsistent
                simplex}. See the definition of inconsistent simplex
                from~\cite{boissonnat2011tancplx}. The time complexity for
                constructing each of these simplices will be $\Omega(\# \pts)$
                if the dimension of the Euclidean space obtained from a
                distance preserving embedding is $\Omega(\# \pts)$.
        \end{enumerate}
        As with the other reconstruction algorithms~\cite{cheng2005,boissonnat2009}, the
        TDC won't be able to handle noisy distance matrix as input or when the
        entries to the distance matrix are geodesic distances on the manifold $\M$.
\end{enumerate}

\section*{Acknowledgements}

This research has been partially supported by the 7th Framework
Programme for Research of the European Commission, under FET-Open
grant number 255827 (CGL Computational Geometry Learning).
Partial support has also been provided by the Advanced Grant of the European
Research Council GUDHI (Geometric Understanding in Higher Dimensions).

Arijit Ghosh is supported by Ramanujan Fellowship (No. SB/S2/RJN-064/2015).
Part of this work was done when Arijit Ghosh was a Researcher
at Max-Planck-Institute for Informatics, Germany supported by
the Indo-German Max Planck Center for Computer Science (IMPECS).
Part of this work was also done when Arijit Ghosh was a
Visiting Scientist
at Advanced Computing and Microelectronics Unit,
Indian Statistical Institute, Kolkata, India.

\appendix

\section{Proof of Theorem~\ref{thm-output-weighted-rDT-eqal-wit}}
\label{ssec-faces-inherit-protection}

This section is devoted to the proof of
Theorem~\ref{thm-output-weighted-rDT-eqal-wit}.
When we talk about
properties P1, P2 and P3 in this section, we are actually
referring to properties introduced in
Section~\ref{ssec-stability-protection-wit-complex}.

Before we go into the detailed calculations, we want to make this
small and obvious observation that directly follows from
triangle inequality.
\begin{observation}
	If $W$ is an $\e$-sample of $\M$ and $L$ a $\lambda$-net
	of $W$ with $\e \leq \lambda$, then $L$ is a $(\lambda, 2\lambda)$-net
	of $\M$.
\end{observation}

\subsection{Proof of properties {\bf P1}, {\bf P2} and {\bf P3}}
\label{appendix-properties-p1-p2-p3}

We will use the following structural result from~\cite{boissonnat2009}.
\begin{lem}\label{lem-bound-cocone-p-voronoi-diagram}
    Let $\theta \in [0, \frac{\pi}{2})$, and $\pts \subset \M$ be an $\epsilon$-sample of $\M$
    with $\epsilon < \frac{1}{9} (1-\sin \theta)^{2} \reach$. For any weight assignment
    ${\varpi} : L \rightarrow [0, \infty)$ with $\widetilde{{\varpi}} < \frac{1}{2}$, for any $p \in \pts$
    and $x \in \vor_{{\varpi}}(p) \cap {\rm K}^{\theta}(p)$, we have
    $$
      \|p-x\| \leq \frac{3\epsilon}{1-\sin \theta} \; .
    $$
\end{lem}

Following lemma is a direct consequence of
Lemma~\ref{lem-bound-cocone-p-voronoi-diagram}.
\begin{lem}\label{lem-property-cocone-complex-basic}
    Let $L$ be an $\epsilon$-sample of $\M$ with $\epsilon < \frac{1}{9}(1-\sin \theta_{0})^{2}\reach$,
    and let ${\varpi} : L \rightarrow [0, \infty)$ be a weight assignment with
    $\widetilde{{\varpi}} < \frac{1}{2}$.
    Let $\sigma \in {\rm K}_{{\varpi}}(L)$.
        \begin{enumerate}
            \item
                Let $p$ be a vertex of $\sigma$ with $\vor_{{\varpi}}(p)\cap {\rm K}^{\theta_{0}}(p)\neq \emptyset$.
                For all vertices $q$ of $\sigma$ and $x \in \vor_{{\varpi}}(p)\cap {\rm K}^{\theta_{0}}(p)$,
                we have $\|x - q\| <4\epsilon$. 
                This implies for all the vertices $q$ of $\sigma$, $\| q - C_{{\varpi}}(\sigma) \| < 4\epsilon$.

            \item
                $\Delta(\sigma) < 8\epsilon$.
		\end{enumerate}
\end{lem}
\begin{proof}
	1. Observe that
	\begin{align*}
		\|q-x\| &= \sqrt{\|p-x\|^{2} - \varpi(p)^{2} + \varpi(q)^{2}}&  \\
		&\leq \sqrt{\left( \frac{3\epsilon}{1-\sin \theta_{0}}\right) + \epsilon^{2}}&
		\mbox{from Lemma~\ref{lem-bound-cocone-p-voronoi-diagram},
		Lemma~\ref{lem-distance-point-on-M-with-sample}~(1) and
		$\widetilde{\varpi} < \frac{1}{2}$}\\
		&< 4\epsilon.&
	\end{align*}
	The bound on $\|q - C_{\varpi}(\sigma)\|$ follows from the fact that
	$\|q-x\| \geq \|q - C_{\varpi}(\sigma)\|$.
	
	2. The bound on $\Delta(\sigma)$ follows from part (1) and triangle inequality.
\end{proof}

The following corollary about witness complex is
from~\cite[Cor.~7.6]{deSilva2008}.

\begin{cor}\label{thm-vin-de-silva-inclusion-witness-del-add}
    For any subsets $W,\, L \subseteq \R^{d}$ with $L$ finite,
    for any ${\varpi} : L \rightarrow
    [0,\infty)$, we have $\wit_{{\varpi}}(L, W) \subseteq \del_{{\varpi}}(L)$.
    Moreover, for any simplex $\sigma$ of $\wit_{{\varpi}}(L, W)$, the weighted
    Voronoi face of $\sigma$ intersects the convex hull of the
    ${\varpi}$-witnesses (among the points of $W$) of
    $\sigma$ and of its subsimplices.
\end{cor}

Following result is a direct consequence of
Lemma~\ref{lem-distance-point-on-M-with-sample}~(2).

\begin{lem}\label{lem-new-properties-witness-complex}
  Let $W \subseteq \M$ be an $\e$-sample of $\M$, $L \subseteq W$
  be a $\lambda$-net of $W$ with $\lambda+\e < \frac{\reach}{4}$,
  and ${\varpi} : L \rightarrow [0, \infty)$ be a weight assignment
  with $\widetilde{{\varpi}} < \frac{1}{2}$.
  \begin{enumerate}
   \item
      For all $pq \in \wit_{{\varpi}}(L, W)$,
      $\|p-q\| \leq (4+10\widetilde{{\varpi}})(\lambda+\e)$.

   \item
      Let $\sigma \in \wit_{{\varpi}}(L, W)$. The distance between
      any vertex $p$ of $\sigma$ and any witness $w$ of $\tau \leq \sigma$
      is at most $(5+12\widetilde{{\varpi}})(\lambda+\e)$.
  \end{enumerate}
\end{lem}
\begin{proof}
	1. First observe that $L$ is a $(\lambda + \e)$-sample of $\M$ via triangle
	inequality.
	Let $w \in W$ be a ${\varpi}$-witness for the edge $pq \in \wit_{{\varpi}}(L, W)$,
	and without loss of generality assume that $p$ is closest neighbor of $w$
	in terms of the weighted distance.
	Then from Lemma~\ref{lem-distance-point-on-M-with-sample}~(2), we have
	$\|w - p\| \leq (1+2\widetilde{{\varpi}})(\lambda+\e)$ and
	$\|w - q\| \leq (3+8\widetilde{{\varpi}})(\lambda+\e)$. Therefore again from triangle
	inequality we get $\|p-q\| \leq (4+10\widetilde{{\varpi}})(\lambda+\e)$.
	
	2. Let $p$ be a vertex of $\sigma$, and
	let $q \in \tau$ is the vertex
	closest to $w$ in terms of the weighted distance. Then from part (1),
	we have $\|p-q\| \leq (4 + 10\widetilde{{\varpi}})(\lambda+\e)$. From
	Lemma~\ref{lem-distance-point-on-M-with-sample}~(2), we have
	$\|w-q\| \leq (1+2\widetilde{{\varpi}})(\lambda + \e)$. From triangle
	inequality, we have $\|p-w\| \leq (5+12\widetilde{{\varpi}})(\lambda+\e)$.
\end{proof}

We will now give the proof of Properties
{\bf P1}, {\bf P2}, and {\bf P3}.
\begin{lem}\label{lem-property-cocone-complex}
    Let $L$ be an $\epsilon$-sample of $\M$ with $\epsilon < \frac{1}{9}(1-\sin \theta_{0})^{2}\reach$,
    and let ${\varpi} : L \rightarrow [0, \infty)$ be a weight assignment with
    $\widetilde{{\varpi}} < \frac{1}{2}$.
    Let $\sigma \in {\rm K}_{{\varpi}}(L)$.
        \begin{enumerate}
%

            \item
                {\bf (Property P1)}
		Assume $\dim \sigma = k \leq m$
                and $\Upsilon(\sigma) \geq \Gamma^{k}_{0}$.
                Additionally, if $\epsilon < \frac{\reach}{8}$ then
                for all vertices $p$ of $\sigma$
                we have
                $$
                    \sin \angle(\aff(\sigma), T_{p}\M) \leq \frac{8\epsilon}{\Gamma^{m}_{0}\reach} .
                $$
            \item
		{\bf (Property P2)}
                If ${\rm K}_{{\varpi}}(L)$ does not contain any $\Gamma_{0}$-slivers
                of dimension $\leq m+1$ and
                $$
                     \epsilon < \frac{\Gamma_{0}^{2m+1}\reach}{12},
                $$
                then dimension of maximal simplices
                in ${\rm K}_{{\varpi}}(L)$ is at most $m$.

	    \item
	      {\bf (Property P3)}
	      Additionally, if $L$ is $\frac{\epsilon}{2}$-sparse
	      and
	      $$
		  \epsilon \leq \frac{\Gamma^{m}_{0} \sin \theta_{0}}{2^{9}} \reach ,
	      $$
	      then
	      for all $m$-simplex
	      $\sigma^{m} \in \del_{\varpi}(L, T\M)$ and $\forall~p\in \sigma$,
	      we have $\vor_{{\varpi}}(\sigma^{m})\cap T_{p}\M \neq \emptyset$.
        \end{enumerate}
\end{lem}
\begin{proof}
%
  1. Follows directly from
  Corollary~\ref{cor-angle-bound-simplex-tangent-space}
  and Lemma~\ref{lem-property-cocone-complex}~(2).

  2. Using part 1 and exactly the proof idea used in the proof
  of~\cite[Lem.~4.9]{boissonnat2011tancplx}, we can show that
  all $m+1$-simplices in ${\rm K}_{{\varpi}}(L)$
  are either $\Gamma_{0}$-bad or have thickness
  $\frac{12\epsilon}{\Gamma^{m}_{0}\reach}$. Using the
  bound on $\epsilon$, we can complete the proof of part 4.

  3. Using the facts that $L$ is $\frac{\epsilon}{2}$-sparse and
  $\widetilde{{\varpi}} < \frac{1}{2}$, we can show that 
  for all simplices $\sigma$ with vertices from $L$, we have 
  $\forall~p \in \sigma$,
  \begin{equation}\label{equation-lem-property-cocone-complex}
  	d(p, N_{{\varpi}}(\sigma^{m})) \geq \frac{3\epsilon}{16}.
  \end{equation}
  
  To reach a contradiction let $p$ be a vertex of
  $\sigma^{m}$ such that $\vor_{{\varpi}}(\sigma^{m})\cap T_{p}\M \neq \emptyset$
  and $q$ be a vertex of $\sigma^{m}$ with $\vor_{{\varpi}}(\sigma^{m})\cap T_{q}\M = \emptyset$.
  Using the bound on $\epsilon$ and part 1 of this lemma, we can show,
  for all $x \in \sigma^{m}$,
  that
  $$
    \sin(\aff\sigma^{m}, T_{x}\M) \leq
    \sin \theta \stackrel{\rm def}{=} \frac{8\epsilon}{\Gamma^{m}_{0}\reach} < \frac{1}{2}
  $$
  and
  $$
    \# (N_{{\varpi}}(\sigma^{m}) \cap T_{x}\M) = 1.
  $$
  Observe that $C_{\varpi}(\sigma^{m})$ is orthogonal projection of $c$
  onto $\aff(\sigma^{m})$ where $c \in \vor_{{\varpi}}(\sigma^{m})\cap T_{p}\M$. 
  Let $c'$ denotes the intersection of $N_{\varpi}(\sigma^{m})$
  and $T_{q}\M$. Note that, by construction, the line segment $cc' \in N_{\varpi}(\sigma^{m})$.
  See Figure \ref{figure-showing-inheritance-of-protection-new}.

  From part (1) of this lemma, we have for all $x \in \sigma^{m}$,
  \begin{eqnarray}
  	\| x - C_{\varpi}(\sigma^{m}) \| \leq \| x - c \| \leq 4 \epsilon.
  \end{eqnarray}

	Therefore
	\begin{eqnarray*}
		\|c - c'\| &\leq& \|c - C_{\varpi}(\sigma^{m})\| + \|C_{\varpi}(\sigma^{m}) - c'\| \\
		&\leq& \|p-c\| \sin \theta + \|q- C_{\varpi}(\sigma^{m})\|\tan\theta \\
		&\leq& 4 \epsilon \sin \theta \left( 1+\frac{1}{\cos \theta}\right) \\
		&<& 12 \epsilon \sin \theta
	\end{eqnarray*}

  Using the facts that $\epsilon \leq \frac{\Gamma_{0}^{m} \sin \theta_{0}}{2^{9}} \reach$
  and $d(p, N_{{\varpi}}(\sigma^{m})) \geq \frac{3\epsilon}{16}$,
  see Equation~\eqref{equation-lem-property-cocone-complex},
  we can show that the line segment $cc' \in {\rm K}(p)$.
  So, $\vor(\sigma^{m})\cap T_{p}\M = \emptyset$ implies there exists a
  $\sigma^{m+1} \in {\rm K}_{{\varpi}}(L)$ with $\sigma^{m} < \sigma^{m+1}$.
  We have reached a contradiction via part 2 of this lemma.
%
%
\end{proof}

\begin{figure}
  \begin{center}
    \includegraphics[width=3.50in]{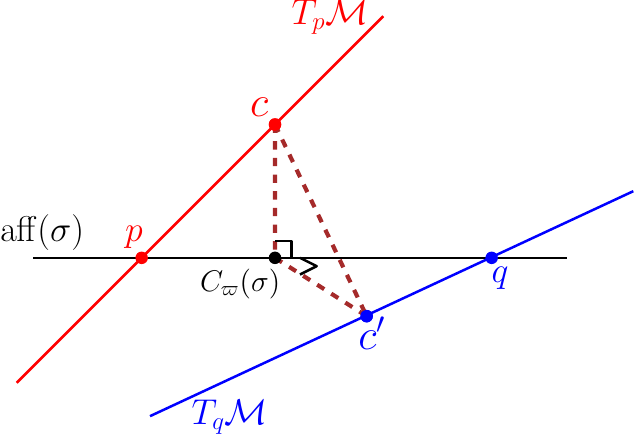}
  \end{center}
  \caption{Diagram for the Lemma~\ref{lem-property-cocone-complex}~(3).}
  \label{figure-for-property-P3-new}
\end{figure}

\subsection{Proof of Theorem~\ref{thm-output-weighted-rDT-eqal-wit}}

Homemorphism and geometric guarantees given in
Theorem~\ref{thm-output-weighted-rDT-eqal-wit}
is a direct consequence of the following
lemma\footnote{Note that this lemma is a special
case of the result proved in~\cite{boissonnat2011tancplx}.}
from~\cite{boissonnat2011tancplx}.

\begin{lem}
    Let $L \subset \M$ be an $(\lambda, 2\lambda)$-net of $\M$,
    and ${\varpi} : L \rightarrow [0, \infty)$
    be a weight assignment satisfying the following properties:
    \begin{enumerate}
        \item
            $\widetilde{{\varpi}} \leq \alpha_{0}$.

        \item
            Dimension of maximal simplices in $\del_{{\varpi}}(L, T\M)$ is equal to $m$

        \item
            All the simplices in $\del_{{\varpi}}(L, T\M)$ are $\Gamma_{0}$-good.

        \item
            For all $\sigma = [p_{0}, \, \dots, \, p_{k}] \in \del_{{\varpi}}(L, T\M)$
            and $\forall~ i \in \{0, \, \dots, \, k \}$,
            $\vor_{{\varpi}}(p_{i})\cap T_{p_{i}}\M \neq \emptyset$.

    \end{enumerate}
    There exists $\lambda_{0} > 0$ that depends only on $\alpha_{0}$, $\Gamma_{0}$
    and $m$ such that for $\lambda \leq \lambda_{0}$, $\del_{{\varpi}}(\pts, T\M)$
    is homeomorphic to and a close geometric approximation of $\M$.
\end{lem}

Rest of this section is devoted to the proof of
$$
	\wit_{\varpi}(L,W) = \del_{\varpi}(L, T\M).
$$
The proof goes through multiple stages.
\begin{description}
	\item[Stage~(1).]
		We will first show, in Lemma~\ref{lem-conditions-weighted-wit-subseteq-tancomplex},
        that no $\Gamma_{0}$-slivers
		of dimension $\leq m+1$
		in ${\rm K}_{\varpi}(L)$
		implies $\wit_{\varpi}(L,W) \subseteq \del_{\varpi}(L, T\M)$.
		
	\item[Stage~(2).]
        In Lemma~\ref{lem-props-we-want-omega-to-satisfy} we prove that
		if $\varpi : L \rightarrow [0,\infty)$ is a stable weight
		assignment then for all $m$-simplices $\sigma \in \del_{\varpi}(L,T\M)$
		and $p \in \sigma$, $\sigma$ will be $\delta^{2}$-power protected
		on $T_{p}\M$.
	
	\item[Stage~(3).]
		Let $\del_{\varpi}(L,T\M)$ does not contain any $\Gamma_{0}$-slivers
		and for all $m$-simplex $\sigma^{m} \in \del_{\varpi}(L,T\M)$	 and
		$\forall\, p \in \sigma^{m}$, $\sigma^{m}$ is $\delta^{2}$-power
		protected on $T_{p}\M$. Then for all $\sigma \in \del_{\varpi}(L,T\M)$,
		$\sigma$ will be $\Omega\left(\frac{\delta^{2}}{m+1}\right)$-power
		protected on $\M$. See, Lemma~\ref{lem-rDt-protected-M}.
	
	\item[Stage~(4).]
		Finally in Lemma~\ref{lem-conditions-rDt-subseteq-witness},
        we will show that conditions in {\bf Stage~(3)} implies
		$$
            \del_{\varpi}(L,T\M) \subseteq \wit_{\varpi}(L,W).
        $$
\end{description}
To see that this will complete the proof, observe that
$\varpi$ being stable weight assignment implies $K_{\varpi}(L)$
contains no $\Gamma_{0}$-slivers of dimension $\leq m+1$. This would
imply dimension of maximal simplices in $K_{\varpi}(L)$ is at most $m$.
Once we have this, simply going through {\bf Stage~(1)} till {\bf Stage~(4)}
will give us the result.

\begin{lem}
    \label{lem-conditions-weighted-wit-subseteq-tancomplex}
    Let $W \subseteq \M$ be a $\e$-sample of $\M$, $L\subset W$ be a $\lambda$-net
    of $W$ with $\e \leq \lambda$, and ${\varpi} : L \rightarrow [0, \, \infty)$ be a weight assignment
    with $\widetilde{{\varpi}} < \frac{1}{2}$ and ${\rm K}_{{\varpi}}(L)$ does not
    contain any $\Gamma_{0}$-sliver of dimension $\leq m+1$.
    If
    $$
        \lambda < \min\left\{
        \frac{3 \sin\theta_{0}}{2^{11}(1+m)}, \,
        \frac{\Gamma_{0}^{2m+1}}{24}
        \right\}\, \reach
    $$
    then
    $$
        \wit_{{\varpi}}(L, W) \subseteq \del_{{\varpi}}(L, T\M) .
    $$
\end{lem}
\begin{proof}
    Note that $L$ is a $(\lambda, 2\lambda)$-net of $\M$.

    To reach a contradiction, let $\sigma^{k}$ be a $k$-simplex in
    $\wit_{{\varpi}}(L, W)$ and $p$ be a vertex of $\sigma$ such that
    $\vor_{{\varpi}}(\sigma^{k})\cap T_{p}\M = \emptyset$.

    Let $w \in W$ be a ${\varpi}$-witness of a subface of $\sigma^{k}$.
    From Lemma~\ref{lem-new-properties-witness-complex}, and the facts
    that
    $\widetilde{{\varpi}}< \frac{1}{2}$ and $\e \leq \lambda$, we have
    $$
      \|p-w\| \leq (5 + 12\widetilde{{\varpi}})(\lambda + \e)< 22\lambda.
    $$
    From Lemma~\ref{lem-sampling-property-manifold}, we have
    $$
      \dist{w}{T_{p}\M} \leq \frac{2\times 11^{2} \lambda^{2}}{\reach}.
    $$

    From Corollary~\ref{thm-vin-de-silva-inclusion-witness-del-add}, there exist
    $c_{k} \in \vor_{{\varpi}}(\sigma^{k})$ that lies in the convex hull of the
    ${\varpi}$-witness of $\sigma^{k}$ (in $W$) and its subfaces. This implies,
    $$
        \rho \stackrel{\rm def}{=} \dist{c_{k}}{T_{p}\M} \leq
	\frac{2\times 11^{2}\lambda^{2}}{\reach} .
    $$
    Note that since $L$ is $\lambda$-sparse
    and $\widetilde{\varpi} < \frac{1}{2}$,
    $$
    		\|p - c_{k}\| \geq d(p, C_{\varpi}(\sigma^{k})) \geq \frac{3\lambda}{8}.
                \footnote{Let $p$ and $q$ be distinct vertices of $\sigma^{k}$,
    and let $x$ be an orthogonal projection of $C_{\varpi}(\sigma^{k})$
    on the line
through
$p$ and $q$. Since $\widetilde{\varpi} < \frac{1}{2}$,
    $x \in pq$.
Using the facts that
    $d(x, p^{\varpi}) = d(x, q^{\varpi})$, $\|p-q\| \geq \lambda$
    and $\varpi(q) < \frac{\|p-q\|}{2}$ (as $\widetilde{\varpi} < 1/2$), we get
    \begin{eqnarray}
    		\|p-x\| \; = \; \frac{\|p-q\|}{2} \left( 1+ \frac{\varpi(p)^{2} - \varpi(q)^{2}}{\| p - q \|^{2}} \right)
    		\; \geq \; \frac{\|p-q\|}{2} \left( 1 - \frac{\varpi(q)^{2}}{\|p-q\|^{2}}\right)
    		\; \geq \; \frac{3\lambda}{8}
    \end{eqnarray}
    The bound on $d(p, C_{\varpi(\sigma^{k})})$ follows from the fact that
    $d(p, x) \geq d(p, C_{\varpi}(\sigma^{k}))$.
    }
	$$

    Note that $\lambda$ is sufficiently small such that
    \begin{equation}\label{equation-1-wit-subset-tancomplex}
        \frac{\rho}{\frac{3\lambda}{8} - 2m\rho} \leq \sin\theta_{0}
    \end{equation}
	and
    \begin{equation}\label{equation-2-wit-subset-tancomplex}
        \sin \theta \stackrel{\rm def}{=}
	\frac{16\lambda}{\Gamma^{m}_{0} \reach} < \frac{\sqrt{3}}{2}.
    \end{equation}

    We will now generate sequence of simplices
    $$
        \sigma^{k} < \sigma^{k+1} < \dots < \sigma^{m} < \sigma^{m+1}
    $$
    and points
    $$
	c_{k}, \, c_{k+1}, \, \dots, \, c_{m}, \, c_{m+1}
    $$
    by walking on $\vor_{{\varpi}}(\sigma^{k})$ satisfying the following properties:
    \begin{description}
        \item[Prop-1.]
            For all $\sigma^{k+i}$, there exists
	    $c_{k+i} \in \vor_{{\varpi}}(\sigma^{k+i})$
            such that
            $$
	      \dist{c_{k+i}}{T_{p}\M} \leq \rho
	    $$
	    and
	    $$
	      \| p - c_{k+i} \| \geq \|p-c_{k}\| - 2i \rho.
	    $$
            From Eq.~\eqref{equation-1-wit-subset-tancomplex},
	    this implies
	    $\sigma^{k+i} \in {\rm K}^{\theta_{0}}_{{\varpi}}(L)$.

        \item[Prop-2.]
            For all $\sigma^{k+i}$, we have
	    $$
	      \vor_{{\varpi}}(\sigma^{k+i})\cap T_{p}\M = \emptyset.
	    $$
    \end{description}
    Note that once we have shown that such sequence of simplices exists, then we would have reached a
    contradiction from Lemma~\ref{lem-property-cocone-complex}~(2).

      We will now show how to generate the above sequence of simplices.
    \begin{description}
        \item[Base case.]
	    From Eq.~\eqref{equation-1-wit-subset-tancomplex},
	    it is easy to see that $\sigma^{k}$
	    and $c_{k}$ satisfy Prop-1 and Prop-2.

        \item[Inductive step.] Wlog lets assume that we have generated
	till $\sigma^{k+i}$,
        satisfying properties {\bf Prop-1} and {\bf Prop-2}, and
        we also assume $k+i \leq m$. Since
	$\sigma^{k+i} \in {\rm K}_{{\varpi}}(L)$, we can
        show, using Lemma~\ref{lem-property-cocone-complex}~(1),
	that
	$$
	  \sin \angle (N_{p}\M, N_{{\varpi}}(\sigma^{k+i})) \leq \sin \theta.
        $$
	From {\bf Prop-1}, we have $\|p - c_{k+i}\| \geq \|p - c_{k} \| - 2i\rho$ and
        $\dist{c_{k+i}}{T_{p}\M} \leq \rho$.
	Therefore, from Eq.~\ref{equation-2-wit-subset-tancomplex},
        there exists
        $\tilde{c}_{k+i} \in T_{p}\M \cap N_{{\varpi}}(\sigma^{k+i})$ such that
        $$
	  \| c_{k+i} - \tilde{c}_{k+i} \| \leq \frac{\rho}{\cos \theta}\leq 2\rho.
	$$
        As $\vor_{{\varpi}}(\sigma^{k+i})\cap T_{p}\M = \emptyset$ hence there exists
        $c_{k+i+1} \in [c_{k+i}, \, \tilde{c}_{k+i})$ such that $c_{k+i+1}\in \vor_{{\varpi}}(\sigma^{k+i+1})$
        with $\sigma^{k+i} < \sigma^{k+i+1}$. Note that, as in the base case, we can show that
        $$
            \dist{c_{k+i+1}}{T_{p}\M} \leq \rho
        $$
        and
        $$
            \| p - c_{k+i+1} \| \leq \| p - c_{k} \|  - 2(i+1)\rho \,.
        $$
    \end{description}
\vskip-1em
\end{proof}

The following lemma connects power protection of $m$-dimensional simplices
in $\del_{{\varpi}}(T\M)$ with stability of ${\varpi}$.
%
\begin{lem}\label{lem-props-we-want-omega-to-satisfy}
    Let $L \subset \M$ be a $(\lambda, 2\lambda)$-net of $\M$
    with
    $$
      \lambda < \min \left\{ \frac{\Gamma^{m}_{0} \sin \theta_{0}}{2^{10}}, \,
      \frac{\Gamma^{2m+1}_{0}}{24} \right\} \reach,
    $$
    and let ${\varpi}: L \rightarrow [0, \, \infty)$,
    with $\widetilde{{\varpi}} \leq \tilde{\alpha}_{0}$,
    be a stable weight assignment.
%
%
%
%
    Then all the $m$-simplices $\sigma \in \del_{{\varpi}}(L, T\M)$
    are $\delta^{2}$-power protected on $T_{p}\M$ for all $p \in {\sigma}$,
    where $\delta = \delta_{0}\lambda$.
\end{lem}
\begin{proof}
    Note that $L$ is $(\lambda, 2\lambda)$-net of $\M$.
    For all $m$-simplices $\sigma$ in $\del_{{\varpi}}(L, T\M)$, we have
    $\vor_{{\varpi}}(\sigma)\cap T_{p}\M \neq \emptyset$ for all $p \in {\sigma}$.
%

    To reach a
    contradiction, lets assume that $\sigma \in
    \del_{\varpi}(L, T\M)$ to be not $\delta^{2}$-power protected
    on $T_{p}\M$ for some $p \in \sigma$.
    Let $c  \in \vor_{{\varpi}}(\sigma)\cap T_{p}\M$
    and $q \in L \setminus \sigma$ such that
    for all $x \in {\sigma}$
    $$
        \|q-c\|^{2} - {\varpi}(q)^{2} - \delta^{2} \leq \|x-c\|^{2}-{\varpi}(x)^{2}.
    $$
    Let $\beta^{2} = \|q-c\|^{2}-\|p-c\|^{2} - ({\varpi}(q)^{2}-{\varpi}(p)^{2})$,
    where $p \in {\sigma}$. Note that $\beta \leq \delta$.

    Let $\xi : L \rightarrow [0, \infty)$
    $$
          \xi(x) = \left\{
        \begin{array}{l l}
            {\varpi}(x) & \quad \text{if $x\neq q$}\\
            \sqrt{{\varpi}(q)^{2}+\beta^{2}} & \quad \text{if $x=q$}\\
        \end{array} \right.
    $$
    Since $L$ is $\lambda$-sparse, $\delta = \delta_{0}\lambda$ and
    $\tilde{\alpha}_{0}^{2}+ \delta^{2}_{0} \leq \alpha^{2}_{0}$,
    we have $\tilde{\xi} \leq \alpha_{0}$. It is easy to see $\xi$ is an
    ewp of ${\varpi}$, and the $(m+1)$-dimensional simplex
    $\tau = q* \sigma \in {\rm K}_{\xi}(L)$.
	As $\varpi$ is stable weight assignment and $\xi$	
	is an ewp of $\varpi$, we get a contradiction from
	Lemma~\ref{lem-property-cocone-complex}~(2) and the fact that
	$\lambda < \min \left\{ \frac{\Gamma^{m}_{0} \sin \theta_{0}}{2^{10}}, \,
      \frac{\Gamma^{2m+1}_{0}}{24} \right\} \reach$.
%
%
%
%
\end{proof}

We will need the
following result due to Boissonnat and Ghosh~\cite[Lem~2.2]{boissonnat2009}.
\begin{lem}\label{lem-restricted-voronoi-lower-dim-flats}
    Let $L \subset \R^{d}$ be a point set, ${\varpi}: L \rightarrow [0, \infty)$
    be a weight distribution, and $H \subseteq \R^{d}$ be a $k$-dimensional flat. Also,
    let $L'$ denotes the projection of the point set $L$ onto $H$, and $p'$
    denotes the projection of $p \in L$ onto $H$. For all $p \in L$, we have
    $$
        \vor_{{\varpi}}(p)\cap H = \vor_{\xi}(p'),
    $$
    where $\xi : L' \rightarrow [0, \infty)$ with
    $$
        \xi(q')^{2} = {\varpi}(q)^{2}-\|q-q'\|^{2}+\max_{x \in \pts} \|x-x'\|^{2}
    $$
    and $\vor_{\xi}(p')$ denotes the Voronoi diagram of $p'$ in $H$ and not in $\R^{d}$.
\end{lem}
From Lemma~\ref{lem-restricted-voronoi-lower-dim-flats},
we have get the following corollary.
\begin{cor}\label{cor-from-lem-restricted-voronoi-lower-dim-flats}
    Let $L \subset \R^{d}$ be a finite set, ${\varpi}: L \rightarrow [0, \infty)$, and let
    $H \subseteq \R^{d}$ be $k$-flat. For a point $p \in L$, if $\vor_{{\varpi}}(p)\cap H$
    is bounded then the dimension of maximal simplices incident to $p$ in
    $\del_{{\varpi}}(L, H) \stackrel{{\rm def}}{=} \left\{ \sigma :
    \vor_{{\varpi}}(\sigma)\cap H \neq \emptyset \right\}$
    is greater than $k$.
\end{cor}

%
%
%

Following lemma connects power protection of $m$-simplices on the tangent space
to that on the manifold.

\begin{lem}\label{lem-rDt-protected-M}
    	Let $L \subset \M$ be a $(\lambda, 2\lambda)$-net of $\M$, and
	$\delta = \delta_{0}\lambda$ with $\delta_{0} < 1$. Let
	${\varpi} : L \rightarrow [0, \infty)$ be a weight
    assignment with $\widetilde{\varpi} < \frac{1}{2}$ and
	satisfying the following properties:
    	\begin{enumerate}
%
%
	  \item
            $\del_{{\varpi}}(\pts, T\M)$ does not contain any $\Gamma_{0}$-sliver, and

	  \item
            $\forall~\sigma^{m} \in \del_{{\varpi}}(\pts, T\M)$,
	    $\sigma^{m}$ is $\delta^{2}$-power protected
            on $T_{p}\M$ for all $p \in \sigma^{m}$.
    	\end{enumerate}
    	If
	$$
	    \lambda \leq \frac{\Gamma^{m}_{0}\reach}{2^{11}},
	$$
	then all $\sigma \in \del_{{\varpi}}(L, T\M)$ are
	$\delta^{2}_{1}$-power protected on $\M$ where
	$$
	  \delta_{1}^{2} = \frac{\delta^{2}}{m+1} - \frac{B\lambda^{3}}{\reach}
	$$
	and $B \stackrel{\rm def}{=} 2^{15}$.
\end{lem}
\begin{proof}
    Let $p$ be a point in $L$, and $L'$ denotes the projection of the point sample
    $L$ onto $T_{p}\man$. For a point $x \in L$, $x'$ is the projection of $x$
    onto $T_{p}\man$ and vise versa, and similarly, let
    $\sigma = [p_{0}, \, \dots, \, p_{k}]$ be a simplex with $p_{i}$'s in $L$
    then $\sigma'$ denotes the simplex $[p_{0}', \, \dots, \, p_{k}']$ and vise
    versa. Note that $p' = p$.

    The weight assignment $\xi: L' \rightarrow [0, \, \infty)$ is defined
    in the following way:
    \begin{equation*}
        \xi(x')^{2} = {\varpi}(x)^{2} - \|x-x'\|^{2} + \max_{y \in L} \|y- y'\|^{2}\, .
    \end{equation*}
    For $\sigma' \subseteq L'$, $\vor_{\xi}(\sigma')$ denotes the Voronoi
    cell in $T_{p}\M$ and not in $\R^{d}$.

    From Lemmas~\ref{lem-restricted-voronoi-lower-dim-flats} and
    \ref{lem-property-cocone-complex-basic}~(1) we have:
    	\begin{description}
        		
	\item[Prop.~(a)]
            For $\sigma \subseteq L$, $\vor_{{\varpi}}(\sigma)\cap T_{p}\M = \vor_{\xi}(\sigma')$.

        \item[Prop.~(b)]
            $\vor_{{\varpi}}(p)\cap T_{p}\M = \vor_{\xi}(p) \subset B(p, 8\lambda)
	    \cap T_{p}\M$.
    	\end{description}
    From Prop.~(a) and the definition of tangential complex, if $\sigma' \in \st(p;\del_{\xi_{p}}(L'))$
    then $\sigma \in \del_{{\varpi}}(L, T\M)$.
    Since all the $m$-simplices of $\del_{{\varpi}}(L, T\M)$ are $\delta^{2}$-power protected
	on the tangent space of the vertices (Hyp.~4), therefore,
    from the definition of $\xi: \rightarrow [0, \infty)$, all the $m$-simplices
    $\sigma'$ incident to $p$ in $\del_{\xi}(L')$ are also $\delta^{2}$-power protected
    on $T_{p}\M$, i.e., there exists $x \in \vor_{\xi_{p}}(\sigma')$ such that for all
    $q' \in {\sigma}'$ and $r' \in L'\setminus {\sigma}'$
    \begin{equation*}
        \|r' - x \|^{2} - \xi(r')^{2}>  \|q' - x \|^{2} - \xi(q')^{2}+ \delta^{2}\, .
    \end{equation*}

    Following properties are a direct consequence of Prop.~(b), and
    Lemmas~\ref{lem:dim-voronoi-protected-simplex}~(1)
    and \ref{thm-protection-lower-dim-simplices}
    \begin{description}
        \item[Prop.~(c)]
            Dimension of maximal simplices incident to $p$ in
	    $\del_{\xi}(L')$ is equal to $m$.

        \item[Prop.~(d)]
            Let $\sigma'$ be a $m$-simplex incident to $p$ in
	    $\del_{\xi}(L')$. Then
            $\vor_{\xi}(\sigma') = C_{\xi}(\sigma')$.

        \item[Prop.~(e)]
            Let $\sigma' \in \del_{\xi}(L')$ be a $j$-simplex incident,
	    with $p \in \sigma'$,
            then $\sigma'$ is $\frac{\delta^{2}}{m-j+1}$-power protected.

    \end{description}
    Note that Prop.~(c) and the definition of tangential complex implies the following
    \begin{description}
        \item[Prop.~(f)]
            Dimension of maximal simplices in $\del_{{\varpi}}(L, T\M)$ is equal to $m$.
    \end{description}

    We will now prove the power protection of simplices in $\del_{{\varpi}}(L, T\M)$
	on the manifold $\M$.
	Let $\sigma \in \del_{{\varpi}}(L, T\M)$ be a $k$-simplex, with $k \leq m$, incident to $p$.
    From Prop.~(e), $\exists ~c' \in \vor_{\xi}(\sigma')$ such that
	$\forall ~ x' \in {\sigma}'$ and $\forall~y' \in L' \setminus {\sigma}'$
	$$
		\|y' - c'\|^{2} - \xi(y')^{2} > \|x' - c'\|^{2} - \xi(x')^{2} + \frac{\delta^{2}}{m+1} .
	$$
	Which, from the definition of $\xi$ and Prop.~(a), implies
	$\forall ~ x \in {\sigma}$ and $\forall ~y \in L \setminus {\sigma}$
	$$
		\|y - c' \|^{2} - {\varpi}(y)^{2} > \|x - c' \|^{2} - {\varpi}(x)^{2} + \frac{\delta^{2}}{m+1} ,
	$$
	and $c' \in \vor_{{\varpi}}(\sigma)$.	

	Let $\hat{c}$ be the point closest to $c'$ on $\M$ and $c$ denotes the point closest to $c'$ in
	$\M \cap N_{{\varpi}}(\sigma) $.

	Using the facts that $\|p-c'\| \leq 8\lambda$
	(from Lemma~\ref{lem-property-cocone-complex-basic}~(1)) and
    \begin{equation}\label{eqn-bound-distance-c'-hat-c}
        \| c' - \hat{c} \|  \leq \frac{2^{7}\lambda^{2}}{\reach}
	\leq \frac{\lambda}{16} < \frac{\reach}{25}
    \end{equation}
	from Lemma~\ref{lem-sampling-property-manifold}~(3) and
	$\lambda \leq \frac{\Gamma_{0}^{m}\reach}{2^{11}} \leq \frac{\reach}{2^{11}}$,
	we get
    \begin{eqnarray}\label{eqn-bound-distance-c'-p}
        \| p - \hat{c} \| \leq \| p - c' \| + \| c' - \hat{c} \|
	\leq \left(8+\frac{1}{16}\right)\lambda
	< \frac{\reach}{4}.
    \end{eqnarray}
    Therefore, using
    $\sin \angle (\aff(\sigma), T_{p}\M) \leq \frac{16\lambda}{\Gamma_{0}^{m}\reach}$
    (from Lemma~\ref{lem-property-cocone-complex}~(1))
    and $\sin \angle (T_{p}\M, T_{\hat{c}}\M) < \frac{6\|p-\hat{c}\|}{\reach}$
    (from Lemma~\ref{lem-sampling-property-manifold}~(3) and
    $\|p-\hat{c}\| < \frac{\reach}{4}$), we have
	\begin{align*}
	  \sin \angle (\aff\sigma, T_{\hat{c}}\M) 	
	  &\leq \sin \angle (\aff\sigma, T_{p}\M) + \sin \angle(T_{p}\M, T_{\hat{c}}\M)& \\
	  &\leq \frac{16\lambda}{\Gamma^{m}_{0}\reach} + \frac{6 \|p-\hat{c}\|}{\reach} &\\
	  &\leq \frac{16\lambda}{\Gamma^{m}_{0}\reach} + \frac{387\lambda}{8\,\reach}&
	  \mbox{as $\|p-\hat{c}\| \leq \frac{129\lambda}{16}$}\\
	  &\leq \frac{1}{4}&\mbox{as $\lambda \leq
	  \frac{\Gamma^{m}_{0}\reach}{2^{11}}$}
	\end{align*}
    Using the above bound on $\sin \angle (\aff\sigma, T_{\hat{c}}\M)$,
    the fact that
    $\|c' - \hat{c}\| \leq \frac{2^{7}\lambda^{2}}{\reach} < \frac{\reach}{25}$
    (Eq.~\eqref{eqn-bound-distance-c'-hat-c})
    and Lemma~\ref{lem:flat.intersect}, we get
    $$
        \|c' - c\| \leq 4 \|c' - \hat{c}\| \leq
        \frac{{2}^{9}\lambda^{2}}{\reach} \stackrel{{\rm def}}{=} \frac{C\lambda^{2}}{\reach}
        \leq \frac{\lambda}{4}.
    $$

    Let $q \in L \setminus {\sigma}$ and $p \in {\sigma}$.
    We will consider the following two cases:
	\begin{description}
		\item[Case-1.]
		  $\|q-c\|^{2} > \|p-c\|^{2} + 2(2\lambda)^{2}$. Using the
		  facts that $ {\varpi}(q) \leq 4\alpha_{0}\lambda$
		  (from part 2(a) of
		  Lemma~\ref{lem-distance-point-on-M-with-sample})
            and $\alpha_{0}< \frac{1}{2}$, we have
			\begin{align*}
			    \|q-c\|^{2} - {\varpi}(q)^{2} - (\|p-c\|^{2} - {\varpi}(p)^{2})
			    &> 8\lambda^{2} - {\varpi}(q)^{2} + {\varpi}(p)^{2}& \\
			    &> 	8\lambda^{2}-{\varpi}(q)^{2}&\\
			    &>  4\lambda^{2}&
			\end{align*}

		\item[Case-2.] $\|q-c\|^{2} \leq \|p-c\|^{2}+ 8\lambda^{2}$.
			This implies $\|q-c\| < \|p-c\| + 3\lambda$.
			
			Using the facts that
			$\|p-c'\| \leq 8\lambda$ (from Lemma~\ref{lem-property-cocone-complex-basic}~(1)),
			$\|c-c'\| \leq \frac{C\lambda^{2}}{\reach} \leq \frac{\lambda}{4}$,
			\begin{eqnarray*}
				\|p-c\| & \leq& \| p-c' \| + \|c-c'|\| \leq
				\left(8+\frac{1}{4}\right)\lambda,~\mbox{and} \\
				\|q-c'\| &\leq& \|q-c \|+ \|c-c'\| \leq \left(8+\frac{13}{4}\right)\lambda,
			\end{eqnarray*}
			we get
			\begin{align*}
				\|q- c \|^{2} - {\varpi}(q)^{2} &\geq (\|q-c'\| - \|c-c'\|)^{2} - {\varpi}(q)^{2}& \\
				&\geq \|q-c'\|^{2} - {\varpi}(q)^{2} - 2\|c-c'\| \|q-c'\|& \\
				&> \|p-c'\|^{2} - {\varpi}(p)^{2} +\frac{\delta^{2}}{m+1}- 2\|c-c'\| \|q-c'\|& \\
				&\geq (\|p-c\| - \|c-c'\|)^{2} -{\varpi}(p)^{2}+\frac{\delta^{2}}{m+1}- 2\|c-c'\| \|q-c'\|&\\
				&\geq \|p-c\|^{2} -{\varpi}(p)^{2} + \frac{\delta^{2}}{m+1} - 2\|c-c'\| (\|q-c'\|+ \|p-c\|)&\\
				&> \|p-c\|^{2} -{\varpi}(p)^{2} + \frac{\delta^{2}}{m+1} - \frac{B\lambda^{3}}{\reach}&
			\end{align*}
			where $B = 2^{15}$.
%
	\end{description}
	From {\bf Case-1} and {\bf 2}, we get
	$$
		\|q- c \|^{2} - {\varpi}(q)^{2} > \|p-c\|^{2} -{\varpi}(p)^{2} +
		\frac{\delta^{2}}{m+1} - \frac{B\lambda^{3}}{\reach}\, .
	$$
\end{proof}

\begin{lem}
    \label{lem-conditions-rDt-subseteq-witness}
    Let $W \subseteq \M$ be an $\e$-sample of $\M$, $L \subseteq W$
    be a $\lambda$-net
    of $W$ with $\e \leq \lambda$, and $\delta = \delta_{0}\lambda$.
    Also, let ${\varpi} : L \rightarrow [0, \infty)$ be a weight
    assignment with $\widetilde{\varpi} < \frac{1}{2}$ and
    satisfying conditions (1) to (2) of Lemma~\ref{lem-rDt-protected-M}.
    If $\delta = \delta_{0}\lambda$,
    $$
      \lambda < \min \, \left\{ \frac{\Gamma^{m}_{0}}{2^{11}}, \,
	  \frac{\delta^{2}_{0}}{B(m+1)} \right\}\, \reach
    $$
    and
    $$
      \e < \frac{\lambda}{24} \left( \frac{\delta^{2}_{0}}{m+1}
      - \frac{B\lambda}{\reach} \right),
    $$
    then
    $$
      \del_{{\varpi}}(L, T\M) \subseteq \wit_{{\varpi}}(L, W).
    $$
\end{lem}
\begin{proof}
    Note that, as $\e \leq \lambda$, $L$ is a $(\lambda, 2\lambda)$-net of
    $\M$.

    Let $\sigma^{k} \in \del_{{\varpi}}(L, \M)$. From Lemma~\ref{lem-rDt-protected-M},
    there exists $c \in \vor_{{\varpi}}(\sigma^{k})\cap \M$ such that $\sigma^{k}$ is
    $\delta^{2}_{1}$-protected at $c$, where
	$$    
    		\delta^{2}_{1}=\frac{\delta^{2}}{m+1}-\frac{B\lambda^{3}}{\reach}.
    $$
    From Lemma~\ref{lem-distance-point-on-M-with-sample}~(2)
    as $c \in \vor_{{\varpi}}(\sigma^{k})\cap \M$, we have for all
    $p \in \sigma^{k}$, $\| p - c\| \leq 4 \lambda$.

    Let $w \in W$ be such that $\|c-w\| \leq \e$. For all $q \in L \setminus \sigma^{k}$
    and $p \in \sigma^{k}$ we have
    \begin{align}\label{eqn-1-witnessing-rDt}
        \|p-w\|^{2} - {\varpi}(p)^{2} &\leq (\|p-c\|+\|c-w\|)^{2} - {\varpi}(p)^{2}& \nonumber\\
        &= \|p-c\|^{2}-{\varpi}(p)^{2}+ \|c-w\|\, (\|c-w\| + 2\|p-c\|)& \nonumber\\
        &\leq \|p-c\|^{2} - {\varpi}(p)^{2} + 9\e\lambda & \nonumber\\
        &< \|q-c\|^{2}-{\varpi}(q)^{2}-(\delta^{2}_{1} - 9\e\lambda)&\nonumber\\
        &\leq \|q-w\|^{2}-{\varpi}(q)^{2}+ \beta -(\delta^{2}_{1} - 9\e\lambda)&
    \end{align}
    Where $\beta = \|w-c\|\, (\|w-c\|+2\|q-w\|)$.

    We have to consider the following two case:
    \begin{enumerate}
        \item
            If $\|q-w\|^{2} > \|p-w\|^{2}+4\lambda^{2}$. Using the fact that
	    ${\varpi}(q) < 2\lambda$, from Lemma~\ref{lem-distance-point-on-M-with-sample}~(1),
	    we get
            \begin{eqnarray*}
                \|q-w\|^{2} -{\varpi}(q)^{2} &>& \|p-w\|^{2}+ 4\lambda^{2} -{\varpi}(q)^{2} \\
                &>& \|p-w\|^{2} \\
                &\geq& \|p-w\|^{2} -{\varpi}(p)^{2}
            \end{eqnarray*}

        \item
            If $\|q-w\|^{2} \leq \|p-w\|^{2}+ 4\lambda^{2}$. This implies
            \begin{eqnarray*}
                \|q-w\| &\leq& \|p-w\|+2\lambda \\ 
                &\leq& \|p-c\|+ \|c-w\| + 2\lambda \\
                &\leq& 7\lambda .
            \end{eqnarray*}
            Now, using Eq.~\eqref{eqn-1-witnessing-rDt} and the facts that
	    $\|q-w\| = 7\lambda$
            and $\|c-w\|\leq \e \leq \lambda$, we get
            \begin{align*}
                \|p-w\|^{2} - {\varpi}(p)^{2} &\leq
                \|q-w\|^{2}-{\varpi}(q)^{2}+ \beta-(\delta^{2}_{1} - 9\e\lambda)&\\
                &\leq \|q-w\|^{2}-{\varpi}(q)^{2} - (\delta^{2}_{1} - 24\e\lambda)&
                \mbox{as $\beta \leq 15\e\lambda$}\\
                &< \|q-w\|^{2}-{\varpi}(q)^{2}&
            \end{align*}
            The last inequality follows from the fact that 
			$$            
            \lambda < \frac{\delta^{2}_{0}\reach}{B\, m} \; \;
            \mbox{and} \; \; 
            \e < \frac{\lambda}{24} \left( \frac{\delta^{2}_{0}}{m} - \frac{B\lambda}{\reach} \right).
            $$
    \end{enumerate}
    This implis $w$ is a witness of $\sigma^{k}$.

    As this is true for all $\sigma^{k} \in \del_{{\varpi}}(L, T\M)$,
    we get $\del_{{\varpi}}(L, T\M) \subseteq \wit_{{\varpi}}(L, W)$.
\end{proof}

\section{Proof of Lemma~\ref{lem-length-forbidden-interval}}
\label{appendix-length-forbidden-interval}


\subsection{Outline of the proof}

We will use a variant of Pumping equation, Lemma~\ref{lem-pumping-equation},
from~\cite{cheng2000} and bound on the height of slivers,
Lemma~\ref{lem-sliver-height-bound}, from~\cite{boissonnat2012csdt}.
Let ${\varpi} : L \rightarrow [0, \infty)$ be a weight assignment
with $\widetilde{{\varpi}} \leq \tilde{\alpha}_{0}$, and $\sigma \subset L$
be a $\Gamma_{0}$-sliver incident to the point $p \in L$.
As in Lemma~\ref{lem-length-forbidden-interval},
${\varpi}_{1}$ is an ewp of ${\varpi}$ such that
$\sigma \in {\rm K}_{{\varpi}}(L)$.
To prove Lemma~\ref{lem-length-forbidden-interval}, we distinguish the
following two cases depending on the point whose weight
is changed when replacing ${\varpi}$ by ${\varpi}_{1}$:
\begin{description}
  \item[Case 1.]
    The point whose weight is changed is $p$. Lemma~\ref{lem-forbidden-interval-cocone}
    takes care of this case and states that
    $$
      {\varpi}(p)^{2} \in J_{{\varpi}}(p, \sigma) =
      \left[ F_{{\varpi}}(p,\sigma)-\frac{\eta_{1}}{2}
      -\delta^{2}_{0}\lambda^{2},\, F_{{\varpi}}(p,\sigma) + \frac{\eta_{1}}{2} \right],
    $$
    for some $\eta_{1} \leq \eta - 2\delta^{2}_{0}\lambda^{2}$.

  \item[Case 2.]
    The point whose weight is changed is {\em not} $p$.
    Lemma~\ref{lem-length-forbidden-interval} takes care of this case
    and states that
    $$
      {\varpi}(p)^{2} \in  I_{{\varpi}}(\sigma,p) =
      \left[ F_{{\varpi}}(p,\sigma) - \frac{\eta}{2},
      \, F_{{\varpi}}(p,\sigma) +   \frac{\eta}{2} \right] .
    $$
\end{description}
Since $J_{{\varpi}}(\sigma,p) \subset I_{{\varpi}}(\sigma,p)$,
Lemma~\ref{lem-length-forbidden-interval} is proved.

The proof of {\bf Case 1} is in the same vein as the proofs
of~\cite[Lem.~10]{cheng2005}
and~\cite[Lem.~4.14]{boissonnat2011tancplx}.

The main technical ingredient in completing the proof of {\bf Case 2}
is in showing that
$$
  \left| F_{{\varpi}}(p,\sigma) - F_{{\varpi}_{1}}(p,\sigma) \right|
  = O\left( \frac{\delta^{2}_{0}\lambda^{2}}{\Gamma_{0}^{m}} \right)
$$	
One way to proving this is by proving
$$
	\max \Big\{
	\left| R_{{\varpi}}(\sigma_{p})^{2} - R_{{\varpi}_{1}}(p,\sigma)^{2}\right|, \,
	\left| d(p,N_{{\varpi}}(\sigma_{p}))^{2} - d(p,N_{{\varpi}_{1}}(p,\sigma))^{2}\right|
	\Big\} =
	O\left( \frac{\delta^{2}_{0}\lambda^{2}}{\Gamma_{0}^{m}} \right),
$$
and this will be done in
Lemma~\ref{lem-forbidden-interval-complicated-forbidden-interval}
using Lemma~\ref{lem:prop-omega-omega1-C-E}
and Corollary \ref{cor-bound-omega-E-R-sigma},


\subsection{Details of the proof}

For the rest of this section we will assume the following hypothesis
\begin{hyp}
  $L \subset \M$ is a $(\lambda, 2\lambda)$-net of $\M$ with
  $$
    \lambda < \frac{1}{18} (1-\sin\theta_{0})^{2} \,\reach.
  $$
\end{hyp}

For a simplex $\sigma$ and a vertex $p \in \sigma$, {\em excentricity}
$H_{{\varpi}}(p, \sigma)$ of $\sigma$ with respect to $p$ is the signed
distance of $C_{{\varpi}}(\sigma)$ from $\aff (\sigma_{p})$, i.e.,
$H_{{\varpi}}(p, \sigma)$ is positive if $C_{{\varpi}}(\sigma)$ and $p$
lie on the same side of $\aff(\sigma_{p})$ and negative if they lie on
different sides of $\aff (\sigma_{p})$.

The following lemma is a variant of the pumping equation
from~\cite{cheng2000,boissonnat2011tancplx,cheng2005}.

\begin{figure}
  \begin{center}
  	\quad\quad\quad\quad\quad\quad
    \includegraphics[width=4.250in]{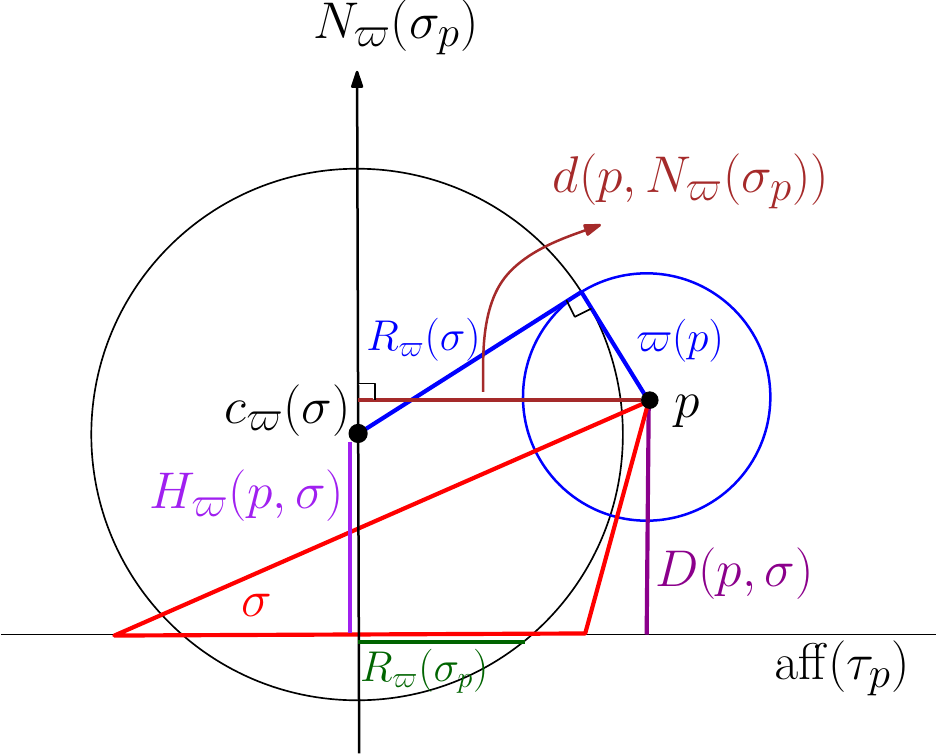}
  \end{center}
  \caption{Diagram for the Lemma~\ref{lem-pumping-equation}.}
  \label{figure-pumping-lemma}
\end{figure}

\begin{lem}[Pumping equation, see Figure~\ref{figure-pumping-lemma}]\label{lem-pumping-equation}
	We will assume that the weight of $p$ is varying and the weight
	of the other vertices of $\sigma$ are fixed. Then
	$$
        2 D(p, \sigma)\, H_{{\varpi}}(p, \sigma)
        = F_{{\varpi}}(p, \sigma) - {\varpi}(p)^{2} .
	 $$	
\end{lem}
The above ``pumping equation'' will be used to bound the length of the
forbidden intervals.

The following result is from~\cite{boissonnat2012csdt}.

\begin{lem}[Sliver altitude bound]\label{lem-sliver-height-bound}
    If a $(k+1)$-simplex $\tau$ is a $\Gamma_{0}$-sliver, then for any vertex $p$ of
    $\sigma$ we have
    $$
        D(p, \sigma) < \frac{2\Gamma_{0}\Delta(\sigma)^{2}}{L(\sigma)} .
    $$
\end{lem}

A variant of the following result can be found
in~\cite[Lem.~10]{cheng2005} and~\cite[Lem.~4.14]{boissonnat2011tancplx}.
We have included the proof for completeness.

\begin{lem}[Case 1]\label{lem-forbidden-interval-cocone}
    Let ${\varpi}: L \rightarrow [0, \infty]$ be a weight assignment
    with $\widetilde{{\varpi}} \leq \tilde{\alpha}_{0}$, and $\sigma \subset L$
    be a $\Gamma_{0}$-sliver incident to the point $p \in L$.
%
%
    Let ${\varpi}_{1}$ be a ewp of ${\varpi}$ satisfying the following
    conditions
    $$
        {\varpi}(q) = {\varpi}_{1}(q), ~\forall~q \in \sigma\setminus p,
        ~\mbox{and}~\sigma \in {\rm K}_{{\varpi}_{1}}(L).
    $$
    If $\lambda$ is sufficiently small then
    $$
        {\varpi}(p)^{2} \in
	\left[ F_{{\varpi}}(p,\sigma) - \frac{\eta_{1}}{2} - \delta_{0}^{2}\lambda^{2},
        \, F_{{\varpi}}(p,\sigma) - \frac{\eta_{1}}{2} \right]
    $$
    where $\eta_{1} \stackrel{\rm def}{=} 2^{14}\Gamma_{0} \lambda^{2}$.
%
\end{lem}
\begin{proof}
%
    Since $|H_{{\varpi}_{1}}(p, \sigma)| \leq \|C_{{\varpi}_{1}}(\sigma) - p\|$, we have from
    Lemma~\ref{lem-property-cocone-complex}~(1)
    $$
        |H_{{\varpi}_{1}}(p, \sigma)| \leq \| C_{{\varpi}_{1}}(\sigma) - p \| < 8\lambda .
    $$
    Since $L$ is $\lambda$-sparse,
    we have from Lemma~\ref{lem-property-cocone-complex}~(2)
    $$
        \lambda \leq L(\sigma) \leq \Delta(\sigma) < 16\lambda
    $$
    From Lemma~\ref{lem-sliver-height-bound}, we have
    $$
        D(p, \sigma) < \frac{2\Gamma_{0}\Delta(\sigma)^{2}}{L(\sigma)} < 2^{9}\Gamma_{0}\lambda.
    $$
    Therefore, using Lemma~\ref{lem-pumping-equation}, we have
    \begin{eqnarray*}
        F_{{\varpi}_{1}}(p, \sigma) - 2D(p, \sigma)|H_{{\varpi}_{1}}(p, \sigma)|
        \leq &{\varpi}_{1}(p)^{2}& \leq  F_{{\varpi}_{1}}(p, \sigma) +
	2D(p, \sigma)|H_{{\varpi}_{1}}(p, \sigma)| \\
        F_{{\varpi}_{1}}(p, \sigma) - 2^{13}\Gamma_{0}\lambda^{2}  \leq
        &{\varpi}_{1}(p)^{2}& \leq  F_{{\varpi}_{1}}(p, \sigma) + 2^{13}\Gamma_{0} \lambda^{2}
    \end{eqnarray*}
    The result now follows from the facts that
    \begin{itemize}
      \item
	$F_{{\varpi}_{1}}(p, \sigma) = F_{{\varpi}}(p, \sigma)$ as,
	from the definition, $F_{{\varpi}_{1}}(p, \sigma)$
	(and $F_{{\varpi}}(p, \sigma)$) depends only on the weights of
	the vertices in $\sigma_{p}$ and for all
	$q \in \sigma\setminus p$, ${\varpi}(q) = {\varpi}_{1}(q)$.

      \item
	${\varpi}_{1}(p)^{2} \in [ {\varpi}(p)^{2}, {\varpi}(p)^{2} + \delta^{2}_{0}\lambda^{2}]$.
    \end{itemize}
\end{proof}


The following lemma show the stability of weighted centers of well
shaped simplices under small perturbations of weight assignments.
The proof is in the same vein as the proof
of~\cite[Lem.~4.1]{boissonnat2012sdt}, and will use singular values
of matrices associated with the simplices.
\begin{lem}\label{lem:prop-omega-omega1-C-E}
    Let $\sigma$ be a simplex with $L(\sigma) \geq \lambda$ and
    $\Upsilon(\sigma)> 0$, and
    $\xi_{i}: \sigma \rightarrow [0, \infty)$,
    with $i \in \{1,\, 2\}$, be weights assignments,
    with $\tilde{\xi}_{i}\leq \alpha_{0}$,
    satisfy the following properties: $\exists~ p \in \sigma$ such that
    \begin{enumerate}
        \item
            $\forall ~q \in \sigma \setminus p$, $\xi_{1}(q) = \xi_{2}(q)$, and

        \item
            $|\xi_{1}(p)^{2} - \xi_{2}(p)^{2}| \leq \delta_{0}^{2} \lambda^{2}$.
    \end{enumerate}
    Then
    $$
        \|C_{\xi_{1}}(\sigma) - C_{\xi_{2}}(\sigma)\|\leq \frac{\delta_{0}^{2}\lambda}{2\Upsilon(\sigma)} ,
    $$
    and for $r \not\in  \sigma$, we have
    $$
	  \big|d(r,N_{\xi_{1}}(\sigma)) - d(r,N_{\xi_{2}}(\sigma)) \big|
	\leq \frac{\delta^{2}_{0} \lambda}{2\Upsilon(\sigma)}
    $$
\end{lem}

\begin{figure}
  \begin{center}
    \includegraphics[width=4.50in]{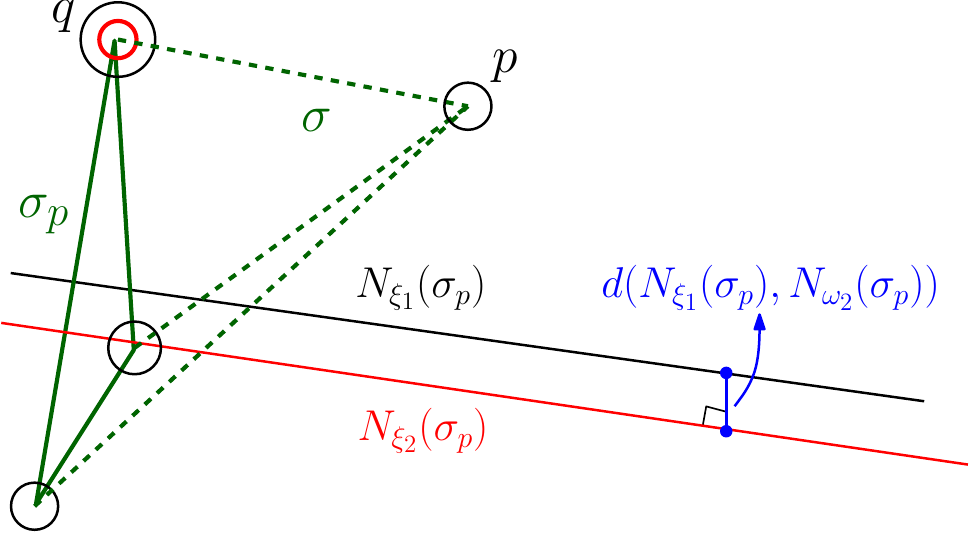}
  \end{center}
  \caption{Diagram for the Lemma~\ref{lem:prop-omega-omega1-C-E}.}
\end{figure}

\begin{proof}
    Let $\sigma = [p_{0}\, \dots \, p_{k}]$, and wlog let $p \neq p_{0}$
    and $\sigma \subset \R^{k}$.
    The ortho-radius of $\sigma$ satisfy the following system of
    $k$-linear equations:
    $$
        (p_{j} - p_{0})^{{\rm T}}C_{\xi_{i}}(\sigma) = \frac{1}{2} (\|p_{j}\|^{2}-\xi_{i}(p_{j})^{2}
        -\|p_{0}\|^{2}+\xi_{i}(p_{0})^{2}).
    $$
    Rewriting the above system of equation we get
    \begin{eqnarray*}
        (p_{j} - p_{0})^{{\rm T}}(C_{\xi_{2}}(\sigma) - C_{\xi_{1}}(\sigma))
        &=& \frac{1}{2}(\xi_{1}(p_{j})^{2} -\xi_{2}(p_{j})^{2}+ \xi_{2}(p_{0})^{2} - \xi_{1}(p_{0})^{2})\\
        &=& \frac{1}{2} (\xi_{1}(p_{j})^{2} - \xi_{2}(p_{j})^{2})~~\mbox{as $p \neq p_{0}$.}
    \end{eqnarray*}
    Letting $P$ be a $k\times k$ matrix whose $j^{th}$ column is $(p_{j} - p_{0})$, we have
    $$
        P^{T}(C_{\xi_{2}}(\sigma) - C_{\xi_{1}}(\sigma)) = \frac{x_{\xi}}{2}
    $$
    where
    $$
        x_{\xi} = (\xi_{1}(p_{1})^{2}-\xi_{2}(p_{1})^{2}, \, \dots, \,
        \xi_{1}(p_{k})^{2}-\xi_{2}(p_{k})^{2})^{{\rm T}}.
    $$
    
    Therefore
    \begin{align*}
        \|C_{\xi_{2}}(\sigma) - C_{\xi_{1}}(\sigma)\| &= \frac{1}{2}\|P^{-{\rm T}}x_{\xi}\| &\\
        &\leq \frac{1}{2} \|P^{-1}\|\, \|x_{\xi}\|& \\
        &\leq s_{1}(P^{-1}) \times \frac{\delta^{2}_{0}\lambda^{2}}{2}&
        \mbox{as $s_{1}(P^{-1}) = \| P^{-1}\|$ and $\|x_{\xi}\| \leq \delta^{2}_{0}\lambda^{2}$}\\
        &= s_{k}(P)^{-1} \times \frac{\delta^{2}_{0}\lambda^{2}}{2}&
        \mbox{as $s_{1}(P^{-1}) = s_{k}(P)^{-1}$, Lemma~\ref{lem-singular-value-1-m-matrix-inverse}}\\
        &\leq \frac{\delta^{2}_{0}\lambda^{2}}{2\sqrt{k} \Upsilon(\sigma) \Delta(\sigma)}&
        \mbox{as $s_{k}(P) \geq \sqrt{k} \Upsilon(\sigma) \Delta(\sigma)$,
        Lemma~\ref{lem:bound.skP}}\\
        &\leq \frac{\delta^{2}_{0}\lambda}{2\Upsilon(\sigma)}&
	\mbox{as $\frac{\Delta(\sigma)}{\lambda} \geq 1$.}
    \end{align*}

    The bound on
    $\big|d(r,N_{\xi_{1}}(\sigma)) - d(r,N_{\xi_{2}}(\sigma))\big|$
    follows directly from the
    part 1 of the lemma and the fact that
    $$
	\big|d(r,N_{\xi_{1}}(\sigma)) - d(r,N_{\xi_{2}}(\sigma))\big|
	\leq
	\|C_{\xi_{1}}(\sigma) - C_{\xi_{2}}(\sigma)\|.
    $$
\end{proof}

\begin{cor}\label{cor-bound-omega-E-R-sigma}
    Let ${\varpi}: L \rightarrow [0, \infty]$ be a weight assignment
    with $\tilde{{\varpi}} \leq \tilde{\alpha}_{0}$, and $\sigma \subset L$
    be a $j$-dimensional
    $\Gamma_{0}$-sliver with $p \in \sigma$ and $j \leq m+1$.
    In addition, we assume
    $$
      \frac{\delta^{2}_{0}}{\Gamma^{m}_{0}} \leq 2.
    $$
%
%
    If ${\varpi}_{1}$ be an ewp of ${\varpi}$
    satisfying the following:
    $\exists~q \in \sigma_{p}$
    such that $\forall~x \in L \setminus \{q\}$, ${\varpi}(x) = {\varpi}_{1}(x)$
    and $\sigma \in {\rm K}_{{\varpi}_{1}}(L, \M)$.
%
%
    Then
    $$
        \left|d(p,N_{{\varpi}}(\sigma_{p}))^{2} - d(p,N_{{\varpi}_{1}}(\sigma_{p}))^{2}
	\right|, \,
        \left| R_{{\varpi}}(\sigma_{p})^{2}-R_{{\varpi}_{1}}(\sigma_{p})^{2}
	\right| \leq
	\frac{49\delta^{2}_{0}\lambda^{2}}{2\Gamma^{m}_{0}} .
    $$
%
%
\end{cor}
\begin{proof}
    Using the fact that $L$ is an $(\lambda, 2\lambda)$-net of $\M$, and
    from Lemmas~\ref{lem-property-cocone-complex}~(1) and (2) we have
    \begin{eqnarray*}
	d(p, N_{{\varpi}_{1}}(\sigma_{p}))
	&\leq&  \|C_{{\varpi}_{1}}(\sigma_{p})-q\| + \|p-q\|\\
        &\leq& 24\lambda.
    \end{eqnarray*}

    From Lemma~\ref{lem:prop-omega-omega1-C-E} we have
    \begin{align*}
	d(p,N_{{\varpi}}(\sigma_{p})) + d(p,N_{{\varpi}_{1}}(\sigma_{p})) &\leq
        2d(p,N_{{\varpi}_{1}}(\sigma_{p}))  + \frac{\delta_{0}^{2}\lambda }{2\Upsilon(\sigma_{p})}& \\
        &\leq 2d(p,N_{{\varpi}_{1}}(\sigma_{p})) + \frac{\delta^{2}_{0} \lambda}{2\Gamma^{m}_{0}}&
	\mbox{as $\sigma$ is a $\Gamma_{0}$-sliver}\\
        &\leq 49\lambda.&
    \end{align*}

    From Lemma~\ref{lem:prop-omega-omega1-C-E} and the fact that
    $d(p, N_{{\varpi}}(\sigma_{p})) + d(p, N_{{\varpi}_{1}}(\sigma_{p}))
    \leq 49\lambda $, we have
    $$
	\big| d(p, N_{{\varpi}}(\sigma_{p}))^{2} -
	d(p, N_{{\varpi}_{1}}(\sigma_{p}))^{2} \big|
        \leq \frac{49\delta^{2}_{0}\lambda^{2}}{2\Gamma_{0}^{m}}
    $$

	Since $\sigma$ is a $\Gamma_{0}$-sliver, $j \geq 2$ (see
	Remark~\ref{remark-good-bad-simplex}).
    As $j \geq 2$, there exists $r \in \sigma_{p} \setminus q$. This implies
    ${\varpi}(r) = {\varpi}_{1}(r)$.

    Using the facts that ${\varpi}_{1}(r) = {\varpi}(r)$,
    $\|C_{{\varpi}_{1}}(\sigma_{p}) -r \| \leq 8\lambda$ (from
    Lemma~\ref{lem-property-cocone-complex}~(1)) and
    $$
      \|C_{{\varpi}}(\sigma_{p}) -C_{{\varpi}_{1}}(\sigma_{p})\| \leq
    \frac{\delta^{2}_{0}\lambda}{2\Gamma_{0}^{m}}
    $$
    (from Lemma~\ref{lem:prop-omega-omega1-C-E} and the fact that
    $\Upsilon(\sigma_{p}) \geq \Gamma_{0}^{j-1} \geq \Gamma_{0}^{m}$), we get
    \begin{eqnarray*}
        R_{{\varpi}}(\sigma_{p})^{2} &=& \|C_{{\varpi}}(\sigma_{p}) -r \|^{2} - {\varpi}(r)^{2}\\
        &\leq& (\|C_{{\varpi}_{1}}(\sigma_{p}) -r \| + \|C_{{\varpi}}(\sigma_{p}) -C_{{\varpi}_{1}}(\sigma_{p})\|)^{2}
        - {\varpi}_{1}(r)^{2}\\
        &\leq& R_{{\varpi}_{1}}(\sigma_{p})^{2}+ \Big(2\|C_{{\varpi}_{1}}(\sigma_{p}) - r\| \\
	&&\quad\quad\quad\quad\quad\quad\quad\quad
	+ \; \|C_{{\varpi}}(\sigma_{p}) - C_{{\varpi}_{1}}(\sigma_{p})\| \Big) \,
        \|C_{{\varpi}}(\sigma_{p}) - C_{{\varpi}_{1}}(\sigma_{p})\|\\
        &\leq& R_{{\varpi}_{1}}(\sigma_{p})^{2}+ \left(16\lambda +
        \frac{\delta^{2}_{0}\lambda}{2\Gamma_{0}^{m}}\right) \, \frac{\delta^{2}_{0}\lambda}{2\Gamma^{m}_{0}}\\
        &\leq& R_{{\varpi}_{1}}(\sigma_{p})^{2} + \frac{17\delta^{2}_{0}\lambda^{2}}{2\Gamma^{m}_{0}}
    \end{eqnarray*}
    and
    \begin{eqnarray*}
        R_{{\varpi}}(\sigma_{p})^{2} &=& \|C_{{\varpi}}(\sigma_{p}) -r \|^{2} - {\varpi}(r)^{2}\\
        &\geq& (\|C_{{\varpi}_{1}}(\sigma_{p}) -r \| - \|C_{{\varpi}}(\sigma_{p}) -C_{{\varpi}_{1}}(\sigma_{p})\|)^{2}
        - {\varpi}_{1}(r)^{2}\\
        &\geq& R_{{\varpi}_{1}}(\sigma_{p})^{2}
        - \Big(2\|C_{{\varpi}_{1}}(\sigma_{p}) - r\| \\
	&&\quad\quad\quad\quad\quad\quad\quad\quad
	-\; \|C_{{\varpi}}(\sigma_{p}) - C_{{\varpi}_{1}}(\sigma_{p})\|\Big)\,
        \|C_{{\varpi}}(\sigma_{p}) - C_{{\varpi}_{1}}(\sigma_{p})\|\\
        &\geq& R_{{\varpi}_{1}}(\sigma_{p})^{2} - 2 \|C_{{\varpi}_{1}}(\sigma_{p}) - r\| \,
        \|C_{{\varpi}}(\sigma_{p}) -C_{{\varpi}_{1}}(\sigma_{p})\| \\
        &\geq& R_{{\varpi}_{1}}(\sigma_{p})^{2} - \frac{8\delta^{2}_{0}}{\Gamma_{0}^{m}} \lambda^{2}
    \end{eqnarray*}
\end{proof}

\begin{lem}[Case 2]\label{lem-forbidden-interval-complicated-forbidden-interval}
%
%
    Assuming the same
    conditions on ${\varpi}$, ${\varpi}_{1}$, $p$, $q$ and $\sigma$ as in
    Corollary~\ref{cor-bound-omega-E-R-sigma}, we get
    $$
        {\varpi}(p)^{2} \in \left[ F_{{\varpi}}(p,\sigma) - \frac{\eta}{2},
        \, F_{{\varpi}}(p,\sigma) + \frac{\eta}{2} \right] .
    $$
\end{lem}
\begin{proof}
    As in the proof of Lemma~\ref{lem-forbidden-interval-cocone}, we can show that
    $|2D(p, \sigma)H_{{\varpi}_{1}}(p, \sigma)| \leq 2^{13}\Gamma_{0}\lambda$.


    From Corollary~\ref{cor-bound-omega-E-R-sigma}, we have
    $$
        \left|d(p, N_{{\varpi}}(\sigma_{p}))^{2} -
	d(p, N_{{\varpi}_{1}}(\sigma_{p}))^{2}\right|, \,
        \left|R_{{\varpi}}(\sigma_{p})^{2} - R_{{\varpi}_{1}}(\sigma_{p})^{2}
	\right|
        \leq \frac{49\delta^{2}_{0}}{2\Gamma^{m}_{0}} \lambda^{2} .
    $$
    This implies, from the definition of $F_{{\varpi}_{1}}(p, \sigma)$,
    $$
        \big| F_{{\varpi}_{1}}(p, \sigma) - F_{{\varpi}}(p, \sigma) \big|
        \leq \frac{49\delta^{2}_{0}}{\Gamma_{0}^{m}} \lambda^{2} .
    $$

    From Lemma~\ref{lem-pumping-equation}, and the above bounds we get
    $$
        {\varpi}_{1}(p)^{2} \in \left[ F_{{\varpi}}(p,\sigma) -
	\frac{\eta}{2},\, F_{{\varpi}}(p,\sigma) + \frac{\eta}{2} \right]\, .
    $$
    The result now follows from the fact that ${\varpi}(p) = {\varpi}_{1}(p)$.
\end{proof}

\begin{remark}\label{remark-eta-eta_1}
    Note that $\eta \geq \eta_{1} + 2\delta^{2}_{0}\lambda^{2}$.
\end{remark}

Combining Lemmas~\ref{lem-forbidden-interval-cocone} and
\ref{lem-forbidden-interval-complicated-forbidden-interval},
completes the proof of Lemma~\ref{lem-length-forbidden-interval}.



\section{Almost normal flats intersecting submanifolds}
\label{appendix-normal-flats-intersecting-submanifolds}

%



\begin{figure}[h]
  \begin{center}
    \includegraphics[width=0.450\textwidth]{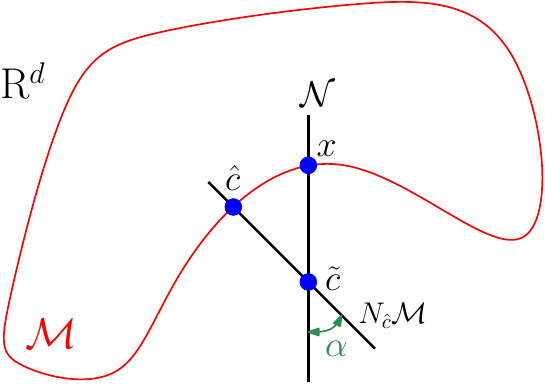}
  \end{center}
  \caption{Almost normal flat ${\mathcal N}$ intersecting the manifold $\M$.}
\end{figure}

The following technical lemma,
which asserts that, for $j\leq m= \dim \man$, if a $(d-j)$-flat, $\EuScript{N}$,
passes through a point $\tilde{c}$ that is close to $\man$, and
the normal space at the point on $\man$ closest to $\tilde{c}$ makes a
small angle with $\EuScript{N}$, then $\EuScript{N}$ must intersect $\man$ in that
vicinity. The technical difficulty stems from the fact that the
codimension may be greater than one.


\begin{lem}
  \label{lem:flat.intersect}
  Let $\tilde{c} \in \rdee$ be such that it has a unique closest point
  $\hat{c}$ on $\man$ and
  $\norm{\tilde{c}- \hat{c}} \leq \rho \leq\frac{\reach}{25}$.
  Let $j \leq m = \dim\man$, and let $\EuScript{N}$ be a
  $(d-j)$-dimensional affine flat passing through $\tilde{c}$ such
  that $\angle (\normspace{\hat{c}}{\man}, \EuScript{N}) \leq \alpha $ with $\sin
  \alpha \leq \frac{1}{4}$. Then there exists an $x \in \EuScript{N} \cap \M$ such
  that $\norm{\tilde{c} - x} \leq 4\rho$.
\end{lem}


The idea of the proof is to consider the $m$-dimensional affine space
$\widetilde{T}_{\hat{c}}\man$ that passes through $\hat{c}$ and is
orthogonal to a $(d-m)$-dimensional affine subspace of
$\EuScript{N}$. We show that the orthogonal projection onto
$\widetilde{T}_{\hat{c}}\man$ induces, in some neighbourhood $V$ of
$\hat{c}$, a diffeomorphism between $\man \cap V$, and
$\widetilde{T}_{\hat{c}}{\man} \cap V$ (\Lemref{lem:man.near.tan}). We use
$\tanspace{\hat{c}}{\man}$ as an intermediary in this calculation
(\Lemref{lem:tan.near.tan}). Then, since $\EuScript{N}$ intersects
$\tanspace{\hat{c}}{\man}$ near $\hat{c}$
(\Lemref{lem-new-section-1}), we can argue that it must also intersect
$\man$ because the established diffeomorphisms make a correspondence
between points along segments parallel to $\EuScript{N}$.

The final bounds are established in \Lemref{lem-structural-result},
from which \Lemref{lem:flat.intersect} follows by a direct
calculation, together with the following observations: If
$\dim \EuScript{N} =
\dim N_{\hat{c}}\man$, then $\angleop{N_{\hat{c}}\man}{\EuScript{N}} =
\angleop{\EuScript{N}}{N_{\hat{c}}\man}$, and if $\dim \EuScript{N} \geq \dim
N_{\hat{c}}\man$, then there is an affine subspace $\widetilde{\EuScript{N}} \subset
\EuScript{N}$, such that $\dim \widetilde{\EuScript{N}} = \dim N_{\hat{c}}\man$, and
$\angleop{N_{\hat{c}}\man}{\widetilde{\EuScript{N}}}
= \angleop{N_{\hat{c}}\man}{\EuScript{N}}$.
Indeed, we may take $\widetilde{\EuScript{N}}$ to be the orthogonal projection of
$N_{\hat{c}}\man$ into $\EuScript{N}$.

We now bound distances to the intersection of $\EuScript{N}$ and
$\tanspace{\hat{c}}{\man}$.
\begin{lem}\label{lem-new-section-1}
  Let $\tilde{c}$, $\hat{c}$ be points in $\R^{d}$ such that the
  projection of $\tilde{c}$ onto $\M$ is $\hat{c}$ and $\|\tilde{c} -
  \hat{c}\| \leq \rho$. Let $\EuScript{N}$ be a $d-m$ dimensional affine flat
  passing through $\tilde{c}$ such that $\angle (\EuScript{N}, N_{\hat{c}} {\cal
    M}) \leq \alpha$. For all $x \in \EuScript{N} \cap T_{\hat{c}}\M$, we have
	\begin{enumerate}
		\item
			$\|\tilde{c} - x\| \leq \frac{\rho}{\cos\alpha}$
		
		\item
			$\|\hat{c} -x \| \leq \left( 1+\frac{1}{\cos \alpha}\right)\, \rho$
	\end{enumerate}
\end{lem}

\begin{figure}[h]
  \begin{center}
    \includegraphics[width=3.50in]{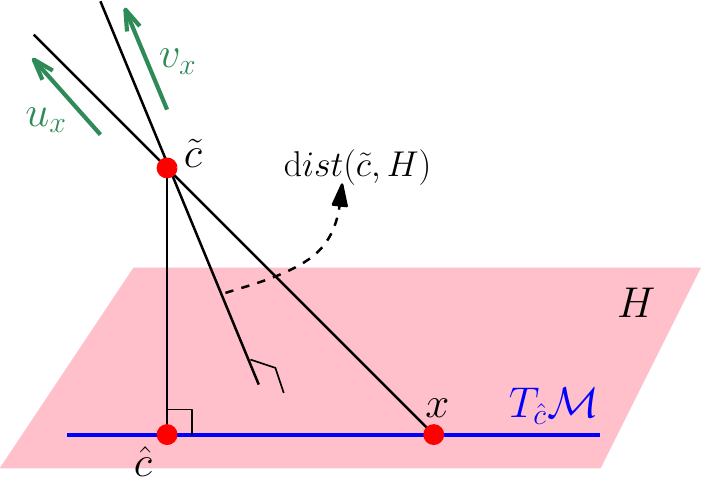}
  \end{center}
  \caption{Diagram for the Lemma~\ref{lem-new-section-1}.}
\end{figure}

\begin{proof}
  For a point $x \in \EuScript{N}\cap T_{\hat{c}}\M$, let $u_{x}$ denote the
  unit vector from $\tilde{c}$ to $x$, and let $v_{x} \in N_{\hat{c}}\M$
  be the unit vector that makes the smallest angle with $u_{x}$. Let
  $H$ denote the hyperplane passing through $\hat{c}$ and
  orthogonal to $v_{x}$.  Since $ \| \tilde{c} - \hat{c} \| \leq \rho $,
  ${\rm dist}(\tilde{c}, H) \leq \rho$. Therefore,
  \begin{equation*}
    \| \tilde{c} - x \| \leq  \frac{{\rm dist}(\tilde{c}, H)}{\cos\alpha}
  \end{equation*}
  and
  \begin{equation*}
    \| \hat{c} - x \| \leq \| \hat{c} - \tilde{c} \| + \| \tilde{c} -
    x  \| \leq \left( 1+ \frac{1}{\cos\alpha}\right) \, \rho
  \end{equation*}
\end{proof}

The following lemma is a direct consequence of the definition of the angle between
two affine spaces.

\begin{lem}
  \label{lem:tan.near.tan}
  Let $p$ be a point in $\M$ and let $\widetilde{T}_{p}\M$ denote a
  $m$-dimensional flat passing through $p$ with $\angle (T_{p}\M,
  \widetilde{T}_{p}\M) \leq \alpha < \frac{\pi}{2}$. If $f_{p}^{\alpha}$
  denote the orthogonal projection of $T_{p}\M$ onto
  $\widetilde{T}_{p}\M$, then
	\begin{enumerate}	
		\item
			The map $f^{\alpha}_{p}$ is bijective.
		\item
			For $r>0$, $f^{\alpha}_{p}(B_{p}(r))
			\supseteq \widetilde{B}_{p}(r\cos \alpha)$ where
			$B_{p}(r) = B(p,r) \cap T_{p}\M$ and
                        $\widetilde{B}_{p}(r) = B(p,r) \cap
                        \widetilde{T}_{p}\M$.
	\end{enumerate}
\end{lem}

\begin{lem}
  \label{lem:man.near.tan}
  Let $p$ be a point in $\M$, and let $\widetilde{T}_{p}\M$ be a
  $m$-dimensional affine flat passing through $p$ with $\angle
  (T_{p}\M, \widetilde{T}_{p}\M) \leq \alpha$. There exists an
  $r(\alpha)$ satisfying :
	\begin{equation*}
	    \frac{7\, r(\alpha)}{\reach} \, +\,  \sin \alpha \, < \, 1
	    \quad \mbox{and}\quad r(\alpha) \leq \frac{\reach}{10}
	\end{equation*}
	such that the orthogonal projection map, $g^{\alpha}_{p}$,  of
	 $B_{\M}(p, r(\alpha)) = B(p,r(\alpha)) \cap \man$ into
	  $\widetilde{T}_{p}\M$ satisfy the following conditions:
	\begin{enumerate}

		\item
			$g^{\alpha}_{p}$ is a diffeomorphism.
		
		\item
			$g^{\alpha}_{p} (B_{\M}(p,r(\alpha)))  \supseteq
			\widetilde{B}_{p}(r(\alpha)\cos\alpha_{1})$ where
			$\sin\alpha_{1}= \frac{r(\alpha)}{2 \reach} + \sin \alpha$.
		
		\item
			Let $x \in g^{\alpha}_{p} (B_{\M}(p,r(\alpha)))$, then
			$\|x-(g^{\alpha}_{p}) ^{-1}(x)\| \leq \|p-x\| \tan \alpha_{1}$.
			
	\end{enumerate}
\end{lem}
\begin{proof}
  1. Let $\pi_{\widetilde{T}_{p}\M}$ denote the orthogonal projection
  of $\R^{d}$ onto $\widetilde{T}_{p}\M$.  The derivative of this map,
  $D\pi_{\widetilde{T}_{p}\M}$, has a kernel of dimension $(d-m)$ that
  is parallel to the orthogonal complement of $\widetilde{T}_{p}\M$ in
  $\R^{d}$.

  We will first show that $D g^{\alpha}_{p}$ is nonsingular for all $x
  \in B_{\M}(p, r(\alpha))$.  From
  Lemma~\ref{lem-sampling-property-manifold}~(3) and the fact that $\angle
  (T_{p}\M, \widetilde{T}_{p}\M) \leq \alpha$, we have
	\begin{eqnarray*}
		\sin \angle (\widetilde{T}_{p}\M, T_{x}\M) & \leq &
                \sin \angle (T_{x}\M, T_{p}\M) + \sin \angle
                (T_{p}\M,\widetilde{T}_{p}\M) \\
		& \leq & \frac{6 r(\alpha)}{\reach} \, +\,  \sin \alpha \, < \, 1.
	\end{eqnarray*}
	Since $g^{\alpha}_{p}$ is the restriction of
        $\pi_{\widetilde{T}_{p}\M}$ to $B_{\M}(p, r(\alpha))$, the
        above inequality implies that $D g^{\alpha}_{p}$ is
        non-singular. Therefore, $g^{\alpha}_{p}$ is a local
        diffeomorphism.
	
	Let $x, \, y \in B_{\M}(p,r(\alpha))$. From
	Lemma~\ref{lem-sampling-property-manifold} part~(1) and
	(3), we have
	\begin{eqnarray*}
		\sin \angle ([x, y], \widetilde{T}_{p}\M) &\leq&
		\sin \angle ([x, y], {T}_{x}\M) + \sin \angle
                (T_{x}\M, T_{p}\M) + \sin \angle (\widetilde{T}_{p},
                T_{p}\M)\\
		&\leq& \frac{\|x-y\|}{2\reach} + \frac{6\|p-x\|}{\reach} + \sin\alpha \\
		&\leq& \frac{7 r(\alpha)}{\reach} + \sin \alpha \; < \;  1.
	\end{eqnarray*}
	This implies $g^{\alpha}_{p}(x) \neq g^{\alpha}_{p}(y)$.
	
	Since $g^{\alpha}_{p}$ is nonsingular and injective on
        $B_{\man}(p,r(\alpha))$, it is a diffeomorphism onto its image.
	
	2. Notice that, for $x \in B_{\M}(p,r(\alpha))$, the angle $\alpha_1$ is
        a bound on the angle between $\seg{p}{x}$ and
        $\tilde{T}_p\man$. The inclusion $g^{\alpha}_{p}
        (B_{\M}(p,r(\alpha)))  \supseteq
        \widetilde{B}_{p}(r(\alpha)\cos\alpha_{1})$ follows since
        $\seg{x}{g^{\alpha}_P(x)}$ is orthogonal to $\tilde{T}_p\man$.
	
	3. Follows similarly.
\end{proof}

\begin{lem}\label{lem-structural-result}
  Let $\tilde{c}$, $\hat{c}$ be points in $\R^{d}$ such that the
  projection of $\tilde{c}$ onto $\M$ is $\hat{c}$ and $\|\tilde{c} -
  \hat{c}\| \leq \rho$. Let $\EuScript{N}$ be a
  $d-m$ dimensional affine flat
  passing through $\tilde{c}$ such that
  $\angle (\EuScript{N}, N_{\hat{c}} {\cal M}) \leq \alpha$. If
	\begin{equation*}
		\rho \leq \frac{r(\alpha) \cos \alpha \cos \alpha_{1}}{1+\cos \alpha}
	\end{equation*}
	Then there exists an $x \in \EuScript{N} \cap \M$ such that
	\begin{equation*}
		\|\tilde{c} -x \| \leq \left( \frac{1}{\cos \alpha} +
                  \left( 1+ \frac{1}{\cos \alpha}\right)(\sin \alpha +
                  \sin \alpha_{1}) \right)\, \rho\, .
	\end{equation*}
\end{lem}
\begin{proof}
  Let $\tilde{T}_{\hat{c}}\M$ denote the orthogonal complement of $\EuScript{N}$ in
  $\R^{d}$ passing through $\hat{c}$. Note that
  $\angle (T_{\hat{c}}\M, \widetilde{T}_{\hat{c}}\M)
  = \angle (\EuScript{N}, N_{\hat{c}}{\cal M})$.
	
  Let $\hat{x} \in \EuScript{N} \cap T_{\hat{c}}\M $ and $\tilde{x} =
  f^{\alpha}_{\hat{c}}(\hat{x})$.  Then from
  Lemma~\ref{lem-new-section-1}, we have
	 \begin{equation*}
	 	\|\tilde{x} - \hat{c}\| \leq \|\hat{x} - \hat{c}\| \leq \left(1+\frac{1}{\cos \alpha}\right)\, 
	 	\rho
	 \end{equation*}
	 and
	\begin{equation*}
		\|\hat{x} - \tilde{x}\| \leq \|\hat{x} -\hat{c}\|\sin \alpha \leq
		\left ( \sin \alpha + \tan \alpha \right) \, \rho\, .
	\end{equation*}
	
	Using the fact that $\rho \leq \frac{r(\alpha) \cos \alpha \cos
          \alpha_{1}}{1+\cos \alpha}$, we have
        $$
	    \| \tilde{x} - \hat{c}\| \leq \left( 1+\frac{1}{\cos\alpha}\right)\, \rho \leq
	    r(\alpha) \cos\alpha_{1}.
        $$
        Therefore, from
        Lemma~\ref{lem-new-section-1}, there exists an $x \in B_{\M}(p,
        r(\alpha))$ such that $g^{\alpha}_{p}(x) = \tilde{x}$ and
	$$
        \|\tilde{x} - x \| \leq \| \tilde{x} - \hat{c} \| \tan
        \alpha_{1} \leq \left( 1+\frac{1}{\cos\alpha}\right) \tan
        \alpha_{1}\,\rho .
        $$
	Therefore
	\begin{eqnarray*}
		\|\tilde{c} - x\| &\leq& \|\tilde{c} -\hat{x}\| +
                \|\hat{x} - \tilde{x}\| + \| \tilde{x} - x\| \\
		& \leq & \frac{\rho}{\cos \alpha} +
                \left(1+\frac{1}{\cos \alpha} \right) (\sin \alpha +
                \tan \alpha_{1}) \, \rho \\
		& = & \left( \frac{1}{\cos \alpha} + \left( 1+
                    \frac{1}{\cos \alpha}\right)(\sin \alpha + \tan
                  \alpha_{1}) \right)\, \rho .
	\end{eqnarray*}
	
	Note that the line segment $\tilde{c}x \in \EuScript{N}$.
\end{proof}

This completes the proof of Lemma~\ref{lem:flat.intersect}.

\phantomsection
\bibliographystyle{alpha}
\addcontentsline{toc}{section}{Bibliography}
\bibliography{biblio}

\end{document}